\documentclass[11pt]{article}

\title{8D $\mathcal{N}=2$ SUGRA}

\global\arraycolsep=1pt
\usepackage[colorlinks=true,backref=true,linkcolor=black,anchorcolor=black,citecolor=black,filecolor=black,menucolor=black,pagecolor=black,urlcolor=black]{hyperref}

\usepackage{xcolor}
\usepackage{tikz}
\usetikzlibrary{arrows}
\usepackage{pgfplots}
\usepackage{amsmath}
\usepackage{amsthm}
\usepackage{amssymb}
\usepackage{mathrsfs}
\usepackage[Symbol]{upgreek}
\usepackage{dsfont}
\usepackage[vcentermath]{youngtab}
\usepackage{textcomp}
\usepackage{bbold}

\usepackage{latexsym}
\usepackage{graphicx}

\newcommand{\eprint}[1]{{\href{http://arxiv.org/abs/#1}{\texttt{[#1}]}}}
\newcommand{\eprintN}[1]{{\href{http://arxiv.org/abs/#1}{\texttt{#1 [hep-th]}}}}

\def\DJo{$\;$\kern-.4em \hbox{D\kern-.8em\raise.15ex\hbox{--}\kern.35em okovi\'c}}

\def\DEVIII#1#2#3#4#5#6#7#8{{\tiny $ { \left[ \begin{array}{ccccccc}  & & \mathfrak{#2} \hspace{-1.2mm}&&&& \vspace{ -1.2mm} \\ \mathfrak{#1}\hspace{-1.2mm} &  \mathfrak{#3} \hspace{-1.2mm}& \mathfrak{#4} \hspace{-1.2mm} & \mathfrak{#5}\hspace{-1.2mm}&\mathfrak{#6}\hspace{-1.2mm}&\mathfrak{#7}\hspace{-1.2mm}&\mathfrak{#8} \end{array}\right] }$}}
\def\DSOXVI#1#2#3#4#5#6#7#8{{\tiny $ {  \vspace{-2mm} \left[ \begin{array}{ccccccccc}  && \mathfrak{#8} \hspace{-1.2mm}&&&&&& \vspace{ -1.3mm} \\ \cdot \hspace{-1.0mm}& \mathfrak{#7}\hspace{-1.2mm} &\mathfrak{#6}\hspace{-1.2mm} &  \mathfrak{#5} \hspace{-1.2mm}& \mathfrak{#4} \hspace{-1.2mm} & \mathfrak{#3}\hspace{-1.2mm}&\mathfrak{#2}\hspace{-1.2mm}&\mathfrak{#1}  \hspace{-1.2mm}\end{array}\right] }$}}

\def\DEVII#1#2#3#4#5#6#7{{\tiny $ { \left[ \begin{array}{cccccc}  & & \mathfrak{#2} \hspace{-1.2mm}&&& \vspace{ -1.0mm} \\ \mathfrak{#1}\hspace{-1.2mm} &  \mathfrak{#3} \hspace{-1.2mm}& \mathfrak{#4} \hspace{-1.2mm} & \mathfrak{#5}\hspace{-1.2mm}&\mathfrak{#6}\hspace{-1.2mm}&\mathfrak{#7} \end{array}\right] }$}}

\def\DSOX#1#2#3#4#5{{\tiny $ {   \biggl[ \begin{array}{ccc}  &&\mathfrak{#3}  \vspace{ -1.5mm} \\  \mathfrak{#1}\hspace{0.2mm}\mathfrak{#2}\hspace{-0.6mm} &\mathfrak{#4} \hspace{-0.9mm}&\vspace{-1.5mm}\\ && \mathfrak{#5}  \end{array}\biggr] }$}}

\def\DSOXII#1#2#3#4#5#6#7{{\tiny $ {   \left[ \begin{array}{ccccccc}  && & \mathfrak{#6} \hspace{-1.2mm}&&& \vspace{ -1.5mm} \\  \mathfrak{#1}\hspace{-1.2mm} &\mathfrak{#2}\hspace{-1.2mm} &  \mathfrak{#3} \hspace{-1.2mm}& \mathfrak{#4} \hspace{-1.2mm} & \mathfrak{#5}\hspace{-1.2mm}&\cdot \hspace{-0.5mm}& \mathfrak{#7} \end{array}\right] }$}}

\def\DSON#1#2#3#4#5#6#7#8{{\tiny $ {   \left[ \begin{array}{cccccccc}  &&&&& & \mathfrak{#7} \hspace{-1.2mm}& \vspace{ -1.5mm} \\  \mathfrak{#1}\hspace{-1.2mm} &\mathfrak{#2}\hspace{-1.2mm} &  \mathfrak{#3} \hspace{-1.2mm}& \mathfrak{#4} \hspace{-1.2mm} & \cdot \hspace{-0.8mm}\cdot\hspace{-0.8mm}\cdot \hspace{-1.2mm} & \mathfrak{#5}\hspace{-1.2mm}& \mathfrak{#6}\hspace{-1.2mm} & \mathfrak{#8} \end{array}\right] }$}}

\def\EiEVII#1{{E_{\fontsize{6.35pt}{6pt}\selectfont   \left[ \begin{array}{cccccc}  & & \mathfrak{0} \hspace{-0.6mm}&&& \vspace{ -1.0mm} \\ #1 \hspace{-0.6mm} &  \mathfrak{0} \hspace{-0.6mm}& \mathfrak{0} \hspace{-0.6mm} & \mathfrak{0}\hspace{-0.6mm}&\mathfrak{0}\hspace{-0.6mm} & \mathfrak{0} \end{array}\right] \fontsize{12.35pt}{12pt}\selectfont }}}

\def\EiEVI#1{{E_{\fontsize{6.35pt}{6pt}\selectfont   \left[ \begin{array}{cccccc}  & & \mathfrak{0} \hspace{-0.6mm}&& \vspace{ -1.0mm} \\ #1 \hspace{-0.6mm} &  \mathfrak{0} \hspace{-0.6mm}& \mathfrak{0} \hspace{-0.6mm} & \mathfrak{0}\hspace{-0.6mm} & \mathfrak{0} \end{array}\right] \fontsize{12.35pt}{12pt}\selectfont }}}

\def\WSOXVI#1#2#3#4#5#6#7#8{{\tiny $ {   \biggl[ \begin{array}{ccc}  &&\mathfrak{#3}  \vspace{ -1.5mm} \\  \mathfrak{#6}\hspace{0.2mm}  \mathfrak{#2}\hspace{0.2mm}  \mathfrak{#3}\hspace{0.2mm}  \mathfrak{#4}\hspace{0.2mm}\mathfrak{#5}\hspace{-0.6mm} &\mathfrak{#7} \hspace{-0.9mm}&\vspace{-1.5mm}\\ && \mathfrak{#8}  \end{array}\biggr] }$}}

\def\WSOXVInk#1#2#3#4#5#6#7#8{{\tiny $ {   \biggl[ \begin{array}{ccc}  &&\mathfrak{#3}  \vspace{ -1.5mm} \\  \mathfrak{#6}\hspace{0.2mm}  \mathfrak{#2}\hspace{0.2mm}  \mathfrak{#3}\hspace{0.2mm}  {#4}\hspace{0.2mm}\mathfrak{#5}\hspace{-0.6mm} &\mathfrak{#7} \hspace{-0.9mm}&\vspace{-1.5mm}\\ && {#8}  \end{array}\biggr] }$}}
\def\DSON#1#2#3#4#5#6#7#8{{\tiny $ {   \left[ \begin{array}{cccccccc}  &&&&& & \mathfrak{#7} \hspace{-1.2mm}& \vspace{ -1.5mm} \\  \mathfrak{#1}\hspace{-1.2mm} &\mathfrak{#2}\hspace{-1.2mm} &  \mathfrak{#3} \hspace{-1.2mm}& \mathfrak{#4} \hspace{-1.2mm} & \cdot \hspace{-0.8mm}\cdot\hspace{-0.8mm}\cdot \hspace{-1.2mm} & \mathfrak{#5}\hspace{-1.2mm}& \mathfrak{#6}\hspace{-1.2mm} & \mathfrak{#8} \end{array}\right] }$}}

\def\SU{SU_{\scriptscriptstyle \rm c}(8)}

\newfont{\bbbold}{msbm10 scaled \magstep1}

\def\cD{{\cal D}}
\def\cE{{\cal E}}
\def\cF{{\cal F}}

\def\cI{{\cal I}}

\def\cL{{\cal L}}

\def\cN{{\cal N}}
\def\cO{{\cal O}}

\def\cU{{\cal U}}
\def\cV{{\cal V}}

\newfont{\goth}{eufm10 scaled \magstep1}

\def\a{\alpha}\def\adt{{\dot \alpha}}
\def\bdt{{\dot \beta}}
\def\cdt{\dot\gamma}

\def\ve{\varepsilon}

\def\be{\begin{equation}}\def\ee{\end{equation}}
\def\bea{\begin{eqnarray}}\def\eea{\end{eqnarray}}
\def\barr{\begin{array}}\def\earr{\end{array}}

\def\hi{\hat{\imath}}\def\hj{\hat{\jmath}}

\def\nn{\nonumber}
\def\bd{\begin{document}}
\def\ed{\end{document}}
\def\ba{\begin{array}}
\def\ea{\end{array}}
\def\bea{\begin{eqnarray}}
\def\eea{\end{eqnarray}}
\def\ft#1#2{{\frac{\scriptstyle #1}{\scriptstyle #2}}}
\def\fft#1#2{\frac{#1}{#2}}
\def\sst#1{{\scriptscriptstyle #1}}
\def\oneone{\rlap 1\mkern4mu{\rm l}}
\def\DEVIII#1#2#3#4#5#6#7#8{{\tiny $ { \left[ \begin{array}{ccccccc}  & & \mathfrak{#2} \hspace{-0.7mm}&&&& \vspace{ -1.5mm} \\ \mathfrak{#1}\hspace{-0.7mm} &  \mathfrak{#3} \hspace{-0.7mm}& \mathfrak{#4} \hspace{-0.7mm} & \mathfrak{#5}\hspace{-0.7mm}&\mathfrak{#6}\hspace{-0.7mm}&\mathfrak{#7}\hspace{-0.7mm}&\mathfrak{#8} \end{array}\right] }$}}
\def\DSOXVI#1#2#3#4#5#6#7#8{{\tiny $ {  \vspace{-2mm} \left[ \begin{array}{ccccccccc}  && \mathfrak{#8} \hspace{-0.7mm}&&&&&& \vspace{ -1.5mm} \\ \cdot \hspace{-0.5mm}& \mathfrak{#7}\hspace{-0.7mm} &\mathfrak{#6}\hspace{-0.7mm} &  \mathfrak{#5} \hspace{-0.7mm}& \mathfrak{#4} \hspace{-0.7mm} & \mathfrak{#3}\hspace{-0.7mm}&\mathfrak{#2}\hspace{-0.7mm}&\mathfrak{#1} \end{array}\right] }$}}

\def\DEVII#1#2#3#4#5#6#7{{\tiny $ { \left[ \begin{array}{cccccc}  & & \mathfrak{#2} \hspace{-0.7mm}&&& \vspace{ -1.5mm} \\ \mathfrak{#1}\hspace{-0.7mm} &  \mathfrak{#3} \hspace{-0.7mm}& \mathfrak{#4} \hspace{-0.7mm} & \mathfrak{#5}\hspace{-0.7mm}&\mathfrak{#6}\hspace{-0.7mm}&\mathfrak{#7} \end{array}\right] }$}}
\def\DSOXII#1#2#3#4#5#6#7{{\tiny $ {   \left[ \begin{array}{ccccccc}  && & \mathfrak{#6} \hspace{-0.7mm}&&& \vspace{ -1.5mm} \\  \mathfrak{#1}\hspace{-0.7mm} &\mathfrak{#2}\hspace{-0.7mm} &  \mathfrak{#3} \hspace{-0.7mm}& \mathfrak{#4} \hspace{-0.7mm} & \mathfrak{#5}\hspace{-0.7mm}&\cdot \hspace{-0.5mm}& \mathfrak{#7} \end{array}\right] }$}}

\def\DSON#1#2#3#4#5#6#7#8{{\tiny $ {   \left[ \begin{array}{cccccccc}  &&&&& & \mathfrak{#7} \hspace{-0.7mm}& \vspace{ -1.5mm} \\  \mathfrak{#1}\hspace{-0.7mm} &\mathfrak{#2}\hspace{-0.7mm} &  \mathfrak{#3} \hspace{-0.7mm}& \mathfrak{#4} \hspace{-0.7mm} & \cdot \hspace{-0.8mm}\cdot\hspace{-0.8mm}\cdot \hspace{-0.7mm} & \mathfrak{#5}\hspace{-0.7mm}& \mathfrak{#6}\hspace{-0.7mm} & \mathfrak{#8} \end{array}\right] }$}}

\def\DlacedLeft{{\fontsize{0.004pt}{0.0005pt}\selectfont  \scriptscriptstyle \mbox{$= \hspace{-2.2mm}  \langle$} \fontsize{12pt}{14.5pt}\selectfont }}

\def\DSpIV#1#2#3#4{{\tiny $ { \left[   \mathfrak{#1}\,  \mbox{-} \mathfrak{#2} \, \mbox{-} \mathfrak{#3} \hspace{-0.2mm}\DlacedLeft \hspace{0.2mm} \mathfrak{#4} \right] }$}}

\def\DSOVIII#1#2#3#4{{\tiny $ {   \biggl[ \begin{array}{ccc}  &&\mathfrak{#3}  \vspace{ -1.5mm} \\  \mathfrak{#1}\hspace{-0.6mm} &\mathfrak{#2} \hspace{-0.9mm}&\vspace{-1.5mm}\\ && \mathfrak{#4}  \end{array}\biggr] }$}}

\newcommand{\eq}[1]{(\ref{#1})}
\newcommand{\w}[1]{\\[0.#1cm]}
\def\eqs#1#2{(\ref{#1}-\ref{#2})}
\def\det{{\rm det\,}}
\def\tr{{\rm tr}}

\newcommand{\hoch}[1]{$\, ^{#1}$}
\newcommand{\imperial}{\it\small Theoretical Physics Group, Imperial College London\\ Prince Consort Road, London SW7 2AZ, UK}
\newcommand{\kings}
{\it\small Department of Mathematics, King's College, University of London\\ Strand, London WC2R 2LS, UK}
\newcommand{\uu}
{\it\small Department of Theoretical Physics, Uppsala, Sweden}
\newcommand{\hip}
{\it\small HIP-Helsinki Institute of Physics, P.O. Box 64 FIN-00014
University of Helsinki, Suomi-Finland}
\newcommand{\stock}
{\it\small Department of Theoretical Physics, Stockholm, Sweden}
\newcommand{\cpht}
{\it\small Centre de Physique Th{\'e}orique, Ecole Polytechnique, CNRS\\ 91128 Palaiseau Cedex, France}
\makeatletter
\renewcommand\theequation{\thesection.\arabic{equation}}
\@addtoreset{equation}{section} \makeatother

\newcommand{\sa}{/ \hspace{-1.2ex}}
\newcommand{\saa}{/ \hspace{-1.4ex}}
\newcommand{\saaa}{\, / \hspace{-1.6ex}}
\newcommand{\Scal}[1]{\Bigl ({#1} \Bigr )}
\newcommand{\scal}[1]{\bigl ({#1} \bigr )}

\newcommand{\CR}{\nonumber \\*}

\newcommand{\trace}{\hbox {tr}~}
\newcommand{\traceS}{\hbox {tr}_{\scriptscriptstyle \mathfrak{S}}~}

\DeclareMathAlphabet{\mathpzc}{OT1}{pzc}{m}{it}
\def\BRST{\,\mathpzc{s}\,}
\def\aBRST{{\scriptstyle (\mathpzc{s})}}
\def\q{{{\scriptscriptstyle (Q)}}}
\def\qs{{\scriptscriptstyle (Q\mathpzc{s})}}
\def\Qsla{{\mathcal{S}_{\q}}}
\def\Slav{{\mathcal{S}_\aBRST}}
\def\epsilonb{{\overline{\epsilon}}}
\def\bulletup{{\scriptstyle \bullet}}

\newcommand{\grad}[3]{{\scriptscriptstyle (#1 , #2, #3 )}}
\newcommand{\gra}[2]{{\scriptscriptstyle (#1 , #2 )}}
\newcommand{\ord}[1]{{\scriptscriptstyle (#1)}}

\def\cL{{\cal L}}
\def\cN{\mathcal{N}}
\def\cO{\mathcal{O}}

\def\ie{{\it i.e.}\ }
\def\eg{{\it e.g.}\ }

\newcommand{\sfrac}[2]{{\scriptstyle \frac{#1}{#2}}}
\newcommand{\stfrac}[2]{{\scriptscriptstyle \frac{#1}{#2}}}

 \def\balpha{{\overline{\alpha}}}
 \def\bbeta{{\overline{\beta}}}
 \def\bgamma{{\overline{\gamma}}}
 \def\bdelta{{\overline{\delta}}}
 \def\bepsilon{{\overline{\epsilon}}}
 \def\bvarepsilon{{\overline{\varepsilon}}}
 \def\bzeta{{\overline{\zeta}}}
 \def\bareta{{\overline{\eta}}}
 \def\btheta{{\overline{\theta}}}
 \def\bvartheta{{\overline{\vartheta}}}
 \def\biota{{\overline{\iota}}}
 \def\bkappa{{\overline{\kappa}}}
 \def\blambda{{\overline{\lambda}}}
 \def\bmu{{\overline{\mu}}}
 \def\bnu{{\overline{\nu}}}
 \def\bxi{{\overline{\xi}}}
 \def\bpi{{\overline{\pi}}}
 \def\brho{{\overline{\rho}}}
 \def\bvarrho{{\overline{\varrho}}}
 \def\bsigma{{\overline{\sigma}}}
 \def\bvarsigma{{\overline{\varsigma}}}
 \def\btau{{\overline{\tau}}}
 \def\bphi{{\overline{\phi}}}
 \def\bvarphi{{\overline{\varphi}}}
 \def\bchi{{\overline{\chi}}}
 \def\bpsi{{\overline{\psi}}}
 \def\bomega{{\overline{\omega}}}

\def\thalf{{\textrm{\tiny\textonehalf}}}
\def\tquarter{{\textrm{\tiny\textonequarter}}}
\def\Ko{{\scriptscriptstyle K}}
\def\tKo{\scriptscriptstyle k }
\def\N{{\mathcal{N}}}
\def\csN{{\fontsize{9.35pt}{9pt}\selectfont \mbox{$\cN$} \fontsize{12.35pt}{12pt}\selectfont }}
\def\cssN{{\fontsize{6.35pt}{6pt}\selectfont \mbox{$\cN$} \fontsize{12.35pt}{12pt}\selectfont }}
\def\csssN{{\fontsize{4.35pt}{4pt}\selectfont \mbox{$\cN$} \fontsize{12.35pt}{12pt}\selectfont }}

\def\ai{{\hat{\imath}}}
\def\aj{{\hat{\jmath}}}
\def\ak{{\hat{k}}}
\def\al{{\hat{l}}}

\def\inv{{ \scriptscriptstyle \rm{\mbox{\tiny-1}}}}
\def\un{{\mathpzc{1}}}
\def\deux{{\mathpzc{2}}}
\def\trois{{\mathpzc{3}}}
\def\quatre{{\mathpzc{4}}}
\def\cinq{{\mathpzc{5}}}
\colorlet{rouge}{red!70!black}

\newcommand{\red}[1]{ {\color{rouge} #1 }} 
\newcommand{\blue}[1]{{\color{blue} #1 }}
\newcommand{\green}[1]{{\color{green} #1 }}
\newcommand{\bleu}[1]{ {\color{cyan} #1 }} 

\renewcommand{\thefootnote}{\arabic{footnote}}

\usepackage{color}
\usepackage{colordvi}
\definecolor{mygreen}{rgb}{0,0.75,0}

\def\guillaume#1#2{{\color{blue} G: #1 \color{black}}{\scriptsize #2}}
\def\valentin#1#2{{\color{mygreen} V: #1 \color{black}}{\scriptsize #2}}

\begin{document}

\renewcommand{\thefootnote}{\arabic{footnote}}
\setcounter{footnote}{0}

\allowdisplaybreaks[1]
\renewcommand{\thefootnote}{\fnsymbol{footnote}}
\def\corr{$\spadesuit $}
\def\trefle{ $\clubsuit$}
\begin{titlepage}
\begin{flushright}
CPHT-RR092.1114\\
\end{flushright}

\bigskip
\bigskip
\centerline{\Large \bf $\cE \nabla^4 R^4$ type invariants and their gradient expansion}
\centerline{\Large \bf }
\bigskip
\bigskip
\centerline{{\bf Guillaume Bossard and Valentin Verschinin}}
\bigskip
\centerline{Centre de Physique Th\'eorique, Ecole Polytechnique, CNRS}
\centerline{91128 Palaiseau cedex, France \footnote{email: bossard@cpht.polytechnique.fr,  valentin.verschinin@cpht.polytechnique.fr}}
\bigskip
\bigskip

\begin{abstract}
We analyse the constraints from supersymmetry on $\nabla^4 R^4$ type corrections to the effective action in $\cN=2$ supergravity in eight dimensions. We prove that there are two classes of invariants that descend respectively from type IIA and type IIB supergravity. We determine the first class as $d$-closed superforms in superspace in eight dimensions, whereas we obtain the second class by dimensional reduction down to four dimensions, in which there is a single class of invariants transforming in the next to minimal unitary representation of $E_{7(7)}$. 
\end{abstract}

\end{titlepage}
\renewcommand{\thefootnote}{\arabic{footnote}}
\setcounter{footnote}{0}

\tableofcontents

\section{Introduction}
The low energy effective action of Type II string theory on $\mathds{R}^{1,9-d} \times T^{d}$ is extremely constrained by supersymmetry and $U$-duality \cite{Green:1997tv,Green:1997as,Kiritsis:1997em}. Although there is no non-perturbative formulation of the theory, duality invariance permits to determine the non-perturbative low energy effective action from perturbative computations in string theory \cite{D'Hoker:2005jc,Green:2008uj,Gomez:2013sla,Green:2014yxa,D'Hoker:2014gfa} and in eleven-dimensional supergravity \cite{Green:1997as,Green:1999pu,Basu:2014uba}. At low orders in the derivative expansion, the effective action is completely determined by the four-graviton amplitude, and one can in principle reconstruct the effective action at these orders from the functions $\cE_\gra{p}{q}$ of the moduli parametrizing the symmetric space  $E_{d(d)}/ K_d$ that define the amplitude \cite{Green:2010wi},
\be \Gamma \sim \int\Scal{  \frac{1}{\kappa^2}  R + \kappa^{2 \frac{d-2}{8-d}}  \cE_\gra{0}{0} R^4  + \kappa^{2 \frac{d+2}{8-d}}  \cE_\gra{1}{0} \nabla^4 R^4  + \kappa^{2 \frac{d+4}{8-d}}  \cE_\gra{0}{1} \nabla^6 R^4 + \dots } \ . \ee 
The functions $ \cE_\gra{0}{0}$, $\cE_\gra{1}{0}$ and $\cE_\gra{0}{1}$ are strongly constrained by supersymmetry, and are in particular eigenfunctions of the Laplace operator on the scalar manifold \cite{Green:1999pu,Green:1998by,Sinha:2002zr,Basu:2011he}. The realisation of these functions as Eisenstein functions \cite{Green:1997tv,Kiritsis:1997em} has been generalised in lower dimensions \cite{Obers:1999um}, and to higher order $\nabla^6 R^4$ type corrections \cite{Green:2005ba}, leading to more developments in lower dimensions \cite{Basu:2007ru,Basu:2007ck,Pioline:2001jn,Kazhdan:2001nx,Pioline:2004xq,Pioline:2010kb,Green:2010kv,Green:2011vz,Fleig:2013psa,Minimal}. 

We have shown in \cite{Minimal} that these functions moreover satisfy to tensorial differential equations that determine their egenvalues for all Casimir operators. The function $\cE_\gra00$ satisfies for example that  its second-order derivative vanishes when restricted to the Joseph ideal \cite{Joseph}, constraining it to lie in the minimal unitary representation of $E_{d(d)}$, in accordance with \cite{Pioline:2001jn,Kazhdan:2001nx,Pioline:2004xq,Pioline:2010kb}. We have shown that $\cE_\gra10$ satisfies to an equivalent equation associated to the next to minimal unitary representation of $E_{7(7)}$ in four dimensions \cite{Minimal}, from the structure of the invariant in the linearised approximation \cite{Galperin:1984av,Berkovits:1997pj,Drummond:2003ex,Bossard:2009sy}. 

This paper extends the analysis of the $\nabla^4 R^4$ type invariant at the non-linear level in eight dimensions.  To carry out this program, we concentrate on terms of maximal R-symmetry weight, similarly as in \cite{Green:1998by,Basu:2011he,Minimal}. We find in this way that the function of the scalar fields must satisfy to a tensorial second-order differential equation consistent with one of the explicit Eisenstein functions conjectured in \cite{Basu:2007ru} to define the non-perturbative threshold function $\cE_\gra10$. The second function does not depend on the type IIB torus complex structure, and is not constrained by this analysis that only considers K\"{a}hler derivatives of the function. However, we prove that the two sets of differential equations satisfied by the two functions defining $\cE_\gra10$, are in the same $E_{7(7)}$ representation in four dimensions. We show moreover that they are the unique differential equations satisfying to this criterium. We conclude therefore  that there is two classes of $\nabla^4 R^4$ invariants in eight dimensions, consistently with the two functions appearing in the string theory effective action. Combining these results with the ones obtains in \cite{Minimal}, we conclude that there is a unique $\nabla^4 R^4$ invariant in dimension five and lower, that splits into two different invariants in dimension $6,\, 7$ and $8$. They descend respectively from type IIA and type IIB  $2$-loop corrections to the supergravity effective action. 

We provide an overview of the results in the first section, that combines results already obtained in \cite{Minimal}, and new ones that are derived in this paper. It exhibits the structure of the $R^4$ and $\nabla^4 R^4$ type invariants as gradient expansions in the covariant derivative of a defining function $\cE$ of the scalar fields parametrising $E_{d(d)}/K_d$. In section \ref{8Dd4R4Explicit} we discuss in details the structure of the $\nabla^4 R^4$ type invariant in eight dimensions that lifts to type IIA in ten dimensions, in the formalism of superforms in superspace  \cite{Voronov,Gates:1997kr,Gates:1997ag}. Because the associated function depends on both the complex scalar parametrising $SL(2)/SO(2)$ and the scalar fields parametrising $SL(3)/SO(3)$, one must consider the gradient expansion in terms of both the K\"{a}hler derivative and the isospin $2$ tangent derivatives on $SL(3)/SO(3)$. This permits to distinguish terms of maximal $U(1)$ weight and isospin, that are uniquely determined as monomials of order twenty-four  in the fermion fields.  

In order to show the existence and the uniqueness of the other class of $\nabla^4 R^4$ type invariants in eight dimensions, we use the uniqueness of the $\nabla^4 R^4$ type invariant in four dimensions, due to the bijective correspondence between supersymmetry invariants and superconformal primaries of Lorentz invariant top component in four dimensions \cite{Drummond:2003ex,Drummond:2010fp}. Any supersymmetry invariant that can be defined in eight dimensions, clearly descends to four dimensions by dimensional reduction on $T^4$. Starting from the type IIA invariant we study in section \ref{8Dd4R4Explicit}, one can consider the corresponding four-dimensional invariant, and the differential equations satisfied by the associated function on $E_{7(7)}/\SU$. Any other solution to these differential equations is also supersymmetric in four dimensions, and for a function defined on $\mathds{R}_+^*\times SL(2)/SO(2) \times SL(3)/SO(3)$ with the appropriate power of the Kaluza--Klein dilaton, it must lift to an invariant in eight dimensions. The invariance of the supersymmetry invariant with respect to the nilpotent subgroup of $E_{7(7)}$ defining the shift of the axions, indeed implies that the dependence in the gauge fields and the axions is defined in such a way as to ensure gauge invariance and diffeormorphism invariance in eight dimensions. 

We show that this line of arguments is indeed valid in section \ref{From4to7to8}, although the proof is not formulated in this order. We rather start by solving the relevant differential equations derived in \cite{Minimal} in four dimensions on a function of the seven-dimensional scalar fields. This way we exhibit the existence of two classes of $\nabla^4 R^4$ type invariants in seven dimensions, which are then shown to lift to corresponding invariants in eight dimensions. We also discuss the properties of the solutions with respect to $E_{d(d)}(\mathds{Z})$ invariance, and we prove that the functions conjectured to define the type II exact low energy effective action components in $R^4$ and $\nabla^4 R^4$ \cite{Obers:1999um,Green:2010kv} are indeed solutions to the equations derived in \cite{Minimal}.

\section{Overview of the results in various dimensions}
In this section we review the $E_{d(d)}$ multiplets of supersymmetric corrections to the supergravity effective action in various dimensions. We will concentrate ourselves on $R^4$ and $\nabla^4 R^4$ type invariants in maximal supergravity. 
\begin{figure}[htbp]
\begin{minipage}[c]{6cm}
\center
\resizebox{6cm}{!}{

 \begin{tikzpicture}

 \def\xshift{- 1}
  \def\xmin{1}
 \def\ymin{0}

\draw[-,draw=black, thick](\xmin - 1,\ymin + 4) -- (\xmin - 1,\ymin + 4.7); \draw[-,draw=black, thick](\xmin,\ymin + 4) -- (\xmin,\ymin + 4.7);
\draw[-,draw=black, thick](\xmin + 2,\ymin + 4) -- (\xmin + 2,\ymin + 4.7); \draw[-,draw=black, thick](\xmin + 3,\ymin + 4) -- (\xmin + 3,\ymin + 4.7);
\draw[-,draw=black, thick](\xmin,\ymin + 4) -- (\xmin - 0.9 ,\ymin + 4.7);
\draw[-,draw=black, thick](\xmin + 3,\ymin + 4) -- (\xmin + 2.1 ,\ymin + 4.7);

\draw[dashed,draw=black, thick](\xmin + 2.5,\ymin - 1) -- (\xmin + 2.5,\ymin - 1.3);
\draw[dashed,draw=black, thick](\xmin - 0.5,\ymin - 1) -- (\xmin - 0.5,\ymin - 1.3);

\draw[<-,draw=black,thick] (\xmin - 3,\ymin + 6) -- (\xmin - 3,\ymin - 2);               
\draw (\xmin - 3,\ymin + 5) node{-};  \draw (\xmin - 1,\ymin + 5) node{IIA};  \draw (\xmin ,\ymin + 5) node{IIB}; \draw (\xmin + 2,\ymin + 5) node{IIA};  \draw (\xmin + 3 ,\ymin + 5) node{IIB};
\draw (\xmin - 3,\ymin + 4) node{-};   \draw (\xmin + \xshift,\ymin + 4) node{\color{rouge} \textbullet};\draw (\xmin + 1 + \xshift,\ymin + 4) node{\textbullet};  \draw (\xmin + \xshift + 3,\ymin + 4) node{\color{rouge}\textbullet};\draw (\xmin + \xshift + 4,\ymin + 4) node{$\circ$};
\draw (\xmin - 3,\ymin + 3) node{-};                \draw (\xmin + \xshift + 0.5,\ymin + 3) node{\textbullet};   \draw (\xmin + \xshift + 3,\ymin + 3) node{\textbullet};\draw (\xmin + 4 + \xshift ,\ymin + 3) node{$\circ$};
\draw (\xmin - 3,\ymin + 2) node{-};                \draw (\xmin + \xshift + 0.5,\ymin + 2) node{\textbullet};   \draw (\xmin + \xshift + 3,\ymin + 2) node{\textbullet};\draw (\xmin + 4 + \xshift,\ymin + 2) node{\color{rouge} \textbullet};
\draw (\xmin - 3,\ymin + 1) node{-};                \draw (\xmin + \xshift + 0.5,\ymin + 1) node{\textbullet};   \draw (\xmin + \xshift + 3.5,\ymin + 1) node{\textbullet};
\draw (\xmin - 3,\ymin) node{-};			    \draw (\xmin + \xshift + 0.5,\ymin) node{\textbullet};         \draw (\xmin + \xshift + 3.5,\ymin) node{\textbullet};
\draw (\xmin - 3,\ymin - 1) node{-};                 \draw (\xmin + \xshift + 0.5 ,\ymin - 1) node{\textbullet};   \draw (\xmin + \xshift + 3.5,\ymin - 1) node{\textbullet};
\draw (\xmin - 3 + 0.3,\ymin + 5) node{$10$};
\draw (\xmin - 3 + 0.3,\ymin + 4) node{$8$};
\draw (\xmin - 3 + 0.3,\ymin + 3) node{$7$};
\draw (\xmin - 3 + 0.3,\ymin + 2) node{$6$};
\draw (\xmin - 3 + 0.3,\ymin + 1) node{$5$};
\draw (\xmin - 3 + 0.3,\ymin) node{$4$};
\draw (\xmin - 3 + 0.3,\ymin - 1) node{$3$};
\draw (\xmin - 3 + 0.5,\ymin + 6) node{dim};

\draw (\xmin + 0.5 + \xshift,\ymin - 1.6) node{$R^4$};
\draw (\xmin + 3.5 + \xshift,\ymin - 1.6) node{$\nabla^4 R^4$};

\draw[-,draw=black, thick](\xmin + 0.5 + \xshift,\ymin -1) -- (\xmin + 0.5 + \xshift,\ymin + 3);
\draw[-,draw=black, thick](\xmin + 3.5 + \xshift ,\ymin -1) -- (\xmin + 3.5 + \xshift,\ymin + 1);
\draw[-,draw=black, thick](\xmin + 3.5 + \xshift,\ymin + 1) -- (\xmin + \xshift + 4,\ymin + 2);  \draw[-,draw=black, thick](\xmin + \xshift + 3.5,\ymin + 1) -- (\xmin + \xshift + 3,\ymin + 2);
\draw[-,draw=black, thick](\xmin + \xshift + 4,\ymin + 2) -- (\xmin + \xshift + 4,\ymin + 2);  \draw[-,draw=black, thick](\xmin + \xshift + 3,\ymin + 2) -- (\xmin + \xshift + 3,\ymin + 2);
\draw[-,draw=black, thick](\xmin + \xshift + 0.5,\ymin + 3) -- (\xmin + \xshift ,\ymin + 4); \draw[-,draw=black, thick](\xmin + \xshift + 0.5,\ymin + 3) -- (\xmin + \xshift + 1,\ymin + 4);

\draw[-,draw=black, thick](\xmin + \xshift + 4,\ymin + 4) -- (\xmin + \xshift + 4,\ymin + 2);  \draw[-,draw=black, thick](\xmin + \xshift + 3,\ymin + 4) -- (\xmin + \xshift + 3,\ymin + 2);
\end{tikzpicture}}
\end{minipage}\hspace{10mm}\begin{minipage}[c]{8cm}
\caption{\small Each node determines an $E_{d(d)}$ multiplet of $R^4$ and $\nabla^4 R^4$ type invariants, respectively. The lines refer to their connection by dimensional reduction. The  \textbullet\,  refers to parity symmetric invariants that can be defined in harmonic superspace in the linearised approximation, while {\color{rouge} \textbullet}\,  indicates that they are complex chiral invariants in the linearised approximation. $\circ$ refers to invariants that cannot be written in harmonic superspace in the linearised approximation.}
\end{minipage}
\end{figure}
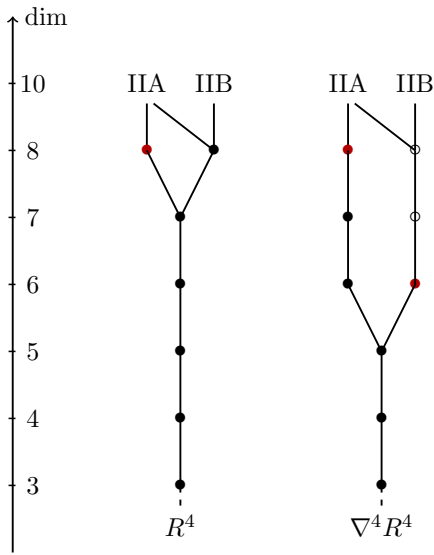

Such corrections are invariant modulo the classical field equations, and are determined by closed superforms within the superform formalism of Bates   \cite{Voronov,Gates:1997kr,Gates:1997ag}. A closed superform depends in general on the scalar fields parametrising $E_{d(d)}/K_d$ through a function $\cE$ and its covariant derivative in tangent frame, and takes the form 
\be \cL[\cE] = \sum_{n,R} \cD^n_R \cE\, \cL^{\bar R} \ , \ee
where $R$ refers to irreducible representations of $K_d$ such that the superforms $\cL^{\bar R}$ are $E_{d(d)}$ invariant and transform with respect to $K_d$ in the conjugate irreducible representation $\bar R$. For BPS protected invariants such as the ones of type $R^4$ and $\nabla^4 R^4$, the appearing irreducible representations $R$ are generally determined from the linearised analysis, and the function $\cE$ satisfies to the constraints that its derivative  $\cD^n_{R^\prime} \cE$ in irreducible representations that do not appear in the gradient expansion either vanish or are related to lower order derivatives of the function in the same representation. 

All along the paper we use the convention that the function $E_w$ for a weight $w$ of $\mathfrak{a}^*(\mathfrak{e}_{d(d)})$ is the Eisenstein function on $E_{d(d)}(\mathds{Z})\backslash E_{d(d)}/K_d$ associated to this weight \cite{Obers:1999um,Green:2010kv}, whereas a function $\cE_w$ refers to any function on $E_{d(d)}/K_d$ solving the same differential equations as $E_w$. Supersymmetry is preserved for any such a solution $\cE_w$ with the appropriate weight $w$, and requiring moreover $E_{d(d)}(\mathds{Z})$ invariance only then distinguishes the  Eisenstein function $E_w$. 

\addtocontents{toc}{\protect\setcounter{tocdepth}{1}}
\subsection{$\cN = 2$ supergravity in eight dimensions}
\addtocontents{toc}{\protect\setcounter{tocdepth}{2}}

In eight dimensions, maximal supergravity admits for duality group $SL(2) \times SL(3)$, and the scalar fields parametrise the symmetric space $SL(2)/SO(2) \times SL(3)/SO(3)$. The K\"ahler derivatives on $SL(2)/SO(2)$  are denoted with $\cD$ and $\bar{\cD}$, while the $SU(2)$ isospin 2 tangent derivatives on  $SL(3)/SO(3)$ are defined as $\cD_{ijkl}$, with $i, j,k,l$ running from $1$ to $2$ of $SU(2)$. The theory includes two $1/2$ BPS $R^4$ type invariants and  two $1/4$ BPS $\nabla^4 R^4$ type invariants, which are supersymmetric up to the classical equations of motion. These invariants decompose in a gradient expansion of a given function  $\cE$ of the scalar fields as follows
\bea\label{7Dresume} 
 R^4&:& \qquad \sum_{n = 0}^{12} \bar{U}^{- 2 n}\bar{\cD}^n \cE_\grad{2}{2}{0} \, \cL^\ord{4n}   \ , \qquad  \sum_{n = 0}^{12} \cD_{[4n]}^n \cE\grad{2}{1}{1}\, \cL^{[4n]} \ , \\
 \nabla^4 R^4&:& \qquad \sum_{n=0}^{14} \biggl( \  \sum_{ k = 0}^{20\mbox{-}n} \bar{U}^{- 2 k} \bar{\cD}^k \cD^{n}_{[4n]} \cE_\grad{2}{1}{0} \, \cL^{\ord{4k} [4n]}  + {U}^{- 2 } {\cD} \cD^{n}_{[4n]} \cE_\grad{2}{1}{0} \, \cL^{\ord{-4} [4n]} \biggr) \ ,  \CR
&&\qquad  \sum_{n = 0}^{14} \cD_{[4n]}^n  \cE'_{\scriptscriptstyle \frac{1}{4}} \, \cL^{[4n]} \ , 
\eea
where the $\cL^{\ord{4k}[4n]}$ are $SL(2) \times SL(3)$ invariant $8$-superforms in the isospin $2n$ representation of $SU(2)$ with $U(1)$ weight $4k$. The indices of the function $\cE_\grad{n}{p}{q}$ refers to the harmonic superspace construction of the associated  invariant in the linearised approximation, whereas the notation $\cE'_{\frac{1}{4}}$ indicates that the corresponding invariant cannot be written as a Lorentz invariant harmonic superspace integral in the linearised approximation. The functions $\cE_\grad{2}{2}{0}$ and $\cE_\grad{2}{1}{1}$ appearing in the $R^4$ invariants satisfy the following equations \cite{Minimal}
\bea
\cD  \cE_\grad{2}{2}{0} & = &  0 \ , \hspace{4cm} \cD_{i j k l} \cE_\grad{2}{2}{0} = 0\ ,  \\
\cD_{i j}{}^{p q} \cD_{k l p q}  \cE_\grad{2}{1}{1}& = & - \frac{3}{12} \cD_{i j k l}  \cE_\grad{2}{1}{1} \ ,  \hspace{2.1cm} \cD \cE_\grad{2}{1}{1} = \bar{\cD } \cE_\grad{2}{1}{1}  = 0\ , 
\eea
in agreement with \cite{Kiritsis:1997em}. The two classes of invariants coincide in trivial topology when the function is a constant, and define the 1-loop counter-term for the supergravity  logarithm divergence \cite{Green:1982sw}. The invariant associated to $\cE_\grad{2}{2}{0}$ is chiral and complex, and its associated complex conjugate associated to the function $\cE_\grad{2}{0}{2}$ satisfies to the complex conjugate constraints. The functions $\cE_\grad{2}{1}{0}$ and $ \cE^\prime_{\scriptscriptstyle \frac{1}{4}}$ defining the $\nabla^4 R^4$ type invariants are discussed in this paper, and are proved to satisfy to
\bea
\Delta_{SL(2)} \cE_\grad{2}{1}{0} &=& 2 \cE_\grad{2}{1}{0} \ ,  \qquad \cD^2 \cE_\grad{2}{1}{0}  = 0 \CR
 \cD_{i j}{}^{p q} \cD_{k l p q}  \cE_\grad{2}{1}{0} & = & \frac{5}{12} \cD_{i j k l}  \cE_\grad{2}{1}{0}  + \frac{1}{9} (\ve_{i k} \ve_{j l} + \ve_{i l} \ve_{j k}) \cE_\grad{2}{1}{0} \ , \\
   \cD_{i j}{}^{p q} \cD_{k l p q}  \cE'_{\scriptscriptstyle \frac{1}{4}} & = &- \frac{7}{12} \cD_{i j k l}  \cE'_{\scriptscriptstyle \frac{1}{4}}  + \frac{5}{18} (\ve_{i k} \ve_{j l} + \ve_{i l} \ve_{j k})\cE'_{\scriptscriptstyle \frac{1}{4}} \ , \CR
   \cD \cE'_{\scriptscriptstyle \frac{1}{4}} & = &0 \ ,\qquad   \bar{ \cD } \cE'_{\scriptscriptstyle \frac{1}{4}} = 0 \ , 
\eea
consistently with \cite{Green:2010wi,Basu:2007ru}. These equations are indeed satisfied by the Eisenstein functions 
\be
 \hat{\cE}_\grad{2}{2}{0}+ \hat{\cE}_\grad{2}{0}{2} = \hat{E}_{[1]} \ , \quad \hat{\cE}_\grad{2}{1}{1}=  \hat{E}_{[\frac{3}{2},0]}\ ,  \quad \cE_\grad{2}{1}{0} +\cE_\grad{2}{0}{1}= E_{[2]} E_{[- \frac{1}{2},0]}\ ,   \quad  \cE'_{\scriptscriptstyle \frac{1}{4}} =  E_{[ \frac{5}{2},0]} \ , 
\ee
which  determine the exact $R^4$ and $\nabla^4 R^4$ thresholds in type II string theory \cite{Kiritsis:1997em,Basu:2007ru}, up two inhomogeneous terms associated to the chiral anomaly and the $SL(3)$ anomaly produced by the 1-loop divergence \cite{Minimal,Green:2010sp}. Here the hat over  $\hat{\cE}_\grad{2}{2}{0}$ and $ \hat{\cE}_\grad{2}{0}{2} $ indicates that their sum satisfies to the inhomogeneous equation with a constant right-hand-side \cite{Green:2010wi}, and similarly for $\hat{\cE}_\grad{2}{1}{1}$. 

\addtocontents{toc}{\protect\setcounter{tocdepth}{1}}
\subsection{$\cN = 2$ supergravity in seven dimensions}
\addtocontents{toc}{\protect\setcounter{tocdepth}{2}}
In seven dimensions maximal supergravity has for duality group $SL(5)$, with maximal compact subgroup $SO(5)$. We label the vector indices $a, b, c$ of $SO(5)$ and the covariant  derivative $\cD_{a b}$ is symmetric traceless, \ie transforms  in the $[0,2]$ of $Sp(2)$. The $R^4$ and $\nabla^4 R^4$ type  invariants  have the following gradient expansion in the function $\cE$ of the scalar fields
\bea
 R^4&:& \qquad \sum_{n = 0}^{12} \cD_{[0,2n]}^n \cE_\gra{4}{2}\,   \cL^{[0,2n]} \ ,  \\
 \nabla^4 R^4&:& \qquad \sum_{n,k = 0}^{n + 2 k \leq 20} \cD_{[4k,2n]}^{n + 2k}\cE_\gra{4}{1}  \, \cL^{[4k,2n]} \ , \qquad \sum_{n = 0}^{20} \cD_{[0,2n]}^n \cE'_{\scriptscriptstyle \frac{1}{4}}  \, \cL^{[0,2n]} \ , \label{D4R48D} 
\eea
where again $ \cL^{[4k,2n]}$ are $SL(5)$ invariant superforms in the $[4k,2n]$ of $Sp(2)$, \ie traceless tensors of $SO(5)$ with $2k$ pairs of antisymmetric indices and $2n$ additional symmetrised indices, while  $(4, p)$ refers to the harmonic superspace construction of the $\frac{p}{4}$ BPS invariant in the linearised approximation. The last invariant depending on $\cE'_{\scriptscriptstyle \frac{1}{4}}$ does not admit a Lorentz invariant harmonic superspace integral form in the linearised approximation. The function $\cE_\gra{4}{2}$ defining the $R^4$ type invariant satisfies to 
\bea
 \cD_{a}{}^{c} \cD_{c}{}^{b} \cE_\gra{4}{2} & = & \frac{3}{20} \cD_{a}{}^{b}\cE_\gra{4}{2}  - \frac{6}{25} \delta_{a}^{b} \cE_\gra{4}{2} \ , \CR
  \Bigl(2 \delta_{[a}^{[c} \cD_{b]}{}^{e} \cD_{e}{}^{d]} + 2 \cD_{[a}{}^{[c} \cD_{b]}{}^{d]} \Bigr) \cE_\gra{4}{2} & = & \frac{1}{20} \delta_{[a}^{[c} \cD_{b]}{}^{d]} \cE_\gra{4}{2}  - \frac{9}{25}  \delta_{ab}^{cd} \cE_\gra{4}{2}\ , 
\eea
consistently with \cite{Kiritsis:1997em}. It is important to remark that the two possible functions multiplying $R^4$ in eight dimensions $\cE_\grad{2}{2}{0}$ and $\cE_\grad{2}{1}{1}$ are related by $SL(5)$ in seven dimensions. The functions appearing in the  $\nabla^4 R^4$ type invariants satisfy instead
\bea
 \cD_{a}{}^{c} \cD_{c}{}^{b} \cE'_{\scriptscriptstyle \frac{1}{4}} & = & \frac{3}{4} \cD_{a}{}^{b}\cE'_{\scriptscriptstyle \frac{1}{4}} \ , \qquad  \Bigl(2 \delta_{[a}^{[c} \cD_{b]}{}^{e} \cD_{e}{}^{d]} + 2 \cD_{[a}{}^{[c} \cD_{b]}{}^{d]} \Bigr) \cE'_{\scriptscriptstyle \frac{1}{4}} = \frac{1}{4} \delta_{[a}^{[c} \cD_{b]}{}^{d]} \cE'_{\scriptscriptstyle \frac{1}{4}} \ ,  \\
  \cD_{a}{}^{c} \cD_{c}{}^{b} \cE_\gra{4}{1} & = & - \frac{1}{4} \cD_{a}{}^{b} \cE_\gra{4}{1}   \ , 
\eea
consistently with \cite{Green:2010wi}. The two invariants coincide for a constant function, and define the counter-term for the 2-loop logarithm divergence in supergravity \cite{Bern:1998ug}. The $SL(5,\mathds{Z})$ invariant Eisenstein functions 
\bea
 \cE_\gra{4}{2} = E_{[\frac{3}{2} 0 0 0]}\ ,  \qquad  \hat{\cE}'_{\scriptscriptstyle \frac{1}{4}}  = \hat{E}_{[\frac{5}{2} 0 0 0]}\ ,  \qquad \hat{\cE}_\gra{4}{1} = \hat{E}_{[ 0 0 \frac{5}{2} 0]} \ , 
\eea
which are conjecture to define the exact low energy effective action in string theory \cite{Kiritsis:1997em,Green:2010wi}, indeed solve these differential equations, up to an inohomogenous right-hand-side for the $\nabla^4 R^4$ type invariants that comes from the anomaly associated to the 2-loop divergence. Once again the hat on the functions refers to these anomalous corrections. 

\addtocontents{toc}{\protect\setcounter{tocdepth}{1}}
\subsection{$\cN = (2,2)$ supergravity in six dimensions}
\addtocontents{toc}{\protect\setcounter{tocdepth}{2}}

In six dimensions the duality group of maximal supergravity is  $SO(5,5)$ with maximal compact subgroup $SO(5) \times SO(5)$. We denote the indices $i, j$ and $\hi, \hj$ running from $1$ to $4$ of the two $Sp(2)$ groups, and respectively $a,b$ and $\hat{a},\hat{b}$ the vector indices of the two $SO(5)\cong Sp(2)/\mathds{Z}_2$. The covariant derivative in tangent frame is a bi-vector of the two $SO(5)$, \ie transforms in the $[0,1] \times [0,1]$ of $Sp(2) \times Sp(2)$. The invariants we discuss in this paper admit the gradient expansion in the function of the scalar fields 
\bea
 R^4 &:& \qquad \sum_{n = 0}^{12} \cD_{[0,n],[0,n]}^n \cE_\grad{4}{2}{2} \, \cL^{[0,n],[0,n]}  \ , \\
\nabla^4 R^4 &:& \qquad \sum_{n, k = 0}^{n + 2k \leq 20} \cD^{n + 2k}_{[0,n],[0,n+2k]} \cE_\grad{4}{2}{0} \, \cL^{[0,n],[0,n + 2k]} \ , \CR
&& \qquad \sum_{n, k = 0}^{n + 2k \leq 20} \cD^{n + 2k}_{[2k,n],[2k,n]} \cE_\grad{4}{1}{1} \, \cL^{[2k,n],[2k,n]} \ , 
\eea
where $\cL^{[2k,n],[2k,m]}$ are $SO(5,5)$ invariant 6-superforms in the corresponding representation of $Sp(2)\times Sp(2)$. The invariant associated to the function $ \cE_\grad{4}{2}{0}$ is complex and chiral, and admits the conjugate invariant of function  $\cE_\grad{4}{0}{2}$ for which the role of the two $Sp(2)$ factors is exchanged. 
The function $\cE_\grad{4}{2}{2}$ satisfies to the following equations \cite{Minimal}
\bea
 \cD_{a}{}^{\hat{a}} \cD_{b \hat{a}} \cE_\grad{4}{2}{2} = - \frac{3}{4} \delta_{a b} \cE_\grad{4}{2}{2}\ ,  \quad \cD^{a}{}_{\hat{a}} \cD_{a \hat{b}} \cE_\grad{4}{2}{2} = - \frac{3}{4} \delta_{\hat{a} \hat{b}} \cE_\grad{4}{2}{2} \ ,  \quad
 \cD_{[a}{}^{[\hat{a}} \cD_{b]}{}^{\hat{b}]} \cE_\grad{4}{2}{2} = 0\ , \qquad 
\eea
whereas  $\cE_\grad{4}{2}{0}$ and $\cE_\grad{4}{1}{1}$ satisfy respectively to
\begin{equation} \begin{aligned}
\cD_{a}{}^{\hat{a}} \cD_{b \hat{a}} \cE_\grad{4}{2}{0} &= - \frac{3}{4} \delta_{a b} \cE_\grad{4}{2}{0} \ , \\
 \cD_{a}{}^{\hat{a}} \cD_{b \hat{a}} \cE_\grad{4}{1}{1} &= - \frac{3}{4} \delta_{a b} \cE_\grad{4}{1}{1}\ , 
 \end{aligned} \qquad \begin{aligned}
\cD_{[a}{}^{[\hat{a}} \cD_{b]}{}^{\hat{b}]} \cE_\grad{4}{2}{0} &= 0\ , \\
\cD^{a}{}_{\hat{a}} \cD{}_{a \hat{b}} \cE_\grad{4}{1}{1} &= - \frac{3}{4} \delta_{\hat{a} \hat{b}} \cE_\grad{4}{1}{1}\  . 
\end{aligned}
\end{equation}
The $SO(5,5,\mathds{Z})$ invariant Eisenstein functions that are conjectured to define the exact string theory low energy effective action \cite{Obers:1999um,Green:2010wi} indeed satisfy to these equations such that 
\be
\cE_\grad{4}{2}{2} = E_{\mbox{\DSOX{\hspace{-0.5mm}\frac{3}{2}}0000}} \ , \qquad \hat{\cE}_\grad{4}{2}{0}+ \hat{\cE}_\grad{4}{0}{2} = \hat{E}_{\mbox{\DSOX{\hspace{-0.5mm}\frac{5}{2}}0000}} \ , \qquad \hat{\cE}_\grad{4}{1}{1} = \hat{E}_{\mbox{\DSOX{\hspace{-0.5mm}0}0003}}\ . 
\ee
up to inohomogenous right-hand-sides associated to the 1-loop divergence of the form factor of the $R^4$ type invariant \cite{Minimal}. 
Although the function $ \cE_\grad{4}{2}{0}$ defined such that $ \cE_\grad{4}{2}{0}+ \cE_\grad{4}{0}{2} = E{\mbox{\DSOX{\hspace{-0.5mm}\frac{5}{2}}0000}}$ is not itself $SO(5,5,\mathds{Z})$ invariant, the associated supersymmetry invariant only depend on the sum $ \cE_\grad{4}{2}{0}+ \cE_\grad{4}{0}{2}$ and the covariant derivative of the individual functions, such that it is duality invariant \cite{Minimal}.

\addtocontents{toc}{\protect\setcounter{tocdepth}{1}}
\subsection{$\cN = 4$ supergravity in five dimensions}
\addtocontents{toc}{\protect\setcounter{tocdepth}{2}}
In five-dimensional supergravity the duality group is $E_{6(6)}$, with maximal compact subgroup $Sp(4)/\mathds{Z}_2$. The covariant derivative in tangent frame is a symplectic traceless rank four antisymmetric tensor of $Sp(4)$, \ie in the  $[0,0,0,1]$ irreducible representation. The  $1/2$ and $1/4$ BPS invariants admit the following gradient expansion in the function of the scalar fields 
\bea
R^4&:& \qquad \sum_{n = 0}^{12} \cD_{[0,0,0,n]}^{n} \cE_\gra{8}{4}\,  \cL^{[0,0,0,n]} \ , \\
\nabla^4 R^4&:& \qquad \sum_{n,k = 0}^{n + 2k \leq 20} \cD_{[0,2k,0,n]}^{n+2k} \cE_\gra{8}{2} \, \cL^{[0,2k,0,n]} \ . 
\eea
The functions $\cE_\gra{8}{2n}$ satisfy to the tensorial equations  
\bea
\cD_{i j p q} \cD^{k l p q} \cE_\gra{8}{4}  &=& - 2 \delta_{i j}^{k l} \cE_\gra{8}{4} \ , \\
\cD_{ijpq} \cD^{pq}{}_{rs} \cD^{rskl}   \cE_\gra{8}{2} + \frac{10}{3} \cD_{ij}{}^{kl}  \cE_\gra{8}{2}&=& -\frac{1}{2} \Scal{ \cD_{ijpq} \cD^{klpq} + \frac{70}{27} \delta_{ij}^{kl} }   \cE_\gra{8}{2} \ , \CR
\cD^3_{[2,0,0,1]} \cE_\gra{8}{2} &=& 0\ , 
\eea
where $\delta_{ij}^{kl}$ is the projector to the antisymmetric symplectic traceless irreducible representation of $Sp(4)$.\footnote{\ie $\delta_{ij}^{kl}= \frac{1}{2} \delta_i^k \delta_j^l -  \frac{1}{2} \delta_i^l \delta_j^k  - \frac{1}{8} \Omega_{ij} \Omega^{kl} $.}  $E_{6(6)}(\mathds{Z})$ invariant solutions to these differential equations are defined by the Eisenstein series of the type
\be
\cE_\gra{8}{4} = \EiEVI{\frac{3}{2}}\ ,  \qquad \cE_\gra{8}{2} = \EiEVI{\frac{5}{2}} \ , 
\ee
that are conjectured to define the non-perturbative low energy effective action in type II string theory \cite{Obers:1999um,Green:2010kv}.

\addtocontents{toc}{\protect\setcounter{tocdepth}{1}}
\subsection{$\cN = 8$ supergravity in four dimensions}
\addtocontents{toc}{\protect\setcounter{tocdepth}{2}}
\label{N8Review}
In $\cN=8$ supergravity the scalar fields parametrise the symmetric space  $E_{7(7)}/\SU$. We denote $i, j\dots$ the indices in the fundamental of $SU(8)$, and the covariant derivative in tangent frame is a complex-selfdual rank four antisymmetric tensor in the $[0,0,0,1,0,0,0]$ representation of $SU(8)$. The invariants of type $R^4$ and $\nabla^4 R^4$  admit respectively the following gradient expansions in the function of the seventy scalar fields
\bea
 R^4&:& \qquad \sum_{n = 0}^{12} \cD^n_{[0,0,0,n,0,0,0]} \cE_\grad{8}{4}{4}\,    \cL^{[0,0,0,n,0,0,0]}\ , \\
 \nabla^4 R^4&:& \qquad  \sum_{n, k = 0}^{n + 2k \leq 20} \mathcal{D}^{n + 2k}_{[0,k,0,n,0,k,0]}\cE_\grad{8}{2}{2}  \, \cL^{[0,k,0,n,0,k,0]}\ .
\eea
The label $(8,4,4)$ and $(8,2,2)$ refer to the harmonic superspace measures that permit to define these invariants in the linearised approximation. The function $\cE_\grad{8}{4}{4}$ was proved to satisfy to 
\be
 \cD_{k l p q} \cD^{i j p q}\cE_\grad{8}{4}{4} = - \frac{9}{2} \delta_{k l}^{i j} \cE_\grad{8}{4}{4}\ , 
\ee
whereas the function defining the $\nabla^4 R^4$ type invariant satisfies to 
\bea
 \cD_{i j p q} \cD^{p q r s} \cD_{r s k l}\cE_\grad{8}{2}{2} &=& - 9 \cD_{i j k l} \cE_\grad{8}{2}{2} \ , \CR
  2 \cD_{j r [kl} \cD^{irmn} \cD_{pq] mn} \cE_\grad{8}{2}{2} &=& - \delta_{j}^{i} \cD_{k l p q} \cE_\grad{8}{2}{2} + 10 \delta^{i}_{[k} \cD_{l p q] j} \cE_\grad{8}{2}{2}\ .  \label{D4R44DInt}
\eea
These differential equations admit as $E_{7(7)}(\mathds{Z})$ invariant solutions the Eisenstein series
\be
 \cE_\grad{8}{4}{4} = \EiEVII{\frac{3}{2}}  \ , \qquad \cE_\grad{8}{2}{2} = \EiEVII{\frac{5}{2}} \ , 
\ee
that are conjectured to define the exact low energy effective action in type II string theory \cite{Obers:1999um,Green:2010kv}.

\addtocontents{toc}{\protect\setcounter{tocdepth}{1}}
\subsection{$\cN = 16$ supergravity in three dimensions}
\addtocontents{toc}{\protect\setcounter{tocdepth}{2}}
In  three dimensions, the duality group is $E_{8(8)}$, of maximal compact subgroup the quotient of $Spin(16)$ by the $\mathds{Z}_2$ kernel of the chiral spinor representation. We denote $i, j$ the $SO(16)$ vector indices and $A, B$ the positive chirality Weyl spinor indices. The covariant derivative in tangent frame is a chiral Weyl spinor, \ie in the  \WSOXVI00000001 representation. In three dimensions there is no four-graviton amplitude, and the corresponding invariants are of type $(\nabla P)^4$ and $\nabla^4 (\nabla P)^4$, where $P_A$ is the scalar momentum of the scalar fields. They admit in this case the following gradient expansion in the function of the 128 scalar fields
\bea
(\nabla P)^4&:& \qquad \sum_{n=0}^{12}  \cD^n_{\mbox{\WSOXVI0000000n} } \cE_\gra{16}{8} \, \cL^{\mbox{\WSOXVI0000000n} }\ , \\
\nabla^4 (\nabla P)^4&:& \qquad \sum_{n, k = 0}^{n + 2k \leq 20}  \cD^{n+2k}_{\mbox{\WSOXVInk000k000n} }  \cE_\gra{16}{4}\, \cL^{\mbox{\WSOXVInk000k000n} }\ , 
\eea
which satisfy respectively to 
\bea
 \Gamma^{i j k l \; A B} \cD_{A} \cD_{B}  \cE_\gra{16}{8}& = & 0 \ , \label{3Ddiff1} \\
 \Gamma^{kl\, AB}  \Gamma_{ijkl}{}^{CD}  \cD_B \cD_C\cD_D\,  \cE_\gra{16}{4} &=& - 168\,  \Gamma_{ij}{}^{AB} \cD_B \, \cE_\gra{16}{4} \ . \label{3Ddiff2}
\eea
The support of  the $E_{8(8)}(\mathds{Z})$ invariant Eisenstein functions conjectured to define the low energy effective action in type II string theory \cite{Green:2010kv}, on BPS instantons  in the decompactification limit  \cite{Green:2011vz}, indicates that they must indeed satisfy to the differential equations (\ref{3Ddiff1},\ref{3Ddiff2}) such that
\bea
 \cE_\gra{16}{8} = E_{\mbox{\DEVIII{\frac{3}{2}}0000000}} \ , \qquad \cE_\gra{16}{4} = E_{\mbox{\DEVIII{\frac{5}{2}}0000000}}\ . 
\eea
\section{The $\nabla^{4} R^{4}$ invariant in eight dimensions}
\label{8Dd4R4Explicit} 
In this paper we investigate the second order corrections of type $S^\ord{5} \sim \int \cE \nabla^{4} R^{4} + \dots $, which can appear in $\cN = 2$ supergravity in eight dimensions
\be S = \frac{1}{\kappa^2} S^\ord{0} + S^\ord{3} + \kappa^{\frac{4}{3}} S^\ord{5} + \kappa^{2} S^\ord{6} + \sum_{n=7}^\infty \kappa^{\frac{2n}{3}-2} S^\ord{n} \ . \ee 
We denote $i, j$ the $SU(2)$ indices, $a,b $ the vector $SO(1,7)$ indices, and $\alpha, \beta$ and $\dot \alpha, \dot \beta$ the Weyl spinor  indices of positive and negative chirality, respectively. The field content of the theory in the linearised approximation is summarised in the following diagram, \cite{Berkovits:1997pj}
\begin{figure}[htbp]
\def\xmin{0}
\def\ymin{0}
\begin{center}
\resizebox{10 cm}{!}{
 \begin{tikzpicture}
\draw[->,draw=black,very thick] (\xmin,\ymin) -- (\xmin + 3,\ymin - 1);
\draw[->,draw=black,very thick] (\xmin + 3,\ymin - 1) -- (\xmin + 6,\ymin - 2); 
\draw[->,draw=black,very thick] (\xmin + 6,\ymin - 2) -- (\xmin + 9,\ymin - 3); 
 
\draw[->,draw=black,very thick] (\xmin + 9,\ymin - 3) -- (\xmin + 12,\ymin - 4); 
\draw[->,draw=black,very thick] (\xmin + 3,\ymin - 3) -- (\xmin + 6,\ymin - 2); 

\draw[->,draw=black,very thick] (\xmin + 3,\ymin - 3) -- (\xmin + 6,\ymin - 4); 
\draw[->,draw=black,very thick] (\xmin + 6,\ymin - 4) -- (\xmin + 9,\ymin - 5); 

\draw[->,draw=black,very thick] (\xmin,\ymin - 4) -- (\xmin + 3,\ymin - 3); 
\draw[->,draw=black,very thick] (\xmin,\ymin - 4) -- (\xmin + 3,\ymin - 5); 
\draw[->,draw=black,very thick] (\xmin + 3,\ymin - 5) -- (\xmin + 6,\ymin - 6); 

\draw[->,draw=black,very thick] (\xmin + 3,\ymin - 5) --  (\xmin + 6,\ymin - 4); 
\draw[->,draw=black,very thick] (\xmin + 6,\ymin - 4) --  (\xmin + 9,\ymin - 3) ; 

\draw[->,draw=black,very thick] (\xmin,\ymin - 8) -- (\xmin + 3,\ymin - 7);
\draw[->,draw=black,very thick] (\xmin + 3,\ymin - 7) -- (\xmin + 6,\ymin - 6);
\draw[->,draw=black,very thick] (\xmin + 6,\ymin - 6) -- (\xmin + 9,\ymin - 5); 
\draw[->,draw=black,very thick] (\xmin + 9,\ymin - 5) -- (\xmin + 12,\ymin - 4);  
 
\draw[->,draw=black,dashed,thick] (\xmin,\ymin - 4) -- (\xmin + 13,\ymin - 4);   
\draw[->,draw=black,dashed,thick] (\xmin,\ymin - 8.5) -- (\xmin,\ymin + 1);

 \draw (\xmin + 1.2,\ymin + 1) node{$U(1)$ weight};
\draw (\xmin - 0.25,\ymin) node{$4$};
\draw (\xmin - 0.25,\ymin - 1) node{$3$};
\draw (\xmin - 0.25,\ymin - 2) node{$2$};
\draw (\xmin - 0.25,\ymin - 3) node{$1$};
\draw (\xmin - 0.25,\ymin - 4) node{$0$};
\draw (\xmin - 0.4,\ymin - 5) node{$-1$};
\draw (\xmin - 0.4,\ymin - 6) node{$-2$};
\draw (\xmin - 0.4,\ymin - 7) node{$-3$};
\draw (\xmin - 0.4,\ymin - 8) node{$-4$};

\draw (\xmin + 3,\ymin - 4 + 0.3) node{$1/2$};
\draw (\xmin + 6,\ymin - 4 + 0.3) node{$1$};
\draw (\xmin + 9,\ymin - 4 + 0.3) node{$3/2$};
\draw (\xmin + 12,\ymin - 4 + 0.3) node{$2$};
\draw (\xmin + 13,\ymin - 4 + 0.3) node{dim};

\draw (\xmin + 3,\ymin - 0.6) node{$\bar{\chi}^{i}_{\dot \alpha}$};
\draw (\xmin + 3,\ymin - 0.6 - 2) node{$\lambda^{i j k}_{\alpha}$};
\draw (\xmin + 3,\ymin - 0.55 - 4) node{$\bar \lambda^{i j k}_{\dot \alpha}$};
\draw (\xmin + 3,\ymin - 0.6 - 6) node{$\chi^{i}_{\alpha}$};

\draw (\xmin + 6 + 0.4,\ymin - 0.6 - 1) node{$\bar{F}^{ij}_{a b}$/$\bar G^{-}_{a b c d}$};
\draw (\xmin + 6,\ymin  - 4.4) node{${H}^{ij}_{a b c}$};
\draw (\xmin + 6 + 0.4,\ymin  - 6.4) node{$F^{ij}_{a b}$/$G^{+}_{a b c d}$};

\draw (\xmin + 9,\ymin  - 5.4) node{$\rho^{i}_{a b \alpha}$};
\draw (\xmin + 9,\ymin  + 0.4 - 3) node{$\bar \rho^{i}_{a b \dot \alpha}$};

\draw (\xmin + 12,\ymin  - 4.4) node{$R_{a b cd}$};

\draw (\xmin + 0.3,\ymin+0.3) node{$\bar{W}$};
\draw (\xmin + 1.5,\ymin- 0.9) node{$\bar{D}^{i}_{\dot \alpha}$};
\draw (\xmin + 1.5,\ymin- 0.9 - 7) node{$D^{i}_{\alpha}$};
\draw (\xmin + 0.5,\ymin+0.5 - 4) node{$L^{i j k l}$};
\draw (\xmin + 0.3,\ymin+0.4 - 8) node{$W$};

\end{tikzpicture}
}
\end{center}
\caption{\small $\cN = 2$ supermultiplet in eight dimensions, defined from the  chiral superfield $W$ and the isospin $2$ real superfield $L^{ijkl}$.}
\label{fig:8DFieldCont}
\end{figure}
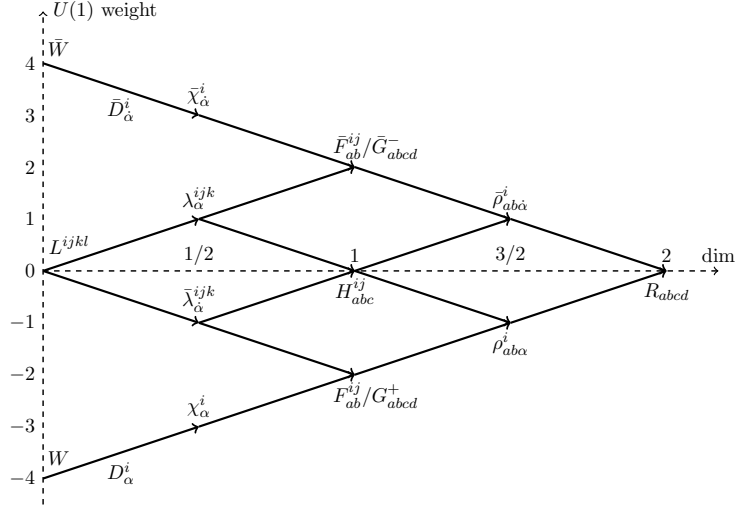

We will perform this analysis within the superform formalism defined in \cite{Gates:1997kr,Gates:1997ag}. In this context, a supersymmetry invariant modulo the classical equations of motion is defined as the integral 
\be \qquad S = \int_{M^8} \iota^* \cL \ , \ee
of the pull-back of a  $d$-closed eight-superform $\mathcal{L}$
\be \label{eq:Closed_SF} d \mathcal{L} = 0 \ , \ee
on an eight-dimensional bosonic subspace $M^8$ embedded in superspace $M^{8|32}$. Because the form is $d$-closed, the integral does not depend on the specific embedding $\iota$, and the integral is supersymmetric modulo the equations of motion. One decomposes the superform $\cL$ in tangent space into $\mathcal{L}_\grad{8 - m - n}{m}{n}$ components, with $8 - m - n$ antisymmetric tangent vector indices $a$, $m$ symmetric pairs of  chiral spinor indices $\alpha,i$ and $n$ symmetric  pairs of anti-chiral indices $\adt,i$. Splitting equation \eqref{eq:Closed_SF} into components we get formally 
\begin{multline} \label{eq:Closed_SF_split} 
\Bigl(D_\grad{1}{0}{0} + T_\grad{1}{1}{0}{}^\grad{0}{1}{0} + T_\grad{1}{0}{1}{}^\grad{0}{0}{1}\Bigr)  \mathcal{L}_\grad{8 - m - n}{m}{n} + T_\grad{2}{0}{0}{}^\grad{0}{0}{1} \mathcal{L}_\grad{7 - m - n}{m}{n + 1} \\ 
+ T_\grad{2}{0}{0}{}^\grad{0}{1}{0} \mathcal{L}_\grad{7 - m - n}{m + 1}{n} + \Bigl(D_\grad{0}{1}{0} + T_\grad{0}{1}{1}{}^\grad{0}{0}{1} + T_\grad{0}{2}{0}{}^\grad{0}{1}{0}  \Bigr)\mathcal{L}_\grad{9 - m - n}{m - 1}{n} \\
+ \Bigl(D_\grad{0}{0}{1} + T_\grad{0}{1}{1}{}^\grad{0}{1}{0}  + T_\grad{0}{0}{2}{}^\grad{0}{0}{1} \Bigr) \mathcal{L}_\grad{9 - m - n}{m}{n - 1} + T_\grad{0}{2}{0}{}^\grad{0}{0}{1} \mathcal{L}_\grad{9 - m - n}{m - 2}{n + 1} \\ 
+ T_\grad{0}{0}{2}{}^\grad{0}{1}{0} \mathcal{L}_\grad{9 - m - n}{m + 1}{n - 2} + T_\grad{1}{1}{0}{}^\grad{0}{0}{1} \mathcal{L}_\grad{8 - m - n}{m - 1}{n + 1}\\
 + T_\grad{1}{0}{1}{}^\grad{0}{1}{0} \mathcal{L}_\grad{8 - m - n}{m + 1}{n - 1}  + T_\grad{0}{1}{1}{}^\grad{1}{0}{0} \mathcal{L}_\grad{10 - m - n}{m - 1}{n - 1} = 0\ , 
\end{multline}
where the torsion components $T_\grad{2-m-n}{m}{n}{}^\grad{1-p-q}{p}{q}$ have their upper indices contracted with the lower ones of the superform component $\cL_\grad{8-m-n}{m}{n}$, with the appropriate combinatoric factor. For a $\nabla^{2k} R^{4}$ type invariant, each component $\mathcal{L}_\grad{8 - m - n}{m}{n}$ has mass dimension $8+2k - m - n$ and $U(1)$ weight $m - n$. We have used the following abbreviations
\be D_\grad{1}{0}{0} \sim D_a \ , \qquad D_\grad{0}{1}{0} \sim D_\alpha^i \ , \qquad D_\grad{0}{0}{1} \sim \bar D_{\adt i} \  ,\ee
as well as
\begin{gather} T_\grad{0}{1}{1}{}^\grad{1}{0}{0} \sim T_\alpha^i{}_{\bdt j}{}^c \ , \CR
T_\grad{0}{2}{0}{}^\grad{0}{0}{1} \sim T_{\alpha\beta}^{ij}{}^{\cdt k} \ , \qquad T_\grad{0}{2}{0}{}^\grad{0}{1}{0} \sim T_{\alpha\beta}^{ij}{}^{\gamma}_k\ , \qquad 
T_\grad{0}{1}{1}{}^\grad{0}{0}{1} \sim T_\alpha^i{}_{\bdt j}{}^{\cdt k} \ , \CR
T_\grad{1}{1}{0}{}^\grad{0}{0}{1} \sim T_a{}_\beta^j{}^{\cdt k}\ ,  \qquad T_\grad{1}{1}{0}{}^\grad{0}{1}{0} \sim T_a{}_\beta^j{}^{\gamma}_k \ , \CR
T_\grad{2}{0}{0}{}^\grad{0}{0}{1} \sim T_{ab}{}^{\cdt k} \ .
 \end{gather}
The explicit action of the covariant derivative and the torsion components have been computed up to mass dimension $3/2$ in \cite{Minimal}.

The complete set of equations \eqref{eq:Closed_SF_split} fixes uniquely the components $\mathcal{L}_\grad{8 - m -n}{m}{n}$ up to $d-$exact terms. But it is enough to enforce some of them to determine the differential equations satisfied by the function of the scalar fields, as was shown for the  $R^{4}$ type invariant in \cite{Minimal}. Here we will extend these results for one class of $\nabla^4 R^4$ type invariants. 
\subsection{Invariant in the linearised approximation} 
In the linearised approximation, the scalar superfields are defined as the chiral superfield $W$ of $U(1)$ weight $-4$ and the isospin $2$ real superfield $L^{ijkl}$ that satisfies to the constraint \cite{Berkovits:1997pj}
\be D_\a^{(i} L^{jklp)} = 0 \ , \qquad D_\adt^{(i} L^{jklp)} = 0 \ , \ee
from which it follows that 
\be D_\a^1 \scal{  (L^{1111})^{2+n} \bar W^{2+m} } =  0 \ . \label{G-analytis8D} \ee
One can therefore define a supersymmetry invariant in the linearised approximation, as
\begin{multline}\label{eq:Lin_Approx}
 \bar D^{16} (D^{2})^{8} \scal{ (L^{1111})^{2 + n} \bar W^{2 + m} }\sim (L^{1111})^{n} \bar W^{m} \left( t_8 t_8 \left( \partial_{a} \partial_{b} R \; \partial^{a} \partial^{b}  R \; R \; R \right)+ \dots \right) + \dots   \\  
+ (L^{1111})^{n - 14} \bar W^{m - 6} (\lambda^{111})^{8} (\bar \lambda^{111})^{8} (\bar \chi^{1})^{8} 
+(L^{1111})^{n - 13} \bar W^{m - 7} (\lambda^{111})^{8} (\bar \lambda^{111})^{7} (\bar \chi^{1})^{8} (\bar \chi^{2})^{1}\\ + \dots 
+(L^{1111})^{n - 6} \bar W^{m - 14} (\lambda^{111})^{8} (\bar \chi^{1})^{8} (\bar \chi^{2})^{8} 
\end{multline}
where the coefficients are not specified, and one understands that the terms in  $\bar W^{m - k} (L^{1111})^{n - l}$ always vanish for $k > m$ or $l > n$. However, this construction cannot be extended to the non-linear level because of the torsion terms 
\be T_\alpha^1{}_\beta^1{}^\gamma_2 = -C_{\alpha \beta} \lambda^{\gamma 111}+ \frac{1}{2} \delta^{\gamma}_{(\alpha} \lambda^{111}_{\beta)}  \, , \qquad T_{\alpha}^1{}_\beta^1{}^{\dot{\gamma} 2} = -2 C_{\alpha \beta}\bar{\chi}^{1 \dot{\gamma}} \ , \ee
that prevent the derivatives $D_{\alpha}^1$ to define vector fields closing among themselves in harmonic superspace. The analysis of these linearised invariants is nonetheless very useful to understand the structure of the corresponding invariant in the full non-linear theory. Considering a linearised invariant defined for an arbitrary analytic function $\cF$ of $L^{1111}$ (which we write $L$ for simplicity) and $\bar W$, we have 
\be \label{FunctionLinear}
 \bar D^{16}  (D^{2})^{8}\cF[\bar W,L]  =  \sum_{p,q} \frac{ \partial^{4+p+q} \cF}{\partial \bar W^{2+q}\partial L^{2+p} }  \mathcal{L}^{\ord{4p}[4q]}_{\rm \scriptscriptstyle lin}\ , \ee
where $\mathcal{L}^{\ord{4p}[4q]}_{\rm \scriptscriptstyle lin}$ are densities of order $4+p+q$ in the fields, that do not depend on the naked scalar fields uncovered by a space-time derivative, as for example 
\bea \cL^{\ord{0}[0]} &\propto&\  t_8 t_8 \left( \partial_{a} \partial_{b} R \; \partial^{a} \partial^{b}  R \; R \; R \right)+ \dots   \CR
\cL^{\ord{8}[8]} &\propto&\  \scal{ t_8  - \tfrac{i}{48} \varepsilon} (\bar F^{11})^4   \scal{t_8  + \tfrac{i}{48} \varepsilon}^2 R^4 +  \dots   \CR
 \mathcal{L}^{\ord{24+4n)}[56-4n)]}_{\rm \scriptscriptstyle lin} &\propto &  (\lambda^{111})^{8} (\bar \lambda^{111})^{8-n} (\bar \chi^{1})^{8} (\bar \chi^{2})^{n}\ . \eea
According to this structure, we expect the non-linear invariant to decompose in the same way in components of $U(1)$ weight multiple of $4$ and even isospin, such that 
\be
  \cL = \sum_{p, q} \bar{U}^{-2 p} \mathcal{F}_{4p [4 q]}(T,\bar T,t)   \mathcal{L}^{\ord{4p} [4 q]}\ ,  \label{Lin800} 
\ee
where $ \bar{U}^{-2 p} \mathcal{F}_{4p [4 q]}$ are tensor functions of the scalar fields $(T,\bar T)\in SL(2)/SO(2)$ and $t\in SL(3)/SO(3)$ of (possibly negative) $U(1)$ weight $-4p$ and isospin $2q$, and $  \mathcal{L}^{\ord{4p} [4 q]}$ are $SL(2)\times SL(3)$ invariant superforms in the dual representation. In the linear approximation, the component $ \mathcal{L}^{\ord{4p} [4 q]}_\grad800$ reduces to $\mathcal{L}^{\ord{4p}[4q]}_{\rm \scriptscriptstyle lin}$, for $p$ prositive. These superforms must satisfy to covariant differential equations in superspace in order for the complete superform $\cL$ to be $d$-closed. Because $ \bar{U}^{-2 p} \mathcal{F}_{p [4 q]}$ are tensor functions of the scalar fields, the only covariant quantities that can enter these equations are the scalar field momenta superforms $P,\, \bar P$ and $P^{ijkl}$. If we assume that there is a unique superform $  \mathcal{L}^{\ord{4p} [4 q]}$ for given $p$ and $q$, as suggested by the linearised analysis, the most general linear  equation consistent with $U(2)$ representation theory is determined up to a rescaling of these superforms as
\begin{multline}  d_\omega \mathcal{L}^{\ord{4p} [4 q]} +2 P^{[4]} \wedge \mathcal{L}^{\ord{4p} [4 q-4]}+ \bar P \wedge  \mathcal{L}^{\ord{4p-4}[4 q]} \\ 
=  a_{p,q} P^{[2]}{}_{ij} \wedge \cL^{\ord{4p}[4q-2]ij} +  b_{p,q}  P_{ijkl} \wedge \cL^{\ord{4p}[4q]ijkl}  + c_{p,q} P \wedge  \mathcal{L}^{\ord{4p +4}[4 q]}\  \end{multline}
for some coefficients $a_{p,q},\, b_{p,q},\, c_{p,q}$. In this notation $[4q]$ refers to $4q$ symmetrised $SU(2)$ indices that are not written explicitly, and identically for a partition $[2][4q-2]$, etc... The closure of the covariant derivative implies moreover the integrability condition \cite{Minimal}
\be   d_\omega^{\; 2}  \cL^{\ord{4p}[4q]}  = - 2 q P^{[1]ijk} \wedge P_{ijkl} \wedge  \cL^{\ord{4p}[4q-1]l} + 2p P\wedge \bar P \wedge\cL^{\ord{4p}[4q]}  \label{Sl3Cs} \ . \ee 
This equation admits for general solution 
\bea && d_\omega \mathcal{L}^{\ord{4p} [4 q]} +2 P^{[4]} \wedge \mathcal{L}^{\ord{4p} [4 q-4]}+ \bar P \wedge  \mathcal{L}^{\ord{4p-4}[4 q]} \CR
&=&   \frac{2q(4 s^\prime-3)}{4q+3} P^{[2]}{}_{ij} \wedge \cL^{\ord{4p}[4q-2]ij}+  \frac{(q+1)(2q+1)(2q+3-2 s^\prime)(2q+2 s^\prime)}{(4q+5)(4q+3)} P_{ijkl} \wedge \cL^{\ord{4p}[4q]ijkl}  \CR
&& \qquad   + \scal{ p(p+1) - s(s-1)}  P \wedge  \mathcal{L}^{\ord{4p +4}[4 q]} \label{RepresentationUSL3}  \ , \hspace{10mm} \eea
for some integration constants $s$ and $s^\prime$. It is natural to define a normalisation of the superform such that the complex conjugate forms do appear with the same coefficient, such as to make manifest the reality condition on the superform. Therefore the definition \eqref{RepresentationUSL3} holds for  strictly positive $p$ only, whereas we will have 
\bea && d_\omega \mathcal{L}^{\ord{0} [4 q]} +2 P^{[4]} \wedge \mathcal{L}^{\ord{0} [4 q-4]} \CR
&=&   \frac{2q(4s^\prime-3)}{4q+3} P^{[2]}{}_{ij} \wedge \cL^{\ord{4p}[4q-2]ij}+  \frac{(q+1)(2q+1)(2q+3-2s^\prime)(2q+2s^\prime)}{(4q+5)(4q+3)} P_{ijkl} \wedge \cL^{\ord{4p}[4q]ijkl}  \CR
&& \qquad  - s(s-1) \bar P \wedge  \mathcal{L}^{\ord{-4}[4 q]} - s(s-1)  P \wedge  \mathcal{L}^{\ord{4}[4 q]}   \ , \hspace{10mm} \eea
for $p=0$ and the complex conjugate condition for strictly negative $-p$, \ie 
\bea && d_\omega \mathcal{L}^{\ord{-4p} [4 q]} +2 P^{[4]} \wedge \mathcal{L}^{\ord{-4p} [4 q-4]}+P \wedge  \mathcal{L}^{\ord{-4p +4}[4 q]} \CR
&=&   \frac{2q(4s^\prime-3)}{4q+3} P^{[2]}{}_{ij} \wedge \cL^{\ord{-4p}[4q-2]ij}+  \frac{(q+1)(2q+1)(2q+3-2s^\prime)(2q+2s^\prime)}{(4q+5)(4q+3)} P_{ijkl} \wedge \cL^{\ord{-4p}[4q]ijkl}  \CR
&& \qquad   + \scal{ p(p+1) - s(s-1)}  \bar P \wedge  \mathcal{L}^{\ord{-4p-4}[4 q]}   \ .  \hspace{10mm} \eea
The range of $p$ can only be bounded if there is a minimal $p$ solution to 
\be p_{\rm \scriptscriptstyle min}(p_{\rm \scriptscriptstyle min}-1) = s(s-1)\  , \ee
such that the exterior differential of the superform set to zero indeed vanishes. This is clearly possible if and only if $s$ is an integer such that $p_{\rm \scriptscriptstyle min}=1-s$ or $s$. For simplicity we will assume that $s$ is indeed a strictly positive integer (we will eventually prove that $s=2$). Because 
\bea && d_\omega \cL^{\ord{4s-4}[4q]}  + 2 P^{[4]} \wedge \cL^{\ord{4s-4}[4q-4]} + \bar P \wedge  \mathcal{L}^{\ord{4s-8}[4 q]} \\
&=&   \frac{2q(4s^\prime-3)}{4q+3} P^{[2]}{}_{ij} \wedge \cL^{\ord{4s-4}[4q-2]ij} +  \frac{(q+1)(2q+1)(2q+3-2s^\prime)(2q+2s^\prime)}{(4q+5)(4q+3)} P_{ijkl} \wedge \cL^{\ord{4s-4}[4q]ijkl} \nn  \eea
it is possible in principle to have $\cL^{\ord{4p}[4q]}=0$ for all $p<s$, but this is generally not the case, and we will see that for the $\nabla^4 R^4$, the gradient expansion rather stops at $p=1-s$.

Using the explicit exterior derivative \eqref{RepresentationUSL3}, one finds that the closed superform is necessarily defined in terms of a unique function such that 
\be 
  \cL[\cE_{s,s^\prime}] =\sum_{q\ge0} \biggl(\   \sum_{p\ge0} \bar{U}^{-2 p}  \bar \cD^p \cD^q_{[4q]} \cE_{s,s^\prime}  \,  \mathcal{L}^{\ord{4p} [4 q]} +  \sum_{p=1}^{s-1} {U}^{-2 p}  \cD^p \cD^q_{[4q]} \cE_{s,s^\prime}  \,  \mathcal{L}^{\ord{-4p} [4 q]}  \biggr) \ , \label{D4R4SuperformGrad}
\ee
and the function must moreover satisfy to 
\bea  \Delta \cE_{s,s^\prime} &=& s(s-1) \cE_{s,t}  \ , \qquad \cD^{s-1} \cE_{s,s^\prime} = 0  \CR
\cD_{ij}{}^{pq} \cD_{klpq} \cE_{s,s^\prime} &=& - \frac{4 s^\prime-3}{12} \cD_{ijkl} \cE_{s,s^\prime}  + \frac{s^\prime ( 2 s^\prime-3)}{18}  ( \varepsilon_{ik} \varepsilon_{jl} +  \varepsilon_{il} \varepsilon_{jk} ) \cE_{s,s^\prime} \label{FunctionConstraint} \ .\eea
In the linearised approximation, \eqref{RepresentationUSL3} reduces to 
\bea  D_\alpha^i   \mathcal{L}^{\ord{4p} [4 q]}_{\rm \scriptscriptstyle lin}  + \partial_a \scal{ (\gamma^{a})_{\alpha \bdt}  \mathcal{L}^{\ord{4p} [4 q]}_{\rm \scriptscriptstyle lin}{}^{\bdt i} }  - 2 \varepsilon^{i[1]}  \lambda_\alpha^{[3]}  \mathcal{L}^{\ord{4p} [4 q-4]}_{\rm \scriptscriptstyle lin}  &=& 0 \ , \CR
\bar  D_{\adt i}   \mathcal{L}^{\ord{4p} [4 q]}_{\rm \scriptscriptstyle lin} +2 \delta_i^{[1]}  \bar \lambda_{\adt}^{[3]}  \mathcal{L}^{\ord{4p} [4 q-4]}_{\rm \scriptscriptstyle lin} +2  \bar \chi_{\adt i}  \mathcal{L}^{\ord{4p-4} [4 q]}_{\rm \scriptscriptstyle lin}  &=& 0 \ , \eea
which is automatically satisfied using the definition \eqref{FunctionLinear} and
\be i  ( \bar  D^{15})^{\adt i}   (D^{2})^{8}\cF[\bar W,L]  =  \sum_{p,q} \frac{ \partial^{4+p+q} \cF}{ \partial \bar W^{2+q}\partial L^{2+p}}  \mathcal{L}^{\ord{4p} [4 q]}_{\rm \scriptscriptstyle lin}{}^{\adt i} \ .  \label{Lin710} \ee

In the next section we will consider the full non-linear superform, concentrating attention on the terms of maximal weight with respect to $U(1)\times SU(2)$.  This will permit to determine the value of the integration constants $s$ and $s^\prime$. Considering the possible terms allowed by representation theory, one obtains that the components of maximal weight are uniquely fixed up to an overall coefficient as
\bea \cL_\grad{8}{0}{0}^{\ord{24}[56]}&\propto& (\bar \chi^{8} )^{[8]} (\lambda^{8})^{[24]} (\bar \lambda^{8})^{[24]} \ , \CR
\cL_\grad{8}{0}{0}^{\ord{28}[52]}&\propto&  (\bar \chi^{9} )^{[7] \dot \alpha} (\lambda^{8})^{[24]} (\bar \lambda^{7})^{[21]}_{\dot \alpha}\ ,  \CR
\cL_\grad{8}{0}{0}^{\ord{32}[48]}&\propto&  (\bar \chi^{10} )^{[6] ab} (\lambda^{8})^{[24]} (\bar \lambda^{6})^{[18]}_{ab} \ , \CR
&\dots&\CR 
\cL_\grad{8}{0}{0}^{\ord{52}[28]}&\propto&  (\bar \chi^{15} )^{[1]\adt} (\lambda^{8})^{[24]} \bar\lambda^{[3]}_{\adt} \ , \CR
\cL_\grad{8}{0}{0}^{\ord{56}[24]}&\propto&  (\bar \chi^{16} ) (\lambda^{8})^{[24]}  \ , \label{MaxWeight}
\eea
where there is always a unique way to define a Lorentz invariant such that the contraction of the indices should not be ambiguous. All these terms already appear in the linearised invariants as depicted in \eqref{eq:Lin_Approx}, suggesting that they multiply the corresponding derivative of the function $\bar D^{6+k} \cD^{14-k}_{[56-4k]} \cE_{s,s^\prime}$ for $k=0$ to $8$, as anticipated in \eqref{D4R4SuperformGrad}.\begin{figure}[htbp]
\center
 \begin{tikzpicture}
 \def\xshift{- 1}
 \def\xmin{0}
 \def\ymin{0}

\draw [dashed] (\xmin + 5/2 , \ymin + 15/2) -- (\xmin + 15/2,\ymin + 5/2);
\draw (\xmin + 14/2,\ymin + 6/2)[rouge] node{\textbullet}; 
\draw (\xmin + 6/2,\ymin + 14/2)[rouge] node{\textbullet};

\draw [dashed] (\xmin + 4/2 , \ymin + 15/2) -- (\xmin + 15/2,\ymin + 4/2);
\draw (\xmin + 14/2,\ymin + 5/2) node{\textbullet}; 
\draw (\xmin + 5/2,\ymin + 14/2) node{\textbullet};

\draw [dashed] (\xmin + 3/2 , \ymin + 15/2) -- (\xmin + 15/2,\ymin + 3/2);
\draw (\xmin + 14/2,\ymin + 4/2) node{\textbullet}; 
\draw (\xmin + 4/2,\ymin + 14/2) node{\textbullet};

\draw [dashed] (\xmin + 2/2 , \ymin + 15/2) -- (\xmin + 15/2,\ymin + 2/2);
\draw (\xmin + 14/2,\ymin + 3/2) node{\textbullet}; 
\draw (\xmin + 3/2,\ymin + 14/2) node{\textbullet};

\draw [dashed] (\xmin + 1/2 , \ymin + 15/2) -- (\xmin + 15/2,\ymin + 1/2);
\draw (\xmin + 14/2,\ymin + 2/2) node{\textbullet}; 
\draw (\xmin + 2/2,\ymin + 14/2) node{\textbullet};

\draw [dashed] (\xmin - 0/2 , \ymin + 15/2) -- (\xmin + 15/2,\ymin - 0/2);
\draw [dashed] (\xmin + 13/2,\ymin - 2/2) -- (\xmin + 15/2,\ymin - 0/2);
\draw (\xmin + 14/2,\ymin + 1/2) node{\textbullet}; 
\draw (\xmin + 1/2,\ymin + 14/2) node{\textbullet};

\draw [dashed] (\xmin - 1/2 , \ymin + 15/2) -- (\xmin + 14/2,\ymin - 0/2);
\draw [dashed] (\xmin + 12/2 , \ymin - 2/2) -- (\xmin + 14/2,\ymin - 0/2);
\draw (\xmin + 14/2,\ymin + 0/2) node{\textbullet}; 
\draw (\xmin + 0/2,\ymin + 14/2) node{\textbullet};
\draw (\xmin + 14/2,\ymin - 1/2) node{\textbullet};

\draw [dashed] (\xmin - 1/2 , \ymin + 14/2) -- (\xmin + 13/2,\ymin - 0/2);
\draw [dashed] (\xmin + 11/2,\ymin - 2/2) -- (\xmin + 13/2,\ymin - 0/2);
\draw (\xmin + 13/2,\ymin - 1/2) node{\textbullet}; 
\draw (\xmin + 0/2,\ymin + 13/2) node{\textbullet};

\draw [dashed] (\xmin - 1/2 , \ymin + 13/2) -- (\xmin + 12/2,\ymin - 0/2);
\draw [dashed] (\xmin + 10/2,\ymin - 2/2) -- (\xmin + 12/2,\ymin - 0/2);
\draw (\xmin + 12/2,\ymin - 1/2) node{\textbullet}; 
\draw (\xmin + 0/2,\ymin + 12/2) node{\textbullet};

\draw [dashed] (\xmin - 1/2 , \ymin + 12/2) -- (\xmin + 11/2,\ymin - 0/2);
\draw [dashed] (\xmin + 9/2,\ymin - 2/2) -- (\xmin + 11/2,\ymin - 0/2);
\draw (\xmin + 11/2,\ymin - 1/2) node{\textbullet}; 
\draw (\xmin + 0/2,\ymin + 11/2) node{\textbullet};

\draw [dashed] (\xmin - 1/2 , \ymin + 11/2) -- (\xmin + 10/2,\ymin - 0/2);
\draw [dashed] (\xmin + 8/2,\ymin - 2/2) --(\xmin + 10/2,\ymin - 0/2);
\draw (\xmin + 10/2,\ymin - 1/2) node{\textbullet}; 
\draw (\xmin + 0/2,\ymin + 10/2) node{\textbullet};

\draw [dashed] (\xmin - 1/2 , \ymin + 10/2) -- (\xmin + 9/2,\ymin - 0/2);
\draw [dashed] (\xmin + 7/2,\ymin - 2/2) -- (\xmin + 9/2,\ymin - 0/2);
\draw (\xmin + 9/2,\ymin - 1/2) node{\textbullet}; 
\draw (\xmin + 0/2,\ymin + 9/2) node{\textbullet};

\draw [dashed] (\xmin - 1/2 , \ymin + 9/2) -- (\xmin + 8/2,\ymin - 0/2);
\draw [dashed]  (\xmin + 6/2,\ymin - 2/2) -- (\xmin + 8/2,\ymin - 0/2);
\draw (\xmin +8/2,\ymin - 1/2) node{\textbullet}; 
\draw (\xmin + 0/2,\ymin + 8/2) node{\textbullet};

\draw [dashed] (\xmin - 1/2 , \ymin + 8/2) -- (\xmin + 7/2,\ymin - 0/2);
\draw [dashed] (\xmin + 5/2,\ymin - 2/2) -- (\xmin + 7/2,\ymin - 0/2);
\draw (\xmin + 7/2,\ymin - 1/2) node{\textbullet}; 
\draw (\xmin + 0/2,\ymin + 7/2) node{\textbullet};

\draw [dashed] (\xmin - 1/2 , \ymin + 7/2) -- (\xmin + 6/2,\ymin - 0/2);
\draw [dashed]  (\xmin + 4/2,\ymin - 2/2) -- (\xmin + 6/2,\ymin - 0/2);
\draw (\xmin + 6/2,\ymin - 1/2) node{\textbullet}; 
\draw (\xmin + 0/2,\ymin + 6/2) node{\textbullet};

\draw [dashed] (\xmin - 1/2 , \ymin + 6/2) -- (\xmin + 5/2,\ymin - 0/2);
\draw [dashed] (\xmin + 3/2,\ymin - 2/2) -- (\xmin + 5/2,\ymin - 0/2);
\draw (\xmin + 5/2,\ymin - 1/2) node{\textbullet}; 
\draw (\xmin + 0/2,\ymin + 5/2) node{\textbullet};

\draw [dashed] (\xmin - 1/2 , \ymin + 5/2) -- (\xmin + 4/2,\ymin -0/2);
\draw [dashed] (\xmin + 2/2,\ymin - 2/2) -- (\xmin + 4/2,\ymin -0/2);
\draw (\xmin + 4/2,\ymin - 1/2) node{\textbullet}; 
\draw (\xmin + 0/2,\ymin + 4/2) node{\textbullet};

\draw [dashed] (\xmin - 1/2 , \ymin + 4/2) -- (\xmin + 3/2,\ymin - 0/2);
\draw [dashed] (\xmin + 1/2,\ymin - 2/2) -- (\xmin + 3/2,\ymin - 0/2);
\draw (\xmin + 3/2,\ymin - 1/2) node{\textbullet}; 
\draw (\xmin + 0/2,\ymin + 3/2) node{\textbullet};

\draw [dashed] (\xmin - 1/2 , \ymin + 3/2) -- (\xmin + 2/2,\ymin - 0/2);
\draw (\xmin + 2/2,\ymin - 1/2) node{\textbullet}; 
\draw (\xmin + 0/2,\ymin + 2/2) node{\textbullet};

\draw [dashed] (\xmin - 1/2 , \ymin + 2/2) -- (\xmin + 1/2,\ymin - 0/2);
\draw (\xmin + 1/2,\ymin - 1/2) node{\textbullet}; 
\draw (\xmin + 0/2,\ymin + 1/2) node{\textbullet};

\draw [dashed] (\xmin  + 0/2 , \ymin - 2/2) -- (\xmin + 2/2,\ymin - 0/2);
\draw (\xmin + 0/2,\ymin + 0/2) node{\textbullet};
\draw (\xmin + 0/2,\ymin - 1/2) node{\textbullet};

\draw [dashed] (\xmin - 1/2 , \ymin - 2/2) -- (\xmin + 1/2,\ymin - 0/2);

\draw[<->,draw=black,thick] (\xmin - 1,\ymin + 8) -- (\xmin - 1,\ymin - 1.5); 
\draw (\xmin - 1,\ymin + 7) node{-}; 
\draw (\xmin - 1,\ymin + 6.5) node{-};
\draw (\xmin - 1,\ymin + 6) node{-};  
\draw (\xmin - 1,\ymin + 5.5) node{-};
\draw (\xmin - 1,\ymin + 5) node{-};
\draw (\xmin - 1,\ymin + 4.5) node{-};
\draw (\xmin - 1,\ymin + 4) node{-};
\draw (\xmin - 1,\ymin + 3.5) node{-}; 
\draw (\xmin - 1,\ymin + 3) node{-}; 
\draw (\xmin - 1,\ymin + 2.5) node{-}; 
\draw (\xmin - 1,\ymin + 2) node{-}; 
\draw (\xmin - 1,\ymin + 1.5) node{-}; 
\draw (\xmin - 1,\ymin + 1) node{-}; 
\draw (\xmin - 1,\ymin + 0.5) node{-}; 
\draw (\xmin - 1,\ymin + 0) node{-};
\draw (\xmin - 1,\ymin - 0.5) node{-};
\draw (\xmin - 1,\ymin - 1) node{-};  

\draw[<-,draw=black,thick] (\xmin + 8,\ymin - 1.5) -- (\xmin ,\ymin - 1.5); 
\draw[-,draw=black,thick] (\xmin + 7,\ymin - 1.5 + 0.07) -- (\xmin + 7,\ymin - 1.5 - 0.07); 
\draw[-,draw=black,thick] (\xmin + 6.5,\ymin -1.5 + 0.07) -- (\xmin + 6.5,\ymin -1.5 - 0.07); 
\draw[-,draw=black,thick] (\xmin + 6,\ymin -1.5 + 0.07) -- (\xmin + 6,\ymin -1.5 - 0.07); 
\draw[-,draw=black,thick] (\xmin + 5.5,\ymin -1.5 + 0.07) -- (\xmin + 5.5,\ymin -1.5 - 0.07); 
\draw[-,draw=black,thick] (\xmin + 5,\ymin -1.5 + 0.07) -- (\xmin + 5,\ymin -1.5 - 0.07); 
\draw[-,draw=black,thick] (\xmin + 4.5,\ymin -1.5 + 0.07) -- (\xmin + 4.5,\ymin -1.5 - 0.07); 
\draw[-,draw=black,thick] (\xmin + 4,\ymin -1.5 + 0.07) -- (\xmin + 4,\ymin -1.5 - 0.07); 
\draw[-,draw=black,thick] (\xmin + 3.5,\ymin -1.5 + 0.07) -- (\xmin + 3.5,\ymin -1.5 - 0.07); 
\draw[-,draw=black,thick] (\xmin + 3,\ymin -1.5 + 0.07) -- (\xmin + 3,\ymin -1.5 - 0.07); 
\draw[-,draw=black,thick] (\xmin + 2.5,\ymin -1.5 + 0.07) -- (\xmin + 2.5,\ymin -1.5 - 0.07); 
\draw[-,draw=black,thick] (\xmin + 2,\ymin -1.5 + 0.07) -- (\xmin + 2,\ymin -1.5 - 0.07); 
\draw[-,draw=black,thick] (\xmin + 1.5,\ymin -1.5 + 0.07) -- (\xmin + 1.5,\ymin -1.5 - 0.07); 
\draw[-,draw=black,thick] (\xmin + 1,\ymin -1.5 + 0.07) -- (\xmin + 1,\ymin -1.5 - 0.07); 
\draw[-,draw=black,thick] (\xmin + 0.5,\ymin -1.5 + 0.07) -- (\xmin + 0.5,\ymin -1.5 - 0.07); 
\draw[-,draw=black,thick] (\xmin,\ymin -1.5 + 0.07) -- (\xmin,\ymin - 1.5 - 0.07); 

\draw (\xmin,\ymin - 2) node{$0$};
\draw (\xmin + 1,\ymin - 2) node{$2$};
\draw (\xmin + 2,\ymin - 2) node{$4$};
\draw (\xmin + 3,\ymin - 2) node{$6$};
\draw (\xmin + 4,\ymin - 2) node{$8$};
\draw (\xmin + 5,\ymin - 2) node{$10$};
\draw (\xmin + 6,\ymin - 2) node{$12$};
\draw (\xmin + 7,\ymin - 2) node{$14$};
\draw (\xmin + 8,\ymin - 2) node{$\mathcal{D}^n_{[4n]}$};

\draw (\xmin - 1.5 ,\ymin - 1) node{$2$};
\draw (\xmin -1.5,\ymin) node{$0$};
\draw (\xmin -1.5,\ymin+1) node{$2$};
\draw (\xmin -1.5,\ymin+2) node{$4$};
\draw (\xmin -1.5,\ymin+3) node{$6$};
\draw (\xmin -1.5,\ymin+4) node{$8$};
\draw (\xmin -1.5,\ymin+5) node{$10$};
\draw (\xmin -1.5,\ymin+6) node{$12$};
\draw (\xmin -1.5,\ymin+7) node{$14$};
\draw (\xmin -1.5,\ymin+8) node{$\bar{\mathcal{D}}^{m}$};
\draw (\xmin -1.5,\ymin - 1.5) node{$\mathcal{D}^{m}$};

\draw (\xmin + 4.5,\ymin+7) node{$\color{rouge} \bar{\mathcal{D}}^{14} \mathcal{D}^{6}_{[24]} \mathcal{E}_\grad{2}{1}{0}$};
\draw (\xmin + 8,\ymin+ 3.4) node{$\color{rouge} \bar{\mathcal{D}}^{6} \mathcal{D}^{14}_{[56]} \mathcal{E}_\grad{2}{1}{0}$};

\draw (\xmin + 13/2,\ymin + 7/2) node{\textbullet};
\draw (\xmin + 12/2,\ymin + 8/2) node{\textbullet}; 
\draw (\xmin + 11/2,\ymin + 9/2) node{\textbullet}; 
\draw (\xmin + 10/2,\ymin + 10/2) node{\textbullet}; 
\draw (\xmin + 9/2,\ymin + 11/2) node{\textbullet}; 
\draw (\xmin + 8/2,\ymin + 12/2) node{\textbullet};
\draw (\xmin + 7/2,\ymin + 13/2) node{\textbullet}; 

\draw (\xmin + 9.1,\ymin+ 2.4) node{$\bar{\mathcal{D}}^{20-n} \mathcal{D}^{n}_{[4 n]} \mathcal{E}_\grad{2}{1}{0}$};
\draw (\xmin + 9.1,\ymin+ 1.4) node{$\bar{\mathcal{D}}^{18-n} \mathcal{D}^{n}_{[4 n]} \mathcal{E}_\grad{2}{1}{0}$};
\draw (\xmin + 9.1,\ymin+ 0.4) node{$\bar{\mathcal{D}}^{16-n} \mathcal{D}^{n}_{[4 n]} \mathcal{E}_\grad{2}{1}{0}$};
\draw (\xmin + 7.8,\ymin - 0.5) node{$\dots$};

\end{tikzpicture}\caption{\small Gradient expansion of the $1/4$ BPS $\nabla^4 R^4$ type invariant in eight dimensions}
\end{figure}
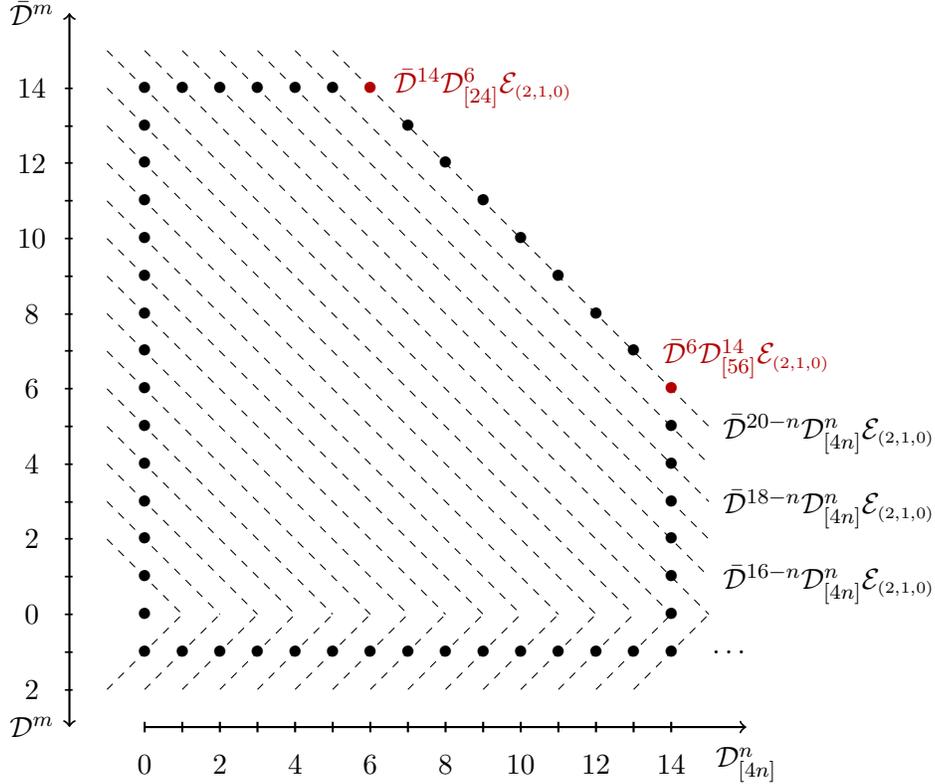

However, in eight dimensions it is not true that all linearised invariants can be written as harmonic superspace integrals, and it is not clear if all linearised invariants do extend to full non-linear invariants. Therefore one cannot rely blindly on the linearised analysis, and we will not assume the closed superform defining the invariant to admit the gradient expansion \eqref{D4R4SuperformGrad} in the following section. Our computation will retrospectively confirm that the structure of the invariant is indeed the one suggested by the linearised analysis, and we will be able to conclude that the invariant admits indeed the gradient expansion \eqref{D4R4SuperformGrad} for $s=1$ and $s^\prime=-\frac{1}{2}$.

\subsection{Constraints on highest R-symmetry weight terms}

We will consider a completely general ansatz for the components of the closed superform 
\be \label{Ansatz} 
 \cL_\grad{8-m- n}{m}{n} = \sum_{\mathpzc{a}, p, q} \bar{U}^{-2 p} \mathcal{F}_{4p [2 q]}^{\mathpzc{a}}(T, \bar T, t) \cI_\grad{8-m- n}{m}{n}^{\mathpzc{a} \, \ord{4p} [2 q]} \ , 
\ee
where $\bar{U}^{-2 p} \mathcal{F}_{4p [2 q]}^{\mathpzc{a}}$ are tensor functions of the scalar fields of $U(1)$ weight $-4p$ and isospin $q$, whereas $\mathpzc{a}$ labels the possible $SL(2) \times SL(3)$ tensors  $ \cI_\grad{8-m- n}{m}{n}^{\mathpzc{a} \, \ord{4p} [2 q]}$ in the appropriate representations of $U(1)\times SU(2)\times Spin(1,7)$ associated to the corresponding grading $(8-m-n,m,n)$. $\mathcal{I}_\grad{8-m- n}{m}{n}^{\mathpzc{a} \, \ord{4p} [2 q]}$ have  $U(1)$ weight  $m-n+4p$ and isospin $j$ such that $q-\tfrac{m+n}{2}\le j\le q+\tfrac{m+n}{2}$, depending of the specific tensor structure for the symmetrised pairs of fermionic indices. Note that we do not assume $q$ to only take even values, as suggested from the linearised analysis in the preceding section, although we will eventually conclude that it must indeed be even. 

We will concentrate on the maximal mass dimension components of the $d$-closure equations \eqref{eq:Closed_SF_split}, \ie 
\begin{multline} \label{eq:DBarI800}
 D_\grad{0}{0}{1} \cL_\grad{8}{0}{0} +  \left( D_\grad{1}{0}{0} + T_\grad{1}{0}{1}{}^\grad{0}{0}{1} \right) \cL_\grad{7}{0}{1} + T_\grad{1}{0}{1}{}^\grad{0}{1}{0} \mathcal{L}_\grad{7}{1}{0}  \\  + T_\grad{2}{0}{0}{}^\grad{0}{0}{1} \mathcal{L}_\grad{6}{0}{2} + T_\grad{2}{0}{0}{}^\grad{0}{1}{0} \mathcal{L}_\grad{6}{1}{1} = 0 
\end{multline}
\vspace{-10mm}
\begin{multline} \label{eq:DI800}
 D_\grad{0}{1}{0} \mathcal{L}_\grad{8}{0}{0} +  \left( D_\grad{1}{0}{0} + T_\grad{1}{1}{0}{}^\grad{0}{1}{0} \right) \mathcal{L}_\grad{7}{1}{0} + T_\grad{1}{1}{0}{}^\grad{0}{0}{1} \mathcal{L}_\grad{7}{0}{1}  \\
 + T_\grad{2}{0}{0}{}^\grad{0}{1}{0} \mathcal{L}_\grad{6}{2}{0}  + T_\grad{2}{0}{0}{}^\grad{0}{0}{1} \mathcal{L}_\grad{6}{1}{1} = 0
\end{multline}
In order to simplify further these equations we will moreover restrict ourselves to the analysis of the terms of highest  $U(1)$ weight and carrying the maximal amount of symmetrised $SU(2)$ indices, which correspond to the terms with maximal value of $p$ and $q$ in \eqref{Ansatz}. 

Let us consider first the components of $\mathcal{L}_\grad{8}{0}{0}$, that are by construction Lorentz scalars of mass dimension $12$. Each $\mathcal{I}_\grad{8}{0}{0}^{\mathpzc{a} \, \ord{4p} [2 q]}$ is therefore a Lorentz scalar of mass dimension $12$,  $U(1)$ weight $4 p$ and isospin $q$. The terms of maximal weight depends only on the fermions fields, because they have the lowest mass dimension while carrying the largest weight representation. However, Fermi statistics requires to limit the number of them to maximise the weight. For example, there are only eight different $\lambda_\alpha^{111}$, so a term in $(\lambda^9)_\alpha$ will necessarily includes at least one $\lambda_\alpha^{112}$, such that the maximal $SU(2)$ representation one obtains for an octic term is $ (\lambda^{8})^{[24]}$ is of isospin 12, while for nine fermions one only gets  $ (\lambda^{9})_\alpha^{[25]}$ of isospin $\frac{25}{2}$. A term with ten fermions $ (\lambda^{10})_{ab}^{[26]}$ has therefore the same mass dimension and $U(2)$ representation as a term in $(\lambda^{8})^{[24]} \bar F_{ab}^{[2]}$. The same argument applies to the sixteen fermion fields $\bar \chi_{\adt}^i$. The terms of maximal weight involving scalar momenta can always be eliminated in favour of lower weight terms through the addition of a $d$-exact term, and will  therefore be disregarded in our analysis. 

The maximal weight terms are therefore the terms of order  24 in the fermions depicted in \eqref{MaxWeight}. We shall here concentrate on the two monomials 
\be \label{eq:8lambda8lambdabar}
 \mathcal{I}^{\mathpzc{1} \, \ord{24} [56]} = (\bar \chi^{8} )^{[8]} (\lambda^{8})^{[24]} (\bar \lambda^{8})^{[24]} \ , \qquad  
 \mathcal{I}^{\mathpzc{2} \, \ord{28} [52]} = (\bar \chi^{9} )^{[7] \dot \alpha} (\lambda^{8})^{[24]} (\bar \lambda^{7})^{[21]}_{\dot \alpha} \ . \ee
The next-to-maximal contribution with a lower isospin could have been $\mathcal{I}_\grad{8}{0}{0}^{\mathpzc{a}\, \ord{24} [54]}$, however the only possible terms must also be of order 24 in the fermions and one checks that there is no Lorentz scalar in this representation. Indeed, lowering the isospin of one of the octic monomial $(\bar \chi^{8} )^{[8]} $, $(\lambda^{8})^{[24]} $ or $(\bar \lambda^{8})^{[24]}$ requires to consider only seven among eight of the $Spin(1,7)$ indices to be antisymmetrised, such that they cannot be scalars. The same reasoning applies to the terms of order nine and seven in $(\bar \chi^{9} )^{[7] \dot \alpha} $ and  $(\bar \lambda^{7})^{[21]}_{\dot \alpha}$, respectively, such that there is no candidate components $\mathcal{I}_\grad{8}{0}{0}^{28 [52]}$ either. It is also clear that one cannot reduce the $U(1)$ weight by $2$ only, since the difference of the $U(1)$ weights of the fermion fields of  identical chirality is zero modulo four. 

Therefore the non-vanishing next to maximal weight terms have $4p+2q=76$. In this case there is always more than one possibility, and one obtains for example three independent $\mathcal{I}_\grad{8}{0}{0}^{\mathpzc{a}\, \ord{20} [56]}$ components 
\bea \label{NextMaxW} 
 \mathcal{I}^{\mathpzc{3} \, \ord{20} [56]} &=& \bar{F}^{[2]}_{a b} (\bar \chi^{6})^{a b \, [6] } (\lambda^{8})^{[24]} (\bar \lambda^{8})^{[24]} \ , \CR
 \mathcal{I}^{\mathpzc{4} \,  \ord{20}  [56]} &=& (\bar \chi^{7})^{[7]}_{\dot \alpha} (\lambda^{8})^{[24]} (\bar \lambda^{9})^{\dot \alpha \; [25]} \ , \CR
 \mathcal{I}^{\mathpzc{5} \,  \ord{20}  [56]} &=& (\bar \chi^{6})^{a b \; [6]} (\lambda^{10})_{a b}^{[26]} (\bar \lambda^{8})^{[24]}\ , 
\eea
where we do not consider the fourth possible component in $P^{[4]} (\bar\chi^{7})^{[7]} (\lambda^{7})^{[21]} (\bar \lambda^{8})^{[24]}$, because such a term can always be eliminated in favour of lower weight terms through the addition of a $d$-exact term in $D_\grad{1}{0}{0} \mathcal{L}_\grad{7}{0}{0}$.  Altogether, we will therefore consider the following ansatz for $\mathcal{L}_\grad{8}{0}{0}$
\begin{multline} \label{eq:L800_Ansatz}
 \mathcal{L}_{abcdefgh} = \ve_{a b c d e f g h}  \biggl( \bar{U}^{-12} \mathcal{F}_{24 [56]}^{\mathpzc{1}} (\bar \chi^{8} )^{[8]} (\lambda^{8})^{[24]} (\bar \lambda^{8})^{[24]}  + \bar{U}^{-14} \mathcal{F}_{28 [52]}^{\mathpzc{2}} (\bar \chi^{9} )^{[7] \dot \alpha} (\lambda^{8})^{[24]} (\bar \lambda^{7})^{[21]}_{\dot \alpha} \\
 + \bar{U}^{-10} \mathcal{F}_{20 [56]}^{\mathpzc{3}}  \bar{F}^{[2]}_{a b} (\bar \chi^{6})^{a b \, [6] } (\lambda^{8})^{[24]} (\bar \lambda^{8})^{[24]} 
+  \bar{U}^{-10} \mathcal{F}_{20 [56]}^{\mathpzc{4}}   (\bar \chi^{7})^{[7]}_{\dot \alpha} (\lambda^{8})^{[24]} (\bar \lambda^{9})^{\dot \alpha \; [25]}  \\
  + \bar{U}^{-10}  \mathcal{F}_{20 [56]}^{\mathpzc{5}}   (\bar \chi^{6})^{a b \; [6]} (\lambda^{10})_{a b}^{[26]} (\bar \lambda^{8})^{[24]}  +  \sum_{ \mathpzc{a}, p \leq 4} \bar{U}^{-2 p} \mathcal{F}_{4p [56]}^{\mathpzc{a}} \mathcal{I}^{\mathpzc{a} \, \ord{4p} [56 ]} \\
   +  \sum_{ \mathpzc{a}, p \leq 6} \bar{U}^{-2 p} \mathcal{F}_{4p [52]}^{\mathpzc{a}} \mathcal{I}^{\mathpzc{a} \, \ord{4p} [52]}  +  \sum_{ \mathpzc{a}, p, q  \leq 25} \bar{U}^{-2 p} \mathcal{F}_{4p [2 q]}^{\mathpzc{a}} \mathcal{I}^{\mathpzc{a} \, \ord{4p} [2 q]} \biggr)
\end{multline}
where the components that are not specified will be irrelevant in our analysis. 

We must also consider the other components of the superform, corresponding to the terms involving naked gravino fields in the formalism in components. The superform component $\mathcal{L}_\grad{7}{1}{0}$ is in the $Spin(1,7)$ representation tensor product of the $7$-form times the positive chirality spinor representation, \ie \DSOVIII1001 or \DSOVIII0010. It has  $U(1)$ weight $1$ and mass dimension $23/2$. The maximal weight term that one can possibly have in this representation is simply obtained by removing one fermion field to the maximal weight term in the component $\cL_\grad800$, and the only possible such term is therefore in the \DSOVIII0010 of $Spin(1,7)$, \ie 
\be
 \mathcal{I}_{a}{}_{\alpha}^{i\,\mathpzc{6} \, 20 [56]} = \ve^{i[1]} (\gamma_a)_{\alpha}{}^{\dot \beta} (\bar \chi^{7})_{\dot \beta}^{[7]} (\lambda^{8})^{[24]} (\bar \lambda^{8})^{[24]}\ . \label{chi7lambda16} 
\ee
We consider therefore the ansatz
\begin{multline}  \label{eq:L710_Ansatz}
 \mathcal{L}_{\alpha a b c d e f g}^{i} = \ve_{a b c d e f g}{}^{h} \biggl( \bar{U}^{-10} \ve^{i [1]}\mathcal{F}_{20 [56]}^{\mathpzc{6}} (\gamma_{h})_{\alpha}{}^{\dot \alpha} (\bar \chi^{7})_{\dot \alpha}^{[7]} (\lambda^{8})^{[24]} (\bar \lambda^{8})^{[24]} + \sum_{\mathpzc{a}, p \leq 4 } \bar{U}^{-2 p} \mathcal{F}_{{4p} [56]}^{\mathpzc{a}} \mathcal{I}_{\alpha h}^{i \; \mathpzc{a} \, \ord{4 p} [56]} \\
 + \sum_{\mathpzc{a}, p, q \leq 27} \bar{U}^{- 2 p} \mathcal{F}_{4p [2 q]}^{\mathpzc{a}} \mathcal{I}_{\alpha h}^{i\;\mathpzc{a} \, \ord{4 p} [2 q]} \biggr)\ . 
\end{multline}
$\mathcal{L}_\grad{7}{0}{1}$ is instead in the direct sum of the \DSOVIII1010 and the  \DSOVIII0001, and admits a $U(1)$ weight $-1$. Because of the chirality of the representations, it cannot admit components in $\bar \chi^7 \lambda^8 \bar \lambda^8$ and the maximal weight components rather include  terms in 
\be
 \mathcal{I}_{a}{}_{\adt}^{i\,\mathpzc{7} \, 20 [56]} = \ve^{i[1]} (\gamma^b)_{\adt}{}^{\beta} (\bar \chi^{6})_{ab}^{[7]} (\lambda^{9})^{[25]}_\beta (\bar \lambda^{8})^{[24]} \ , 
\ee
and others in the same representation of $U(2)$, such that the general ansatz for $\mathcal{L}_\grad{7}{0}{1}$ takes the form 
\be  \label{eq:L701_Ansatz}
 \mathcal{L}_\grad{7}{0}{1} = \sum_{ \mathpzc{a}, p \leq 5} \bar{U}^{-2 p} \mathcal{F}_{4p [56]}^{\mathpzc{a}} \mathcal{I}_\grad{7}{0}{1}^{\mathpzc{a} \, \ord{4p} [56]}  +  \sum_{ \mathpzc{a}, p, q  \leq 27} \bar{U}^{-2 p} \mathcal{F}_{4p [2 q]}^{\mathpzc{a}} \mathcal{I}_\grad{7}{0}{1}^{\mathpzc{a} \, \ord{4p} [2 q]}\ . 
\ee
We will not need to specify any of these terms in our analysis. The $\mathcal{L}_\grad{6}{1}{1}$ component is of mass dimension $11$ and $U(1)$ weight $0$. Chirality implies that the highest weight terms one can build in the relevant representations of the Lorentz group are in $\bar \chi^5 \lambda^9 \bar \lambda^8$ or $\bar F \bar \chi^4\lambda^8 \bar \lambda^8$, as for example 
\be
 \mathcal{I}_{ab}{}_\alpha^i{}_{\bdt}^j{}^{\mathpzc{8} \, 16 [56]} = \ve^{i[1]} \ve^{j[1]} (\gamma_{[a})_{\alpha \bdt}   (\gamma_{b]})^{\cdt \delta}  (\bar \chi^{5})_{\cdt}^{[5]} (\lambda^{9})^{[25]}_\delta  (\bar \lambda^{8})^{[24]} \ , 
\ee
and other terms of the same weight, such that the ansatz is of the form 
\be  \label{eq:L611_Ansatz}
 \mathcal{L}_\grad{6}{1}{1} = \sum_{ \mathpzc{a}, p \leq 4} \bar{U}^{-2 p} \mathcal{F}_{4p [56]}^{\mathpzc{a}} \mathcal{I}_\grad{6}{1}{1}^{\mathpzc{a} \, \ord{4p} [56]} +  \sum_{ \mathpzc{a}, p, q  \leq 27} \bar{U}^{-2 p} \mathcal{F}_{4p [2 q]}^{\mathpzc{a}} \mathcal{I}_\grad{6}{1}{1}^{\mathpzc{a} \, \ord{4p} [2 q]} \ . 
\ee
Note moreover that terms of odd isospin are expected to vanish. Finally the $\mathcal{L}_\grad{6}{0}{2}$ component has mass dimension $11$ and $U(1)$ weight $-2$. The maximal isospin terms are in  $\lambda^{8} \bar \lambda^8 \bar \chi^6$ as for example 
\be
 \mathcal{I}_{ab}{}_\alpha^i{}_{\beta}^j{}^{\mathpzc{10} \, 20 [56]} = \ve^{i[1]} \ve^{j[1]}C_{\alpha\beta} (\bar \chi^{6})_{ab}^{[6]} (\lambda^{8})^{[24]} (\bar \lambda^{8})^{[24]} \  , 
\ee
and other Lorentz tensor combinations such that the ansatz is 
\be  \label{eq:L602_Ansatz}
 \mathcal{L}_\grad{6}{0}{2} = \sum_{ \mathpzc{a}, p \leq 5} \bar{U}^{-2 p} \mathcal{F}_{4p [56]}^{\mathpzc{a}} \mathcal{I}_\grad{6}{0}{2}^{\mathpzc{a} \, \ord{4p} [56]}  +  \sum_{ \mathpzc{a}, p, q  \leq 27} \bar{U}^{-2 p} \mathcal{F}_{4p [2 q]}^{\mathpzc{a}} \mathcal{I}_\grad{6}{0}{2}^{\mathpzc{a} \, \ord{4p} [2 q]} \ . 
\ee

Let us now describe the action of the fermionic covariant derivatives on a general tensor function $\bar{U}^{-2 p} \mathcal{F}_{4p [2 q]}^{\mathpzc{a}}$. Since the tensor transforms covariantly with respect to $U(2)$, one obtains 
\bea \label{eq:D_on_Equations}
 \bar D_{\dot \alpha i} \left(\bar{U}^{-2 p} \mathcal{F}_{4p [2 q]}^{\mathpzc{a}}(T, \bar T, t) \right) &=& \bar{U}^{-2 (p+1)} \bar{\mathcal{D}} \mathcal{F}_{4p [2 q]}^{\mathpzc{a}} \bar P_{\dot \alpha i} +
2 \bar{U}^{-2 p} \mathcal{D}_{j k l m}\mathcal{F}_{4p [2 q]}^{\mathpzc{a}} \bar P_{\dot \alpha i}^{j k l m} \ , \CR
 D_{\alpha}^{i} \left(\bar{U}^{-2 p} \mathcal{F}_{4p [2 q]}^{\mathpzc{a}}(T, \bar T, t) \right) &=& \bar{U}^{-2 (p -1)}  (1 - T \bar{T} )^2 \mathcal{D}  \mathcal{F}_{4p  [2 q]}^{\mathpzc{a}} P_{\alpha}^{i} +
2 \bar{U}^{-2 p} \mathcal{D}_{j k l m}\mathcal{F}_{4p [2 q]}^{\mathpzc{a}} P_{\alpha}^{i \; j k l m} \ ,  \CR
 D_{a} \left(\bar{U}^{-2 p} \mathcal{F}_{4p [2 q]}^{\mathpzc{a}}(T, \bar T, t) \right) &=& \bar{U}^{-2 (p -1)} (1 - T \bar{T} )^2 \mathcal{D} \mathcal{F}_{4p [2 q]}^{\mathpzc{a}} P_{a} + \bar{U}^{-2 (p+1)} \bar{\mathcal{D}} \mathcal{F}_{4p [2 q]}^{\mathpzc{a}} \bar{P}_{a} \CR
  && \hspace{30mm} + 2 \bar{U}^{-2 p} \mathcal{D}_{ij k l }\mathcal{F}_{4p [2 q]}^{\mathpzc{a}} P_{a}^{ij k l}\ , 
\eea
where the field $T$ is the unit disk coordinate on $SL(2)/SO(2)$, and $U$ is the $U(1)$ weight $-2$ variable satisfying to
\be U \bar U (1-T \bar T ) = 1\ . \ee
The momentum components were derived in \cite{Minimal} to be
\be
 P_{\alpha}^{i} = 2 \chi_{\alpha}^{i} \ , \quad \bar P_{\dot \alpha i } = 2 \bar \chi_{\dot \alpha i } \ , \quad P_{\alpha}^{i \; j k l m} = - \ve^{i (j} \lambda_{\alpha}^{k l m)} \ , \quad \bar P_{\dot \alpha i}^{j k l m} = \delta_{i}^{(j} \bar \lambda_{\dot \alpha}^{k l m)} \ . 
\ee
It is helpful to decompose $\mathcal{D}_{j k l m} \mathcal{F}_{4p [2 q]}^{\mathpzc{a}}$ into irreducible representations, as
\begin{multline}   \label{eq:Irr_representations}
 \mathcal{D}_{i j k l}\mathcal{F}_{4p [2 q]}^{\mathpzc{a}} P^{i j k l} = \cD_{(i j k l} \cF_{4p [2 q])}^{\mathpzc{a}} P^{i j k l} +\frac{4 q}{(q + 2)}  \ve_{i [1]} \cD_{(j k l}{}^{m} \cF_{4p [2 q - 1] )m}^{\mathpzc{a}} P^{i j k l} \\
+  \frac{6 (2 q -1) q }{(q+1) (2 q+3)} \ve_{i [1]}  \ve_{j [1]} \cD_{(k l}{}^{m n} \cF_{4p [2 q - 2] )m n}^{\mathpzc{a}} P^{i j k l}  \\
 +  \frac{4 (q-1) (2 q - 1)}{(2 q + 1)(q+1)} \ve_{i [1]}  \ve_{j [1]}  \ve_{k [1]} \cD_{(l}{}^{m n r} \cF_{4p [2 q - 3]) m n r}^{\mathpzc{a}} P^{i j k l} \\
+  \frac{6}{(2 q+1) (q- 1) (2 q- 1) q} \ve_{i [1]}  \ve_{j [1]}  \ve_{k [1]} \ve_{l [1]}\cD^{m n r s} \cF_{4p [2 q - 4] m n r s}^{\mathpzc{a}} P^{i j k l}  \ , 
\end{multline}
where we denote with $(i_{1} \dots i_{2n})$ the symmetrisation of $2n$ indices, while the numbers into brackets sum up to the total number of symmetrised indices $i_1,\, i_{2q}$ that are not written explicitly. One understands that the uncontracted indices of the terms in $\cD \cF_{4p[2q]}$ are symmetrised first, and all the indices  $i_1,\, i_{2q}$ that are not written explicitly are symmetrised afterward.

We are now ready to solve equation  \eqref{eq:DBarI800}  in terms of our ansatz (\ref{eq:L800_Ansatz},\ref{eq:L710_Ansatz},\ref{eq:L701_Ansatz}), \ie 
\begin{multline} \label{eq:DBarI800_FullForm}
 \bar{D}_{\dot \alpha i} \mathcal{L}_{a b c d e f g h} +  8 D_{[a} \mathcal{L}_{b c d e f g h] \dot \alpha i} + 8 T_{\dot \alpha i [a}{}^{\dot \beta j}  \mathcal{L}_{\dot \beta j | b c d e f g h]}  
 + 8 T_{\dot \alpha i [a}{}^{\beta}_{j} \mathcal{L}_{\beta | b c d e r g h]}^{j} \\
  + 28 T_{[ a b}{}^{\dot \beta j} \mathcal{L}_{\dot \beta j c d e f g h] \dot \alpha i}  + 28 T_{[ a b}{}^{\beta}_{j} \mathcal{L}_{\beta c d e f g h] \dot \alpha i}^{j} = 0 \ . 
\end{multline}
We shall only consider the mixings between the terms involving tensor functions of $U(1)$ weight lower or equal to $-24$ and of isospin $28$. As a consequence of \eqref{eq:L701_Ansatz}, there is no mixing contribution coming from $T_\grad{1}{0}{1}{}^\grad{0}{0}{1} \mathcal{L}_\grad{7}{0}{1}$ at this weight, and these terms can be disregarded. However, there are  contributions from $D_\grad{1}{0}{0} \mathcal{L}_\grad{7}{0}{1}$, because the application of the derivative to the tensor functions can increase the weight. Those mixings are nonetheless either proportional to $\bar P$ or to $P^{i j k l}$, and we can neglect them as long as one does consider terms involving explicitly the scalar momenta. Disregarding these terms will allow us to simplify drastically the computation in the following. Because the maximal weight terms in the ansatz  \eqref{eq:L710_Ansatz} are associated to tensor functions of $U(1)$ weight $-20$ and isospin $28$, the terms in  $T_\grad{1}{0}{1}{}^\grad{0}{1}{0} \mathcal{L}_\grad{7}{1}{0}$ do not contribute either in the computation. Because the isospin 28 terms in $\mathcal{L}_\grad{6}{1}{1}$ and $\mathcal{L}_\grad{6}{0}{2}$ are all associated to tensor function of $U(1)$ weight greater than $-20$, we can also disregard the terms in $T_\grad{2}{0}{0}{}^\grad{0}{1}{0} \mathcal{L}_\grad{6}{1}{1}$ and $T_\grad{2}{0}{0}{}^\grad{0}{0}{1} \mathcal{L}_\grad{6}{0}{2}$ in equation \eqref{eq:DBarI800_FullForm}.

We get therefore that equation \eqref{eq:DBarI800_FullForm} simplifies drastically to
\be \label{eq:DBarI800_Simplified}
 \bar{D}_{\dot \alpha i} \mathcal{L}_{a b c d e f g h}  \approx  0 \ , 
\ee
when restricted to the terms involving tensors functions of isospin  $28$ and of $U(1)$ weight less or equal to $ - 24$.

In order to solve \eqref{eq:DBarI800_Simplified}, it will be convenient to define an explicit basis of fermion fields monomials as follows
\be
\begin{split}
(\bar \chi^{6})^{(i_1 \dots i_6)}_{a b} &\equiv \frac{1}{6!} (\gamma_{a b})^{\dot \alpha \dot \beta} \ve_{\dot \alpha \dot \beta}{}^{\dot \gamma \dots \dot \delta} \bar \chi^{(i_1}_{\dot \gamma} \hspace{-1mm}\dots \bar \chi^{i_6)}_{\dot \delta} \ , \\
 \left(\bar \chi^{7} \right)^{(i_1 \dots i_{7})}_{\dot \alpha} &\equiv  \frac{1}{7!} \ve_{\dot \alpha}{}^{\dot \beta \dots \dot \gamma} \bar \chi_{\dot \beta}^{(i_1} \dots \bar \chi_{\dot \gamma}^{i_7)} \ , \\
\left(\bar\chi^{8} \right)^{(i_1\dots i_8)} &\equiv \frac{1}{8!} \ve^{\dot \alpha \dots \dot \beta} \bar \chi_{\dot \alpha}^{( i_1} \dots \bar \chi_{\dot \beta}^{i_8)} \ , \\
\left(\bar\chi^{9} \right)^{(i_1\dots i_7)}_{\dot \alpha} &\equiv \left(\bar\chi^{8} \right)^{(i_1\dots i_7) j} \bar \chi_{\dot \alpha j} \ , \\
\left(\bar\chi^{10} \right)^{(i_1\dots i_6)}_{a b} &\equiv (\gamma_{a b})^{\dot \alpha \dot \beta} \left(\bar\chi^{9} \right)^{(i_1\dots i_6) j}_{\dot \alpha} \bar \chi_{\dot \beta j} \ , 
\end{split} \hspace{2 mm}
\begin{split}
\left(\bar \lambda^{6} \right)^{(i_1 \dots i_{18})}_{a b}  & \equiv \frac{1}{6!}  (\gamma_{a b})^{\dot \alpha \dot \beta} \ve_{\dot \alpha \dot \beta}{}^{\dot \gamma \dots \dot \delta} \bar \lambda_{\dot \gamma}^{(i_1 i_2 i_3} \hspace{-2mm}\dots \bar \lambda_{\dot \delta}^{i_{16} i_{17} i_{18})}  , \\
\left(\bar \lambda^{7} \right)^{(i_1 \dots i_{21})}_{\dot \alpha} & \equiv  \frac{1}{7!} \ve_{\dot \alpha}{}^{\dot \beta \dots \dot \gamma} \bar \lambda_{\dot \beta}^{(i_1 i_2 i_3} \dots \bar \lambda_{\dot \gamma}^{i_{19} i_{20} i_{21})} \ , \\
\left(\bar \lambda^{8} \right)^{(i_1 \dots i_{24})} & \equiv \frac{1}{8!} \ve^{\dot \alpha \dots \dot \beta}\bar \lambda_{\dot \alpha}^{(i_1 i_2 i_3} \dots \bar \lambda_{\dot \beta}^{i_{22} i_{23} i_{24})} \ , \\
\left(\bar \lambda^{9} \right)^{(i_1 \dots i_{25})}_{\dot \alpha} &\equiv \left( \bar \lambda^{8}\right)^{j(i_1 \dots i_{23}} \bar \lambda_{\dot \alpha }^{i_{24} i_{25})}{}_j \ , \\
\left(\bar \lambda^{10} \right)^{(i_1 \dots i_{26})}_{a b} &\equiv (\gamma_{a b})^{\dot \alpha \dot \beta}\left( \bar \lambda^{9}\right)^{j(i_1 \dots i_{24}}_{\dot \alpha} \bar \lambda_{\dot \beta }^{i_{25} i_{26})}{}_j \ . 
\end{split}
\ee
Let us first consider the action  of the fermionic derivative $\bar{D}_{\dot \alpha i }$ on the tensor function  $\bar U^{-12} \mathcal{F}_{24 [56]}^{\mathpzc{1}} $ in $\mathcal{L}_\grad{8}{0}{0}$, \ie 
\bea \label{EquaDF56} 
&& \bar{D}_{\dot \alpha i }\left(\bar{U}^{-12} \mathcal{F}_{24 [56]}^{\mathpzc{1}}\right)  (\bar \chi^{8} )^{[8]} (\lambda^{8})^{[24]} (\bar \lambda^{8})^{[24]} \\
 &=& 2 \Scal{ \bar{U}^{-14} \bar{\mathcal{D}} \mathcal{F}_{24 [56]}^{\mathpzc{1}}\,  \bar \chi_{\dot \alpha i} + \bar{U}^{-12} \cD_{ijkl} \mathcal{F}_{24 [56]}^{\mathpzc{1}}\, \bar \lambda_{\adt}^{jkl}  }(\bar \chi^{8} )^{[8]} (\lambda^{8})^{[24]} (\bar \lambda^{8})^{[24]}   \CR
& =& \frac{16}{9} \Scal{ \bar{U}^{-14} \bar{\mathcal{D}}\mathcal{F}_{24 [55]i}^{\mathpzc{1}} \,  (\bar \chi^{9} )^{[7]}_{\dot \alpha} (\bar \lambda^{8})^{[24]} -3 \bar{U}^{-12} \cD_{i[2]}{}^j \mathcal{F}_{24 [55]j}^{\mathpzc{1}}\, (\bar \chi^{8} )^{[8]} (\bar \lambda^{9})^{[25]}_{\adt} } ( \lambda^{8})^{[24]} + \dots \nn 
\eea
where the dots state for lower isospin terms in $(\bar\lambda^9)^{[23]}$ and $(\bar\lambda^9)^{[21]}$ that we neglect at this order. The first term can only be canceled by the one coming from the application of $\bar{D}_{\dot \alpha i }$ on $\bar{U}^{-14} \mathcal{F}_{28 [52]}^{\mathpzc{2}}$, leading to
\bea
&&  \bar{D}_{\dot \alpha i } \left(\bar{U}^{-14} \mathcal{F}_{28 [52]}^{\mathpzc{2}}\right)  (\bar \chi^{9} )^{[7]\,\dot \beta} (\lambda^{8})^{[24]} (\bar \lambda^{7})^{[21]}_{\dot \beta} \CR
&=&  2 \bar{U}^{-14} \mathcal{D}_{i j k l}\mathcal{F}_{28 [52]}^{\mathpzc{2}} (\bar \chi^{9})^{[7]\,\dot \beta} (\lambda^{8})^{[24]} (\bar \lambda^{7})^{[21]}_{\dot \beta} \bar \lambda^{j k l}_{\adt} +\dots \CR
 &=&  - 2\bar{U}^{-14} \mathcal{D}_{[i3]} \mathcal{F}_{28 [52]}^{\mathpzc{2}} (\bar \chi^{9})^{[7]}_{\dot \alpha} (\lambda^{8})^{[24]} (\bar \lambda^{8})^{[24]} + \dots 
\eea
where the dots state for lower isospin terms that we neglect at this order. We conclude that the two tensor functions must be related through 
\be
\frac{16}{9} \bar{\mathcal{D}}\mathcal{F}_{24 [56]}^{\mathpzc{1}} = 2 \mathcal{D}_{[4]} \mathcal{F}_{28 [52]}^{\mathpzc{2}} \ . 
\ee
It means that $\mathcal{F}_{24 [56]}^{\mathpzc{1}}$ can be written as the covariant derivative on $SL(3)/SO(3)$ of a given tensor function $\mathcal{F}_{24 [52]}$. Therefore we have 
\be \label{eq:1_solution}
\mathcal{F}_{24 [56]}^{\mathpzc{1}} =  \mathcal{D}_{[4]}  \mathcal{F}_{24 [52]} \ , \qquad  \mathcal{F}_{28 [52]}^{\mathpzc{2}} = \frac{8}{9} \bar{\mathcal{D}} \mathcal{F}_{24 [52]}\ . 
\ee
Note that in principle the two tensor functions $\mathcal{F}_{24[52]}$ could differ by an inhomogeneous term such that  $\bar \cD \cD_{[4]}  c_{24 [52]}= 0$. However one argues that the equation 
\be
\mathcal{D}_{(i_1 i_2 i_3 i_4} \mathcal{G}_{n \, i_4 i_5 i_6 \dots i_{4 m}) } = 0
\ee
admits no solution, and such inhomogeneous term can only be a holomorphic tensor on the symmetric space  $SL(2)/SO(2)$, \ie  $c_{\mathpzc{2}}(T,\bar T) = ( 1- T \bar T)^{- 12} \tilde{c}_{\mathpzc{2}}(T)$. Considering other constraints from supersymmetry one would get to the conclusion that such  inhomogeneous terms must vanish because the supersymmetry constraint is linear in the tensor functions. For simplicity we shall assume from the beginning that all such terms vanish. 

The second terms in \eqref{EquaDF56}, decomposes as
\bea \label{EquaDF56L9} 
& &- \frac{16}{3}  \bar{U}^{-12} \cD_{i[2]}{}^j \mathcal{F}_{24 [55]j}^{\mathpzc{1}}\, (\bar \chi^{8} )^{[8]} (\bar \lambda^{9})^{[25]}_{\adt} ( \lambda^{8})^{[24]} \CR
&=& -  \frac{8}{3}  \bar{U}^{-12}  \Scal{ \cD_{(i[2]}{}^j \mathcal{F}_{24 [55])j}^{\mathpzc{1}}\, (\bar \lambda^{9})^{[25]}_{\adt}  + \frac{55}{29}  \mathcal{D}_{[2]}{}^{j k} \mathcal{F}_{24 [54] j k}^{\mathpzc{1}} (\bar \lambda^{9})^{[24]}_{\adt}{}_i }  (\bar \chi^{8} )^{[8]} ( \lambda^{8})^{[24]} \ ,  \eea
where $ \cD_{(i[2]}{}^j \mathcal{F}_{24 [55])j}^{\mathpzc{1}}$ is of isospin $58$, and therefore cannot be canceled by any other term since there is no components with a tensor function $\bar U^{-12} \mathcal{F}^{\mathpzc{a}}_{24[54]}$ as we  discussed above. Therefore we conclude that 
\be \label{eq:Function_Const}
 \mathcal{D}_{[3]}{}^{j} \mathcal{F}_{24 [55]j}^\un  = \mathcal{D}_{[3]}{}^{j} \biggl( \frac{1}{14} \cD_{[3] j}\mathcal{F}_{24 [52]} + \frac{13}{14} \cD_{[4]} \mathcal{F}_{24 [51]j} \biggr)  = 0\ . 
\ee
The first term vanishes because the commutator of two covariant derivative involves the contraction of three of their respective indices, such that 
\be \cD_{[4]} \mathcal{D}_{[3]}{}^{j} \mathcal{F}_{24 [51]j} = 0 \ , \ee
and therefore
\be  \mathcal{D}_{[3]}{}^{j} \mathcal{F}_{24 [51]j} = 0\ . \label{oddIso}  \ee
Now it remains to cancel the second term in \eqref{EquaDF56L9}, for which we will need to consider the action of the covariant derivative on next to maximal weight terms \eqref{NextMaxW}, \ie 
\begin{multline}\label{DL=0}
 \bar{D}_{\dot \alpha i }  \bigl( \bar{U}^{-12} \mathcal{F}^{\mathpzc{1}}_{24 [56]} \bigr) (\bar \chi^{8} )^{[8]} (\lambda^{8})^{[24]} (\bar \lambda^{8})^{[24]} +  \bar{D}_{\dot \alpha i } \left(\bar{U}^{-14} \mathcal{F}_{28 [52]}^{\mathpzc{2}}\right)  (\bar \chi^{9} )^{[7]\,\dot \beta} (\lambda^{8})^{[24]} (\bar \lambda^{7})^{[21]}_{\dot \beta} \\  +\bar{U}^{-12} \mathcal{F}^{\mathpzc{1}}_{24 [56]}  \bar{D}_{\dot \alpha i }\bigl(  (\bar \chi^{8} )^{[8]} (\lambda^{8})^{[24]} (\bar \lambda^{8})^{[24]} \bigr) \\ +
 \bar{D}_{\dot \alpha i }  \bigl(\bar{U}^{-10}  \mathcal{F}_{20 [56]}^{\mathpzc{3}}\bigr)  \bar{F}^{[2]}_{a b} (\bar \chi^{6})^{a b \, [6] } (\lambda^{8})^{[24]} (\bar \lambda^{8})^{[24]} 
 + \bar{D}_{\dot \alpha i }  \bigl( \bar{U}^{-10} \mathcal{F}_{20 [56]}^{\mathpzc{4}}\bigr)   (\bar \chi^{7})^{[7]}_{\dot \alpha} (\lambda^{8})^{[24]} (\bar \lambda^{9})^{\dot \alpha [25]}\\   +  \bar{D}_{\dot \alpha i } \bigl( \bar{U}^{-10}  \mathcal{F}_{20 [56]}^{\mathpzc{5}}\bigr)   (\bar \chi^{6})^{a b \; [6]} (\lambda^{10})_{a b}^{[26]} (\bar \lambda^{8})^{[24]}    + \dots = 0 
\end{multline}
where we have already computed the two first terms to simplify to the second term in \eqref{EquaDF56L9}. The corresponding tensor function has $U(1)$ weight $28$ and isospin $56$. Therefore it also gets contributions from the action of the covariant derivative on tensor functions of $U(1)$ weight $28$ and isospin $52$. However, there is a large number of terms like that, and analysing them all would be rather cumbersome. In order to bypass this difficulty, we remark that their contributions only arise as an isospin $56$ tensor function times a combination of the  fields of isospin $55$, whereas the term we want to cancel in \eqref{EquaDF56L9} includes a combination of the fields of isospin $57$. Therefore we will be able to neglect the contribution from the isospin $52$ terms in $\cL_\grad800$. In the same way, the action of the covariant derivative in the order 24 term in the fermions of maximal isospin decomposes into a term of isospin $57$ and a term of isospin $55$ that we will neglect, \ie 
\be
  \bar{D}_{\dot \alpha i } \left((\bar \chi^{8} )^{[8]} (\lambda^{8})^{[24]} (\bar \lambda^{8})^{[24]} \right) =\ve_{i j} \mathcal{J}^{[j 56]}_{\dot \alpha} + \delta_{i}^{[1]} \mathcal{J}^{[55]}_{\dot \alpha}\ . 
\ee
To carry out this computation, we need the explicit action of the fermionic covariant derivative on the fermions derived in \cite{Minimal}, and their complex conjugate
\bea \label{eq:DFermions}
 \bar D_{\dot \alpha i} \bar \chi_{\dot \beta}^j &=&  -\frac{1}{8} (\gamma^{ab} )_{\dot \alpha \dot \beta} \Scal{ \bar F^{\;\;\;\;\;\;j}_{ab\;i}- \frac{1}{4} \scal{\lambda_{ikl}  \gamma_{ab} \lambda^{jkl} }} 
  +\frac{1}{192}  (\gamma^{abcd})_{\dot \alpha \dot \beta}  \delta_i^j \bar G^{-}_{a b c d}- \frac{1}{4} \bar \lambda_{\adt ki}{}^j \bar \chi_\bdt^k \CR
  && \hspace{60mm} -C_{\dot \alpha \dot \beta}\Scal{ \frac{3}{32}  \delta_i^j ( \lambda \lambda) +\frac{1}{2}   \scal{\bar \chi^k \bar \lambda_{ki}{}^j }}\ \CR
D_\alpha^i \bar \lambda_\bdt^{jkl} &=& ( \gamma^a)_{\alpha\bdt} \Scal{ \hspace{-0.5mm}- i P_a^{ijkl} \hspace{-0.5mm}+ \frac{1}{2} ( \lambda^{p(ij} \gamma_a \bar \lambda^{kl)}{}_p) \hspace{-0.5mm}- \varepsilon^{i(j} ( \chi^k \gamma_a \bar \chi^{l)})}  + \frac{i}{12} (\gamma^{abc})_{\alpha\bdt} \varepsilon^{i(j} H_{abc}^{kl)} - \frac{3}{4} \lambda^{pi(j}_\alpha \bar \lambda_\bdt^{kl)}{}_p\CR
D_\alpha^i \lambda_\beta^{jkl} &=& - \frac{1}{4} (\gamma^{ab})_{\alpha\beta} \varepsilon^{i(j} \Scal{ \bar F_{ab}^{kl)} + ( \bar \chi_p \gamma_{ab}\bar  \lambda^{kl)p} )}  + \frac{1}{4} \lambda_\alpha^{pi(j} \lambda_\beta^{kl)}{}_p - \frac{1}{2} C_{\alpha\beta} ( \lambda^{p(ij} \lambda^{kl)}{}_p) \CR
&& \hspace{90mm} + (\gamma^a)_{\alpha\bdt} \bar \chi^{\adt i} \bar \lambda^{\bdt\, jkl} (\gamma_a)_{\adt\beta} \ , 
 \eea
where the term in  $P_a^{ijkl}$ will be neglected to avoid considering contributions from $D_a \cL_\grad701$. 

Using these expression in \eqref{DL=0}, substituting the two first terms by the second of \eqref{EquaDF56L9}, and including the covariant derivative on the tensor functions 
\be\bar{D}_{\dot \alpha i} \bar{U}^{-10}  \mathcal{F}_{20 [56]}^{\mathpzc{a}} =  2 \bar{U}^{-12} \bar{\mathcal{D}} \mathcal{F}_{20 [56]}^{\mathpzc{a}} \bar{\chi}_{\dot \alpha i} +2 \bar{U}^{-10}  \cD_{ijkl}\mathcal{F}_{20 [56]}^{\mathpzc{a}} \bar\lambda_\adt^{jkl}  \ , 
 \ee
 while neglecting the second term of larger $U(1)$ weight, one obtains after some algebra the constraint 
\begin{multline}
 \bar{D}_{\dot \alpha i } \mathcal{L}_{abcdefgh} \approx  \ve_{a b c d e f g h} \Biggl(
 \bar{U}^{-12} \left(- \frac{1}{8} \mathcal{F}_{24 [56]}^{\mathpzc{1}}  - 4 \bar{\mathcal{D}} \mathcal{F}_{20 [56]}^{\mathpzc{3}} \right) \ve_{i j} (\gamma^{ab})_{\dot \alpha}{}^{\dot \beta} \bar{F}^{[2]}_{a b} (\bar \chi^{7})^{[j6]}_{\dot \beta} (\lambda^{8})^{[24]} (\bar \lambda^{8})^{[24]}  \\
 +  \bar{U}^{-12} \left(- 2 \mathcal{F}_{24 [56]}^{\mathpzc{1}}  - \frac{440}{87} \mathcal{D}_{[2]}{}^{j k} \mathcal{F}_{24 [54]jk}^{\mathpzc{1}} +  2 \bar{\mathcal{D}}  \mathcal{F}_{20 [56]}^{\mathpzc{4}} \right) \ve_{i j} (\bar \chi^{8})^{[j 7]} (\lambda^{8})^{[24]} (\bar \lambda^{9})^{[25]}_{\dot \alpha}   \\
  + \bar{U}^{-12} \left( - \frac{25}{504} \mathcal{F}_{24 [56]}^{\mathpzc{1}}  - 4 \bar{\mathcal{D}} \mathcal{F}_{20 [56]}^{\mathpzc{5}}\right) \ve_{i j} (\gamma^{ab})_{\dot \alpha}{}^{\dot \beta} (\bar \chi^{7})^{[j 6]}_{\dot \beta} (\lambda^{10})_{a b}^{[26]} (\bar \lambda^{8})^{[24]} \Biggr)  = 0 \ . 
\end{multline}
In order for $\cL$ to satisfy the $d$-closure equation, each of these terms  must cancel separately, and we get
\be
 \bar{\mathcal{D}} \mathcal{F}_{20 [56]}^{\mathpzc{3}} = - \frac{1}{32} \mathcal{F}_{24 [56]}^{\mathpzc{1}}  \ , \quad \bar{\mathcal{D}}  \mathcal{F}_{20 [56]}^{\mathpzc{4}} = \mathcal{F}_{24 [56]}^{\mathpzc{1}}  +  \frac{220}{87} \mathcal{D}_{[2]}{}^{j k} \mathcal{F}_{24 [54]jk}^{\mathpzc{1}} \ , \quad  \bar{\mathcal{D}} \mathcal{F}_{20 [56]}^{\mathpzc{5}} = - \frac{25}{2016} \mathcal{F}_{24 [56]}^{\mathpzc{1}}  \ . 
\ee
Using  \eqref{eq:1_solution} in the third equation, we solve similarly these equations by defining the tensor functions in terms of a lower weight tensor function $\mathcal{F}_{20 [52]}$, such that 
\bea\label{ReduceTensor} 
 \mathcal{F}_{24 [52]} &=& \bar{\mathcal{D}} \mathcal{F}_{20 [52]}\ ,  \CR
 \mathcal{F}_{24[56]}^{\mathpzc{1}} &=& \mathcal{D}_{[4]} \bar{\mathcal{D}} \mathcal{F}_{20[52]} \ , \qquad \hspace{5mm} \mathcal{F}_{28 [52]}^{\mathpzc{2}} = \frac{8}{9} \bar{\mathcal{D}}^2 \mathcal{F}_{20 [52]} \ ,  \CR
  \mathcal{F}_{20[56]}^{\mathpzc{3}} &=& - \frac{1}{32} \mathcal{D}_{[4]}  \mathcal{F}_{20[52]}\ , \qquad  \mathcal{F}_{20 [56]}^{\mathpzc{5}} = - \frac{25}{2016} \mathcal{D}_{[4]} \mathcal{F}_{20 [52]}  \ , 
\eea
whereas 
\bea
&&  \mathcal{F}_{20 [56]}^{\mathpzc{4}} \CR
 &=& \mathcal{D}_{[4]}  \mathcal{F}_{20[52]} +  \frac{220}{87} \left(\frac{3}{770} \mathcal{D}_{[2]}{}^{j k} \mathcal{D}_{j k [2]} \mathcal{F}_{20 [52]} + \frac{52}{385} \mathcal{D}_{[2]}{}^{j k} \mathcal{D}_{j [3]} \mathcal{F}_{20 [51]k} +  \frac{663}{770} \mathcal{D}_{[2]}{}^{j k} \mathcal{D}_{[4]} \mathcal{F}_{20 [50]jk} \right) \CR
 &=& \mathcal{D}_{[4]}  \mathcal{F}_{20[52]} +  \frac{220}{87} \left(\frac{1}{14} \mathcal{D}_{[2]}{}^{j k} \mathcal{D}_{j k [2]} \mathcal{F}_{20 [52]} +  \frac{663}{770} \mathcal{D}_{[2]}{}^{j k} \mathcal{D}_{[4]} \mathcal{F}_{20 [50]jk} \right)\ ,  \label{ReduceF5} \eea
 where we have reduced the two-derivative term with
\be
 D_{[2]}{}^{ij} D_{[3]j} = \frac{1}{2} \delta_{[1]}^{i} D_{[2]}{}^{jk} D_{j k [2]}  \ . 
\ee
Again we neglected the possible holomorphic inhomogeneous solutions to these equations, because they must all cancel at the end by unicity and linearity of the equations in the tensor functions. Note moreover that \eqref{oddIso} together with \eqref{ReduceTensor} imply that the tensor function $ \mathcal{F}_{20 [52]}$ also satisfies to the same constraint 
 \be \cD_{[3]}{}^j \mathcal{F}_{20 [51]j} = 0 \ . \label{evenIsospin}\ee

To summarise the results obtained so far, the expression of $\mathcal{L}_\grad{8}{0}{0}$ subject to these constraints takes the following form \begin{multline} 
 \mathcal{L}_{abcdefgh} = \ve_{a b c d e f g h}  \biggl( \bar{U}^{-12} \bar{\mathcal{D}} \mathcal{D}_{[4]} \mathcal{F}_{20 [52]} (\bar \chi^{8} )^{[8]} (\lambda^{8})^{[24]} (\bar \lambda^{8})^{[24]}  \\ 
 + \frac{8}{9} \bar{U}^{-14} \bar \cD^2 \mathcal{F}_{20 [52]} (\bar \chi^{9} )^{[7] \dot \alpha} (\lambda^{8})^{[24]} (\bar \lambda^{7})^{[21]}_{\dot \alpha}
  - \frac{1}{32} \bar{U}^{-10} \mathcal{D}_{[4]} \mathcal{F}_{20 [52]}  \bar{F}^{[2]}_{a b} (\bar \chi^{6})^{a b \, [6] } (\lambda^{8})^{[24]} (\bar \lambda^{8})^{[24]}  \\
 +  \bar{U}^{-10} \Biggl(\mathcal{D}_{[4]}  \mathcal{F}_{20[52]} +  \frac{220}{87} \biggl(\frac{1}{14} \mathcal{D}_{[2]}{}^{j k} \mathcal{D}_{j k [2]} \mathcal{F}_{20 [52]} +  \frac{663}{770} \mathcal{D}_{[2]}{}^{j k} \mathcal{D}_{[4]} \mathcal{F}_{20 [j k 50]} \biggr)  \Biggr)  (\bar \chi^{7})^{[7]}_{\dot \alpha} (\lambda^{8})^{[24]} (\bar \lambda^{9})^{\dot \alpha [25]}   \\
  - \frac{25}{2016} \bar{U}^{-10} \mathcal{D}_{[4]} \mathcal{F}_{20 [52]}   (\bar \chi^{6})^{a b \; [6]} (\lambda^{10})_{a b}^{[26]} (\bar \lambda^{8})^{[24]}  +  \sum_{ \mathpzc{a}, p \leq 4} \bar{U}^{-2 p} \mathcal{F}_{p [56]}^{\mathpzc{a}} \mathcal{I}^{\mathpzc{a} \, 4p [56 ]} \\ +  \sum_{ \mathpzc{a}, p, q  \leq 25} \bar{U}^{-2 p} \mathcal{F}_{p [2 q]}^{\mathpzc{a}} \mathcal{I}^{\mathpzc{a} \, 4p [2 q]} \biggr)
\end{multline}
We will now constrain the superform to satisfy equation \eqref{eq:DI800}, \ie 
\begin{multline} \label{eq:DI800New}
D_{\alpha}^{i} \mathcal{L}_{abcdefgh} + 8 D_{[a} \mathcal{L}_{bcdefgh]}{}_{\alpha}^{i} + 8 T_{\alpha [a}^{i}{}^{\dot \beta j} \mathcal{L}_{\dot \beta j bcdefgh]} 
+ 8 T_{\alpha [a}^{i}{}^{\beta}_{j} \mathcal{L}_{\beta bcdefgh]}^{j} \\ 
+ 28 T_{[ab}{}^{\dot \beta j} \mathcal{L}_{\dot \beta j c d e f g h]}{}_{\alpha}^{i} + 28 T_{[ab}{}^{\beta}_{j} \mathcal{L}_{\beta c d e f g h]}^{j} {}_{\alpha}^{i} = 0 \ . 
\end{multline}
Again we will start from the action of the covariant derivative on the maximal weight term, and we will then consider all the terms that are needed to cancel this derivative. Similarly as in (\ref{EquaDF56},\ref{EquaDF56L9}) and using the constraint \eqref{oddIso}, one obtains that 
\bea
&& {D}_{\alpha}^i\left(\bar{U}^{-12} \mathcal{F}_{24 [56]}^{\mathpzc{1}}\right)  (\bar \chi^{8} )^{[8]} (\lambda^{8})^{[24]} (\bar \lambda^{8})^{[24]} \\
 &=& 2 \Scal{ (1-T \bar T)^2\bar{U}^{-10} {\mathcal{D}} \mathcal{F}_{24 [56]}^{\mathpzc{1}}\,  \chi_{\alpha}^i - \bar{U}^{-12} \cD^{ijkl} \mathcal{F}_{24 [56]}^{\mathpzc{1}}\,  \lambda_{\alpha jkl}  }(\bar \chi^{8} )^{[8]} (\lambda^{8})^{[24]} (\bar \lambda^{8})^{[24]}   \CR
& =& 2 \Scal{ (1-T \bar T)^2\bar{U}^{-10} {\mathcal{D}} \mathcal{F}_{24 [56]}^{\mathpzc{1}}\,  \chi_{\alpha}^i (\lambda^{8})^{[24]} +   \frac{220}{87} \bar{U}^{-12} \mathcal{D}_{[2]}{}^{j k} \mathcal{F}_{24 [54] j k}^{\mathpzc{1}}  \ (\lambda^{9})^{[24]i}_\alpha }(\bar \chi^{8} )^{[8]}  (\bar \lambda^{8})^{[24]} + \dots\nn
 \eea
Using moreover \eqref{ReduceTensor} and the same steps as in \eqref{ReduceF5}, one obtains moreover 
\bea  \label{EquabDF56} 
&& {D}_{\alpha}^i\left(\bar{U}^{-12} \mathcal{F}_{24 [56]}^{\mathpzc{1}}\right)  (\bar \chi^{8} )^{[8]} (\lambda^{8})^{[24]} (\bar \lambda^{8})^{[24]} \\
& =& 2  (1-T \bar T)^2\bar{U}^{-10} {\mathcal{D}} \bar \cD \cD_{[4]} \mathcal{F}_{20 [52]}\,  \chi_{\alpha}^i(\bar \chi^{8} )^{[8]} (\lambda^{8})^{[24]} (\bar \lambda^{8})^{[24]} \CR
&&  +   \frac{440}{87} \bar{U}^{-12} \bar{\mathcal{D}}  \mathcal{D}_{[2]}{}^{jk} \left( \frac{1}{14}  \mathcal{D}_{[2]jk}\mathcal{F}_{20[52]}  +  \frac{663}{770} \mathcal{D}_{[4]}\mathcal{F}_{20[50]jk} \right) \ (\lambda^{9})^{[24]i}_\alpha (\bar \chi^{8} )^{[8]}  (\bar \lambda^{8})^{[24]} + \dots\nn
 \eea
where the dots state for some lower isospin terms in $(\lambda^9)^{[23]}_\alpha$ and $(\lambda^9)^{[21]}_\alpha$ that we disregard in this computation. 

After investigation, it turns out that the only terms that can contribute to cancel the terms of isospin $57/2$ in $\bar \chi^8 \lambda^9 \bar \lambda^8$ in \eqref{EquabDF56} are the ones coming from the action of the covariant derivative on the fermions of the maximal weight term itself. Using the action of the covariant derivative on the fermion $\lambda$ and $\bar \lambda$ \eqref{eq:DFermions}, as well as
\be
 D_{\alpha}^i \bar \chi_{\dot \beta}^j =  \frac{1}{2} (\gamma^a)_{\alpha \dot \beta} \Scal{  -i  \ve^{i j}  \bar P_a  +  \scal{\bar\chi_k \gamma_a \lambda^{ijk}}} + \frac{3}{4} \lambda_\alpha^{ijk} \bar \chi_{\bdt k}\ ,  
\ee
one obtains finally 
\bea && {D}_{\alpha}^i \Scal{ \bar{U}^{-12} \bar{\mathcal{D}} \cD_{[4]} \mathcal{F}_{20 [52]} \,   (\bar \chi^{8} )^{[8]} (\lambda^{8})^{[24]} (\bar \lambda^{8})^{[24]} } \CR
\hspace{-0.5mm}&=& \bar{U}^{-12}   \bar{\mathcal{D}} \biggl(  \frac{440}{87} 
  \mathcal{D}_{[2]}{}^{jk}\hspace{-0.8mm} \left( \frac{1}{14}  \mathcal{D}_{[2]jk}\mathcal{F}_{20[52]} \hspace{-0.2mm}+ \hspace{-0.2mm}  \frac{663}{770} \mathcal{D}_{[4]}\mathcal{F}_{20[50]jk}\hspace{-0.5mm} \right) \hspace{-0.8mm}- \frac{10}{9}  \mathcal{D}_{[4]}\mathcal{F}_{20[52]} \biggr)  (\bar \chi^{8})^{[8]}
 (\lambda^{9})^{[24 i]}_{\alpha} (\bar \lambda^{8})^{[24]}\CR
 && \quad + \dots \eea
so we conclude that supersymmetry implies the tensor function $\mathcal{F}_{20[52]}$ to satisfy to  
\be \label{eq:SL3_space}
 \mathcal{D}_{[2]}{}^{jk} \left( \frac{1}{14}  \mathcal{D}_{[2]jk}\mathcal{F}_{20[52]} + \frac{663}{770} \mathcal{D}_{[4]}\mathcal{F}_{20[50]jk} \right) = { \frac{29}{132}} \mathcal{D}_{[4]}\mathcal{F}_{20[52]} \ . 
\ee
This equation is one of the main results of this section, that will allow us to determine the differential equation satisfied by the function that defines the invariant. To summarise the results obtained so far, the expression of $\mathcal{L}_\grad{8}{0}{0}$ subject to these constraints takes the following form 
\begin{multline} \label{eq:L800_1}
 \mathcal{L}_{abcdefgh} = \ve_{a b c d e f g h}  \biggl( \bar{U}^{-12} \bar{\mathcal{D}} \mathcal{D}_{[4]} \mathcal{F}_{20 [52]} (\bar \chi^{8} )^{[8]} (\lambda^{8})^{[24]} (\bar \lambda^{8})^{[24]} \\ + \frac{8}{9} \bar{U}^{-14} \bar \cD^2 \mathcal{F}_{20 [52]} (\bar \chi^{9} )^{[7] \dot \alpha} (\lambda^{8})^{[24]} (\bar \lambda^{7})^{[21]}_{\dot \alpha} 
 - \frac{1}{32} \bar{U}^{-10} \mathcal{D}_{[4]} \mathcal{F}_{20 [52]}  \bar{F}^{[2]}_{a b} (\bar \chi^{6})^{a b \, [6] } (\lambda^{8})^{[24]} (\bar \lambda^{8})^{[24]}  \\
 + \frac{14}{9}  \bar{U}^{-10} \mathcal{D}_{[4]} \mathcal{F}_{20 [52]} \, (\bar \chi^{7})^{[7]}_{\dot \alpha} (\lambda^{8})^{[24]} (\bar \lambda^{9})^{\dot \alpha [25]}   
  - \frac{25}{2016} \bar{U}^{-10} \mathcal{D}_{[4]} \mathcal{F}_{20 [52]}   (\bar \chi^{6})^{a b \; [6]} (\lambda^{10})_{a b}^{[26]} (\bar \lambda^{8})^{[24]}  \\ +  \sum_{ \mathpzc{a}, p \leq 4} \bar{U}^{-2 p} \mathcal{F}_{p [56]}^{\mathpzc{a}} \mathcal{I}^{\mathpzc{a} \, 4p [56 ]}  +  \sum_{ \mathpzc{a}, p, q  \leq 25} \bar{U}^{-2 p} \mathcal{F}_{p [2 q]}^{\mathpzc{a}} \mathcal{I}^{\mathpzc{a} \, 4p [2 q]} \biggr) \ . 
\end{multline}
We recover already here the structure of the gradient expansion anticipated in \eqref{D4R4SuperformGrad}, such that all the tensor functions are related to each other via covariant derivatives maximising the isospin, \ie such that all $SU(2)$ indices are symmetrised.

Now it remains to cancel the first term in \eqref{EquabDF56} in order to deduce a differential equation with respect to the $SL(2)/SO(2)$ scalar fields. As before, we shall concentrate on terms of maximal weight, so for $ \chi_{\alpha}^{[1]} (\bar \chi^{8} )^{[8]} (\lambda^{8})^{[24]} (\bar \lambda^{8})^{[24]}$ we have in principle to consider all the terms of $U(1)$ weight $21$ and isospin $57/2$. In order to avoid considering the terms in $D_a \cL_\grad710$ we shall disregard terms involving the scalar momenta, and the remaining possible field combinations are 
\bea  \label{eq:term_mixings1}
&&  \chi (\bar  \chi^{8})^{[8]} (\bar \lambda^{8})^{[24]} (\lambda^{8})^{[24]} \ ,  \quad H^{[2]}  (\bar  \chi^{7})^{[7]} (\bar \lambda^{8})^{[24]} (\lambda^{8})^{[24]}  \ , \\
 &&  (\bar \chi^{7})^{[7]} (\bar \lambda^{9})^{[25]}  (\lambda^{9})^{[25]} \ , \quad   \bar{F}^{[2]} (\bar \chi^{6})^{[6]} (\bar \lambda^{8})^{[24]}  (\lambda^{9})^{[25]}  \ , \quad  (\bar \chi^{6})^{[6]} (\bar \lambda^{8})^{[24]}  (\lambda^{11})^{[27]} \ .   \label{eq:term_mixings2}
\eea
To simplify further the computation, we note that the torsion 
\begin{multline} \label{eqn:D1Torsion}
T_{a \alpha}^{\;\;i\;\;\dot\beta j} =\frac{i}{24}  (\gamma^{bcd})_{\alpha}^{\;\;\dot \beta}  \varepsilon^{ij} \left( \bar G^{-}_{a b c d} - \frac{1}{24}  \left(\lambda^{klp} \gamma_{abcd} \lambda_{klp} \right) \right) \\
 + \frac{i}{24} \scal{ \gamma_{a}^{\;\;bc} + 4 \delta_a^{[b} \gamma^{c]}}_{\alpha}^{\;\;\dot \beta}  \left(  \bar{F}_{bc}^{ij} - \frac{1}{4}   \left(\bar \chi_k\gamma_{bc} \bar \lambda^{ijk}\right) + 2 \left(\lambda^i \gamma_{bc}  \lambda^j \right)\right)  + \frac{i}{4} (\gamma^{b})_{\alpha}^{\;\;\dot \beta} \bar{F}_{ab}^{ij} \ . 
\end{multline}
is such that the contribution of maximal isospin coming from $T_\grad{1}{1}{0}{}^\grad{0}{0}{1} \mathcal{L}_\grad{7}{0}{1}$, only produces terms listed in \eqref{eq:term_mixings2}, such that restring attention to the terms listed in \eqref{eq:term_mixings1} we can neglect this contribution. Moreover, because the term of maximal isospin in $\cL_\grad710$ proportional to $\bar \chi^7 \lambda^8 \bar \lambda^8$ has isospin $55/2$ \eqref{eq:L710_Ansatz}, the contribution of $D_\grad100 \cL_\grad710$ independent of the scalar momenta has itself maximal isospin  $55/2$, and will not contribute to the terms we are concentrating on. Therefore we only need to analyse the two following terms in \eqref{eq:DI800New}
\be \label{eq:L710_Ansatz_simp}
 D_{\alpha}^{i} \mathcal{L}_{abcdefgh}  + 8 T_{\alpha [a}^{i}{}^{\beta}_{j} \mathcal{L}_{\beta bcdefgh]}^{j} + \dots = 0 \ee
 proportional to the two field combinations listed in \eqref{eq:term_mixings1}.  In the first term in $\eqref{eq:L710_Ansatz_simp}$ we shall only need the contributions 
\begin{multline} \label{DLHH} 
 D_{\alpha}^{i} \mathcal{L}_{abcdefgh} = \ve_{a b c d e f g h}  \biggl( 2 \bar{U}^{-10} (1 - T \bar T)^2  \mathcal{D} \bar{\mathcal{D}} \mathcal{D}_{[4]} \mathcal{F}_{20 [52]} \, \chi^{i}_{\alpha} (\bar \chi^{8} )^{[8]} (\lambda^{8})^{[24]} (\bar \lambda^{8})^{[24]} \\ 
 - \frac{1}{32} \bar{U}^{-10} \mathcal{D}_{[4]} \mathcal{F}_{20 [52]} \bigl(D_{\alpha}^{i} \bar{F}^{[2]}_{a b}\bigr) (\bar \chi^{6})^{a b \, [6] } (\lambda^{8})^{[24]} (\bar \lambda^{8})^{[24]}  \\
 -  \frac{14}{9} \bar{U}^{-10} \mathcal{D}_{[4]}  \mathcal{F}_{20[52]} (\bar \chi^{7})^{[7]}_{\dot \alpha} (\lambda^{8})^{[24]} \bigl(D_{\alpha}^{i} (\bar \lambda^{9})^{\dot \alpha  [25]} \bigr)  + \dots \biggr)\ . 
\end{multline}
We need therefore the explicit action of the covariant derivative on $\bar \lambda$ already displayed in \eqref{eq:DFermions}  and the one on $\bar F$ also computed in \cite{Minimal}, 
\bea
D_{\alpha}^{i} \bar F^{j k}_{a b} &=&  (\gamma_{ab})_{\alpha}{}^{\beta} \Bigl(  - \frac{i}{9} H^{(i j}_{cde} (\gamma^{cde} \bar \chi^{j)})_{\beta} - \frac{1}{4} (\bar \chi^{(i} \gamma_{cd} \bar \chi^{j}) (\gamma^{cd}\chi^{k)} )_{\alpha}
\Bigr) - \frac{4 i}{3} H^{(i j}_{abc} (\gamma^{c} \bar \chi^{k)})_{\alpha}  \CR
&& +  (\gamma_{[a})_{\alpha}{}^{\dot \alpha} \Bigl( \frac{i}{3} H^{(i j}_{b]cd} (\gamma^{cd} \bar \chi^{k)})_{\dot \alpha} + \frac{7}{3} (\bar \chi^{(i} \gamma_{b]c} \bar \chi^{j}) (\gamma^{c}\chi^{k)})_{\dot \alpha} \Bigr) + \dots
\eea
where the dots state for terms of isospin $1/2$ in $H \bar \chi$ and $\bar \chi \chi^2$ as well as many terms in  $\bar F \lambda, \bar \chi \bar \lambda \lambda, \lambda^3, D \bar \chi, \bar P \bar \lambda, P^{[4]} \bar \chi, \bar G \lambda$, that are irrelevant in our computation. At the end of the computation we get that all the terms in  $H \bar \chi^6 \bar \lambda^8 \lambda^8$ cancel out in \eqref{DLHH}, and the expression simplifies to 
\be
 D_{\alpha}^{i} \mathcal{L}_{abcdefgh} = 2 \ve_{a b c d e f g h} \bar{U}^{-10} \bigl((1 - T \bar T)^2  \mathcal{D} \bar{\mathcal{D}}  + 28 \bigr) \mathcal{D}_{[4]} \mathcal{F}_{20 [52]} \chi^{i}_{\alpha} (\bar \chi^{8} )^{[8]} (\lambda^{8})^{[24]} (\bar \lambda^{8})^{[24]}  + \dots 
 \ee
The second contribution from \be
 8 T_{\alpha [a}^{i}{}^{\beta}_{j} \mathcal{L}_{\beta bcdefgh]}^{j} = 8 \ve_{[abcdefg}{}^{a_1} \bar{U}^{-10}
 \mathcal{F}_{20 [j 55]}^{\mathpzc{6}} T_{\alpha h]}^{i}{}^{\beta j} (\gamma_{a_1})_{\beta}{}^{\dot \alpha} (\bar \chi^{7})_{\dot \alpha}^{[7]} (\lambda^{8})^{[24]} (\bar \lambda^{8})^{[24]}\ , 
\ee
is evaluated using the expression of the torsion $T_\grad{1}{1}{0}{}^\grad{0}{1}{0}$
\begin{multline}
T_{a \alpha \; j}^{\; i \;\, \beta} = \ve_{j k} (\gamma^{b c})_{\alpha}{}^{\beta} \Bigl( - \frac{1}{6} H_{a b c}^{i k} + \frac{i}{8} (\chi^{(i} \gamma_{abc}\bar \chi^{k)}) \Bigr) +
\ve_{j k} (\gamma_{a}{}^{bcd})_{\alpha}{}^{\beta} \Bigl( - \frac{1}{36} H_{b c d}^{i k} + \frac{i}{24} (\chi^{(i} \gamma_{b c d}\bar \chi^{k)}) \Bigr) \\
+ \frac{5 i}{12} \delta_{\alpha}{}^{\beta} \ve_{jk} (\chi^{(i} \gamma_{a}\bar \chi^{k)}) + \frac{i}{12} (\gamma_{a}{}^{b})_{\alpha}{}^{\beta} \ve_{jk} (\chi^{(i} \gamma_{b}\bar \chi^{k)}) +\dots
\end{multline}
as 
\be
 8 T_{\alpha [a}^{i}{}^{\beta}_{j} \mathcal{L}_{\beta bcdefgh]}^{j} = \epsilon_{abcdefgh} \bar{U}^{-10} \mathcal{F}_{20 [56]}^{\mathpzc{6}} \biggl( \frac{88 i}{3} \chi_{\alpha}^{i} (\bar \chi^{8})^{[8]}  + \frac{1}{36} H^{i[1]}_{a b c} (\gamma^{a b c})_{\alpha}{}^{\dot \alpha} (\bar \chi^{7})^{[7]}_{\dot \alpha} \biggr) 
 (\bar \lambda^{8})^{[24]} (\lambda^{8})^{[24]}\, .
\ee
The sum of the two contributions finally gives the equation 
\begin{multline}
\Bigl(2(1 - T \bar T)^2  \mathcal{D} \bar{\mathcal{D}} \mathcal{D}_{[4]} \mathcal{F}_{20 [52]}  + 56 \mathcal{D}_{[4]} \mathcal{F}_{20 [52]}  + \frac{88 i}{3} \mathcal{F}_{20 [56]}^{\mathpzc{6}}\Bigr) \chi^{i}_{\alpha} (\bar \chi^{8} )^{[8]} (\lambda^{8})^{[24]} (\bar \lambda^{8})^{[24]} \\
 + \frac{1}{36}  \mathcal{F}_{20 [56]}^{\mathpzc{6}} H^{i[1]}_{a b c} (\gamma^{a b c})_{\alpha}{}^{\dot \alpha} (\bar \chi^{7})^{[7]}_{\dot \alpha} (\bar \lambda^{8})^{[24]} (\lambda^{8})^{[24]} = 0\ . 
\end{multline}
Because the two terms are clearly linearly independent, the tensor function  $\mathcal{F}_{20 [56]}^{\mathpzc{6}} $ must vanish, such that there is finally no contribution from the torsion term, and we obtain the following differential equation for  $\mathcal{F}_{20 [52]}$
\be \label{eq:SL2_space}
 (1 - T \bar T)^{2} \mathcal{D} \bar{\mathcal{D}} \mathcal{F}_{20 [52]} = - 28 \mathcal{F}_{20 [52]} \ . 
\ee
Note that one might have expected to have a non-trivial term \eqref{chi7lambda16} from the linearised analysis because such a term does appear in \eqref{Lin710}. However the linearised $\cL_\grad800$ component \eqref{Lin800} also includes a term in $P^{[4]} (\bar\chi^{7})^{[7]} (\lambda^{7})^{[21]} (\bar \lambda^{8})^{[24]}$ that we have disregarded in our analysis, and one checks that they are tight together in the linearised approximation such that removing the second through the addition of the exterior derivative $d$ of the $(7,0,0)$ superform  
\be \cL_{abcdefg} =  \varepsilon_{abcdefgh} (\gamma^h)^{\adt\beta}  \cF_{20[52]} ( \bar\chi^7)_\adt^{[7]} ( \lambda^7)_\beta^{[21]} ( \bar \lambda^8)^{[24]}  \ , \ee
one also remove the former. 

\subsection{The gradient expansion of the invariant}
The structure of the maximal weight terms of $\cL_\grad800$ derived in the preceding section \eqref{eq:L800_1}, together with the constraint \eqref{evenIsospin} reproduces precisely the structure of the invariants defined in the linearised approximation, such that we conclude that we can indeed trust the gradient expansion \eqref{D4R4SuperformGrad}. Extending the computation of the last section indeed necessarily implies that the tensor function $\cF_{20[52]}$ is itself determined as the covariant derivative of a lower weight tensor functions, according to the constraints implied by supersymmetry in the linearised approximation \eqref{FunctionLinear}. We conclude therefore that there is a function  $\cE_\grad{2}{1}{0}$ of the complex scalar field $T$ and the five scalars $t^\upmu$ parametrising $SL(3)/SO(3)$, such that
\be
 \mathcal{F}_{20 [52]}(T, \bar T, t) = \bar{\cD}^{5} \cD_{[52]}^{13} \cE_\grad{2}{1}{0}(T, \bar T, t) \ , \label{DerivativeSingle}
\ee
where the function $\cE_\grad{2}{1}{0}$ multiplies the singlet superform $\cL^{\ord{0}[0]}$ including the $\nabla^4 R^4$ type term. 
The subscript $(2,1,0)$ denotes the analytic superspace including only half of the positive chirality fermionic coordinates, on which one can integrate the function \eqref{G-analytis8D} to define the invariant in the linearised approximation.

By construction, \eqref{DerivativeSingle} implies that \eqref{evenIsospin} is automatically satisfied, and using the property that the covariant derivative on $SL(2)/SO(2)$ and $SL(3)/SO(3)$ commute, we deduce from \eqref{eq:SL3_space} and \eqref{eq:SL2_space} that the function  $\cE_\grad{2}{1}{0}$ satisfies to 
\be \label{eq:SUSY_const}
  (1 - T \bar T)^{2} \mathcal{D} \bar{\cD}^{6} \cE_\grad{2}{1}{0} = - 28 \bar{\cD}^{5} \cE_\grad{2}{1}{0} \ , \qquad 
  \mathcal{D}_{[2]}{}^{jk} \mathcal{D}^{14}_{([54] j k)} \cE_\grad{2}{1}{0} = \frac{29}{132} \mathcal{D}^{14}_{[56]} \cE_\grad{2}{1}{0}\ . 
\ee
Using the commutation relation between $\cD$ and $\bar \cD$, one derives the standard formula \cite{Green:1998by}
\bea
 (1 - T \bar T)^{2} \mathcal{D} \bar{\cD}^{n} \cE_\grad{2}{1}{0} & =& - n (n - 1) \bar{\cD}^{n-1} \cE_\grad{2}{1}{0}  + (1 - T \bar T)^{2} \bar{\cD}^{n} \cD \cE_\grad{2}{1}{0} \CR
& = &\bar  \cD^{n-1} ( \Delta_{SL(2)} - n (n - 1)) \cE_\grad{2}{1}{0}\ , 
\eea
which one uses to prove that the first equation in \eqref{eq:SUSY_const} implies that the function  $\cE_\grad{2}{1}{0}$ is an eigen function of the Laplace operator, \ie 
\be
 \Delta_{SL(2)} \cE_\grad{2}{1}{0} = 2 \cE_\grad{2}{1}{0} \ . \label{LaplaceSL2} 
\ee
Note that the general solution to this equation can be obtained from an anti-holomorphic function $ \cF[\bar \tau] $ and its complex conjugate as
\be  - (\tau - \bar \tau )^2 \partial \bar \partial \Scal{  \Scal{ \bar \partial + \frac{2}{\tau - \bar \tau }} \cF[\bar \tau]} = 2 \Scal{ \bar \partial + \frac{2}{\tau - \bar \tau }} \cF[\bar \tau]   \ , \ee
where $\tau$ is the upper complex half plan coordinate $\tau = i\frac{1-T}{1+T}$.  
One computes that 
\be \cD^2\Scal{  \Scal{ \bar \partial + \frac{2}{\tau - \bar \tau }} \cF[\bar \tau]}  = -   \partial \Scal{  (\tau - \bar \tau )^2  \partial \Scal{  \Scal{ \bar \partial + \frac{2}{\tau - \bar \tau }} \cF[\bar \tau]}} = 0 \ ,  \ee
which implies that the terms in $\cD^n\cE_\grad{2}{1}{0}  $ only depend on the holomorphic function of $\tau$ for $n\ge 2$, whereas the terms in $\bar \cD^n\cE_\grad{2}{1}{0}  $ only depend on the anti-holomorphic function $\cF[\bar \tau]$. Altogether with the structure \eqref{D4R4SuperformGrad} described in the preceding section, we conclude that the function $\cE_\grad{2}{1}{0}  $ only include the anti-holomorphic part such that it satisfies moreover to 
\be \cD^2  \cE_\grad{2}{1}{0} = 0 \ . \label{AntiHolo2}  \ee
For the differential equation on $SL(3)/SO(3)$, one uses equivalently the commutation relations  
\be
[\cD_{i j k l}, \cD^{p q r s}] \cF_{(i_1 \dots i_n)} = \frac{n}{4} \delta^{p q r s}_{i j k) ( i_1} \cF_{i_2 \dots i_n) (l} - \frac{n}{8} \delta^{pqrs}_{i j k l} \cF_{(i_1 \dots i_n)} 
\ee
to prove that 
\be   \mathcal{D}_{[2]}{}^{jk} \mathcal{D}^{n}_{([4n-2]j k)} \cE_\grad{2}{1}{0} = \frac{2n+1}{4n-1} \cD^{n-1}_{[4n-4]}  \mathcal{D}_{[2]}{}^{jk} \mathcal{D}_{[2]j k} \cE_\grad{2}{1}{0} \ , \ee
such that the second equation in \eqref{eq:SUSY_const} reduces to 
\be \frac{29}{55} \cD^{13}_{[52]} \Scal{  \mathcal{D}_{[2]}{}^{jk} \mathcal{D}_{[2]j k} - \frac{5}{12} \cD_{[4]} } \cE_\grad210= 0  \ , \ee
so that 
\be \cD_{(i j}{}^{pq} \cD_{kl) pq} \cE_\grad210 = \frac{5}{12}  \cD_{i j k l}  \cE_\grad210\ . \ee
As explained in \cite{Minimal}, this equation moreover implies that $ \cE_\grad210$ is an eigen function of the Laplace operator 
\be
\Delta_{SL(3)} \cE_\grad{2}{1}{0} = \frac{4}{3} \cE_\grad{2}{1}{0}  \ ,
\ee
such that 
\be\label{SL3Equation}
 \cD_{i j}{}^{p q} \cD_{k l p q} \cE_\grad{2}{1}{0} = \frac{5}{12} \cD_{i j k l}  \cE_\grad{2}{1}{0}  + \frac{1}{9} (\ve_{i k} \ve_{j l} + \ve_{i l} \ve_{j k}) \cE_\grad{2}{1}{0}
\ee
which is precisely equation \eqref{FunctionConstraint} for $s^\prime=-\frac{1}{2}$.

The closed-superform defining the invariant, admits therefore the gradient expansion 
\be 
  \cL[\cE_\grad210] =\sum_{q\ge0} \biggl(\   \sum_{p\ge0} \bar{U}^{-2 p}  \bar \cD^p \cD^q_{[4q]}\cE_\grad210  \,  \mathcal{L}^{\ord{4p} [4 q]} +   {U}^{-2}  \cD \cD^q_{[4q]}\cE_\grad210 \,  \mathcal{L}^{\ord{-4} [4 q]}  \biggr) \ , \label{D4R4SuperformGradFinal}
\ee
for an arbitrary solution to \eqref{AntiHolo2} and \eqref{SL3Equation}. Of course one has the complex conjugate invariant, defined such that 
\be 
  \bar \cL[\cE_\grad201] =\sum_{q\ge0} \biggl(\   \sum_{p\ge0} {U}^{-2 p}  \cD^p \cD^q_{[4q]}\cE_\grad201  \,  \bar{\mathcal{L}}^{\ord{-4p} [4 q]} +   \bar {U}^{-2}  \bar \cD \cD^q_{[4q]}\cE_\grad201 \,  \bar{ \mathcal{L}}^{\ord{4} [4 q]}  \biggr) \ , \label{D4R4SuperformGradbar}
\ee
and the associated function multiplying $\nabla^4 R^4$ is $\cE_\grad210 + \cE_\grad201$, which is defined to be a real function of $\tau$ and $\bar \tau$. This is consistent with the appearance of the threshold function 
\be \cE_\grad210(T, \bar T, t)  + \cE_\grad201(T, \bar T, t)  = E_{[2]}(\tau,\bar \tau)  E_{[-\frac{1}{2},0]}(t) \ , \ee
in the low energy effective action of type II string theory compactified on $T^2$ \cite{Green:2010wi,Basu:2007ru}. The Eisenstein function  $E_{[s,0]}$ satisfies in general to the differential equation \cite{Minimal}
\be \cD_{ij}{}^{pq} \cD_{klpq} \cE_{s} = - \frac{4 s-3}{12} \cD_{ijkl} \cE_{s}  + \frac{s ( 2 s-3)}{18}  ( \varepsilon_{ik} \varepsilon_{jl} +  \varepsilon_{il} \varepsilon_{jk} )  \cE_{s} \ , \label{EisenQuadra} \ee
such that $ E_{[-\frac{1}{2},0]}$ is indeed a solution to \eqref{SL3Equation}, whereas $E_{[2]}$ solves \eqref{LaplaceSL2}. Using the explicit expansion of the Eisenstein series $E_{[2]}$, 
\be E_{[2]} = 2 \zeta(4) \tau_2^{\; 2}  + \pi \zeta(3) \tau_2^{\; -1} +\frac{\pi}{2}  \sum_{N=1}^\infty \sum_{r|N} \Scal{  \frac{1}{r^3}  } \frac{  1+ 2 N \pi \tau_2}{\tau_2} \scal{ e^{2\pi i N \tau }
 + e^{-2\pi i N \bar  \tau }} \ , \ee
one finds indeed that $E_{[2]} = \cE_{[2]} + \overline \cE_{[2]}$  for the complex function 
\be  \cE_{[2]}  = - \frac{1}{2}  \zeta(4) \frac{ 3 \tau \bar \tau^2 - \bar \tau^3}{\tau - \bar \tau}  + \frac{\pi}{2} \zeta(3) \tau_2^{\; -1} +\frac{\pi}{2}  \sum_{N=1}^\infty \sum_{r|N} \Scal{  \frac{1}{r^3}  } \frac{  1+ 2 N \pi \tau_2}{\tau_2} e^{-2\pi i N \bar  \tau } \ , \ee
that satisfies to 
\be \cD^2  \cE_{[2]}  = 0 \ . \ee
However this complex function is not modular invariant, and in order for the supersymmetry invariant to preserve $SL(2,\mathds{Z})$, it is necessary that 
\be \bar{\mathcal{L}}^{\ord{0} [4 q]}  = {\mathcal{L}}^{\ord{0} [4 q]} \ , \qquad \bar{\mathcal{L}}^{\ord{\pm4} [4 q]}  = {\mathcal{L}}^{\ord{\pm4} [4 q]}\ ,\label{RealConstraint}\ee
such that the whole invariant only depends on the gradient expansion of the modular invariant function $E_{[2]}$. This reality condition is indeed compatible with the linearised analysis, because there is only one linearised invariant for each values of $p$ and $q$, and \eqref{RealConstraint} must therefore be satisfied in the linearised approximation. We know indirectly that this reality condition must be satisfied at the non-linear level, because the term in $ 2 \zeta(4) \tau_2^{\; 2} $ lifts to type IIA supergravity in ten dimensions \cite{Green:2010wi}, where it is known to appear in the 2-loop string theory effective action \cite{D'Hoker:2014gfa}, which is by construction invariant with respect to the $B$ field gauge transformations. 
\section{Decompactification limit in lower dimensions}
\label{From4to7to8}
We have derived in the last section the structure of the chiral $\nabla^4 R^4$ type invariant in eight dimensions, however the same analysis does not apply directly to the second real $\nabla^4 R^4$ type invariant \eqref{D4R48D}. To understand the two invariants, we are going to analysis the corresponding invariant obtained by dimensional reduction in four dimensions. We will see that these two invariants are related through the action of $E_{7(7)}$ in four dimensions. Solving the differential equation satisfied by the function $\cE_\grad422$ defining the $\nabla^4 R^4$ type invariant \eqref{D4R44DInt} in four dimensions in the decompactification limit, we will indeed obtain  that it lifts to the two independent invariants \eqref{D4R48D} in eight dimensions. 

We must warn the reader that considering explicit decompositions of $E_{7(7)}$ and $SL(5)$ forced us to use the same indices for various representations. Each subsection in this section uses a different definition of the indices that is recalled in the beginning.
\subsection{$R^4$ and $\nabla^4 R^4$ type invariants in four dimensions}
In this subsection we shall review the results displayed in section \ref{N8Review}, which were originally derived in \cite{Minimal}. In $\cN=8$ supergravity the scalar fields parametrise the symmetric space $E_{7(7)}/\SU $, where $\SU$ is the quotient of $SU(8)$ by the $\mathds{Z}_2$ kernel of the antisymmetric rank two tensor representation,
and the covariant derivative $\cD_{ijkl}$ on $E_{7(7)}/\SU $ in tangent frame are in the rank four antisymmetric complex selfdual representation of $SU(8)$, \be \cD^{ijkl} = \frac{1}{24} \varepsilon^{ijklmnpq} \cD_{mnpq} \ , \ee
with $i,j,k,l$ running from $1$ to $8$ are in the fundamental representation of $SU(8)$. 

In four dimensions there is a bijective correspondence between the supersymmetry invariants and the linearised invariants defined as superspace integrals in harmonic superspace, due to the enhanced superconformal symmetry $SU(2,2|8)$ of the theory in the linearised approximation \cite{Drummond:2003ex,Drummond:2010fp}. 
\addtocontents{toc}{\protect\setcounter{tocdepth}{1}}
\subsubsection{The $R^4$ type invariant}
\addtocontents{toc}{\protect\setcounter{tocdepth}{2}}
One defines $R^4$ type invariants in the linearised approximation using harmonic variables $u^{r}{}_{i}$ and $u^{\hat{r}}{}_{i}$ parametrising  $SU(8)/S(U(4) \times U(4))$, where $r$ runs from $1$ to $4$, and $\hat{r}$ from $5$ to $8$. One defines the G-analytic superfield \cite{Hartwell:1994rp}
\be W = u^1{}_i u^2{}_j u^3{}_k u^4{}_l W^{ijkl} \ , \ee
satisfying to 
\be u^r{}_i D^i_\alpha \, W = 0 \ , \qquad u^i{}_{\hat{r}} \bar D_{\adt i} \, W = 0 \ , \ee
such that one can define the supersymmetric Lagrangians 
\bea&&  \int d^8\theta d^8 \bar \theta du  F_u^{[0,0,0,n,0,0,0]} \, W^{4+n} \CR
&\sim &W^{n\,  [0,0,0,n,0,0,0]} R^4 + \dots +W^{n-12 \,  [0,0,0,n-12,0,0,0]}   \chi^{8 \, [0,0,0,6,0,0,0]} \bar \chi^{8 \, [0,0,0,6,0,0,0]} \ , \eea
with
\be F_u^{[0,0,0,n,0,0,0]} \sim \prod_{k=1}^n ( u^{[i_k}{}_1  u^{j_k}{}_2  u^{k_k}{}_3  u^{l_k]}{}_4 ) \ . \ee
Using the bijective correspondence one concludes that the $R^4$ type invariant is unique in four dimensions, and admits the following gradient expansion in a function $\cE_\grad844$ 
\be \cL[\cE_\grad{8}{4}{4}] =  \sum_{n = 0}^{12} \cD^n_{[0,0,0,n,0,0,0]} \cE_\grad{8}{4}{4}\,    \cL^{[0,0,0,n,0,0,0]}\ , \ee
which satisfies to the constraint that its second derivative restricted to the $[0,1,0,0,0,1,0]$ irreducible representation of $SU(8)$ vanishes, \ie 
\be \Scal{28 \cD_{ijpq} \cD^{klpq}- 3 \delta_{ij}^{kl} \Delta }  \, \cE_\grad844 = 0 \ . \label{Quadratic} \ee
This constraint implies by consistency that all the higher order derivatives in representations that do not belong to the $[0,0,0,n,0,0,0]$ irreducible representations vanish. In particular, the third derivative in the $[0,1,0,1,0,1,0]$ also vanishes, \ie 
\be \Scal{ 4 \cD_{ijpq} \cD^{pqmn} \cD_{mnkl} - \cD_{ijkl} \scal{ \Delta + 24} } \cE_\grad844 = 0 \ .  \label{CubicC} \ee
The linear term in the differential operator in this formula comes from the symmetrisation of the cubic term, using the commutation relation
\be [ \cD^{ijkl} , \cD_{pqrs} ]  \cD_{tuvw}= - 24 \delta^{ijkl}_{qrs][t} \cD_{uvw][p} + 3 \delta^{ijkl}_{pqrs} \cD_{tuvw} \ .  \label{Comut} \ee
It follows from representation theory that the quadratic constraint \eqref{Quadratic} implies the cubic constraint \eqref{CubicC} and its complex conjugate, and using \eqref{Quadratic} in \eqref{CubicC}, one obtains 
\be - 16\,  \cD_{ijkl} \scal{ \Delta + 42} \cE_\grad844= 0 \ , \ee
so we conclude that the function $\cE_\grad844$ defining the $R^4$ type invariant satisfies moreover to 
\be \cD_{ijpq} \cD^{klpq} \cE_\grad844  = - \frac{9}{2} \delta_{ij}^{kl} \cE_\grad844 \ , \label{QuadraR4}  \ee
such that 
\be \Delta \cE_\grad844 = \frac{1}{3} \cD_{ijkl} \cD^{ijkl} \cE_\grad844 = - 42 \, \cE_\grad844\ . \ee
In the following, it will be convenient to rewrite this constraint in terms of  the $\mathfrak{e}_{7(7)}$ valued differential operator ${\bf D}_{56}$ in the fundamental representation \cite{Minimal}
\be
 {\bf D}^{2}_{56}  \ \cE_\grad844 = - \frac{9}{2} \mathds{1}_{56}  \cE_\grad844\ . \label{R44DEq} 
\ee

\addtocontents{toc}{\protect\setcounter{tocdepth}{1}}
\subsubsection{The $\nabla^4 R^4$ type invariant}
\addtocontents{toc}{\protect\setcounter{tocdepth}{2}}

One defines $\nabla^4 R^4$ type invariants in the linearised approximation using harmonic variable parametrising   $SU(8)/S(U(2) \times U(4) \times U(2))$. We define the G-analytic superfield \cite{Hartwell:1994rp}
\be
W^{rs} = u^{1}{}_{i} u^{2}{}_{j} u^{r}{}_{k} u^{s}{}_{l} W^{ijkl} \ , 
\ee
where $r, s$ are now $SU(4)$ indices running from $1$ to $4$ and $W^{rs}$ is in the $[0,1,0]$ representation. Since $SU(4) \simeq SO(6)$, $W^{rs}$ is a vector of $SO(6)$, and the general monomials in $W^{rs}$ are the symmetric traceless monomials times an arbitrary power of the scalar product of $W^{rs}$  with itself.  The general invariant Lagrangian is defined as the harmonic superspace integral over $24$ Grassmann variables of such monomials as
\bea
&& \int d^{8} \theta d^{8} \bar \theta d u F^{[0,k,0,n,0,k,0]}_{u \; r_1 s_1 \dots r_n s_n} (W^{rs} W_{rs})^{2 + k} W^{r_1 s_1} W^{r_2 s_2} \dots W^{r_n s_n} \\
&\sim &W^{n+2k\,  [0,k,0,n,0,k,0]} \nabla^4 R^4 + \dots +W^{n+2k-20 \,  [0,k-6,0,n-8,0,k-6,0]}   \chi^{12 \, [0,4,0,4,0,2,0]} \bar \chi^{12 \, [0,2,0,4,0,4,0]} \nn \ . \eea
Using the bijective correspondence, one concludes that the non-linear invariant admits the following gradient expansion in the function $\cE_\grad822$ 
\be
 \cL[\cE_\grad{8}{2}{2}] =   \sum_{n, k = 0}^{n + 2k \leq 20} \mathcal{D}^{n + 2k}_{[0,k,0,n,0,k,0]}\cE_\grad{8}{2}{2}  \, \cL^{[0,k,0,n,0,k,0]}\ , 
\ee
which satisfies to the constraint that its third derivative restricted to the $[0,2,0,0,0,0,0]\oplus[1,0,0,1,0,0,1]\oplus[0,0,0,0,0,2,0]$ representation of $SU(8)$ vanishes, \ie 
\bea
\Scal{ 4 \cD_{ijpq} \cD^{pqmn} \cD_{mnkl} - \cD_{ijkl} \scal{ \Delta + 24} } \cE_\grad{8}{2}{2} &=& 0  \ , \\
 \Scal{ 36 \cD_{jr[kl} \cD^{irmn} \cD_{pq]mn} - \delta^i_j \cD_{klpq} ( \Delta + 42)   + \delta^i_{[k} \cD_{lpq]j} ( \Delta-120)}  \cE_\grad{8}{2}{2} &=&0   \label{CubicR} \ , \\
\Scal{ 4 \cD^{ijpq} \cD_{pqmn} \cD^{mnkl} - \cD^{ijkl} \scal{ \Delta + 24} } \cE_\grad{8}{2}{2} &=& 0\  \label{CubicCb}. 
\eea
One computes similarly that \eqref{CubicR} implies 
\be  \Scal{ 24 \cD_{r[klp} \cD^{irmn} \cD_{q]jmn} - \delta^i_j \cD_{klpq} ( \Delta -12)   + \delta^i_{[k} \cD_{lpq]j} ( \Delta+96)} \cE_\grad{8}{2}{2} = 0 \ . \label{CubicR2}  \ee
Using the property that the function satisfies to all (\ref{CubicR},\ref{CubicCb},\ref{CubicR2}), one gets the following  integrability condition in the $[0,1,0,0,0,1,0]$, 
\bea &&  \cD^{[i|rpq} \Bigl( 36 \cD_{rs[kl} \cD^{j]smn} \cD_{pq]mn} - \delta^{j]}_r \cD_{klpq} ( \Delta + 42)   + \delta^{j]}_{[k} \cD_{lpq]r} ( \Delta-120)\Bigr . \CR
&& \hspace{15mm}  \Bigl . +48 \cD_{s[klp} \cD^{j]smn} \cD_{q]rmn} - 2 \delta^{j]}_r \cD_{klpq} ( \Delta -12)   + 2 \delta^{j]}_{[k} \cD_{lpq]r} ( \Delta+96)  \Bigr)  \cE_\grad{8}{2}{2}  \CR
&=&  -9   \cD_{klrs} \cD^{rspq} \cD_{pqmn} \cD^{mnij}  \cE_\grad{8}{2}{2}   + \frac{3}{2} \cD^{ijpq} \cD_{klpq}  \scal{ 5\Delta +246}\cE_\grad{8}{2}{2}  - \frac{9}{16} \delta^{ij}_{kl}  \Delta \scal{ \Delta + 60 } \cE_\grad{8}{2}{2}  \CR
&=&  \frac{21}{4} \Scal{  \cD^{ijpq} \cD_{klpq} - \frac{3}{28} \delta^{ij}_{kl} \Delta } \scal{ \Delta + 60 } \cE_\grad{8}{2}{2} \ ,
\eea
where we only used \eqref{CubicCb} in the last step. Because the function  $ \cE_\grad{8}{2}{2}$ does not satisfy to the quadratic constraint \eqref{Quadratic}, we conclude that it must  satisfy instead 
\be \Delta  \cE_\grad822  = - 60 \cE_\grad822  \ . \label{60} \ee
Therefore the constraints (\ref{CubicR},\ref{CubicC}) simplify to
\bea
 \cD_{i j p q} \cD^{p q r s} \cD_{r s k l}\cE_\grad{8}{2}{2} &=& - 9 \cD_{i j k l} \cE_\grad{8}{2}{2} \ , \\
  2 \cD_{j r [kl} \cD^{irmn} \cD_{pq] mn} \cE_\grad{8}{2}{2} &=& - \delta_{j}^{i} \cD_{k l p q} \cE_\grad{8}{2}{2} + 10 \delta^{i}_{[k} \cD_{l p q] j} \cE_\grad{8}{2}{2}\ . 
\eea
These constraints can be rewritten in terms of the $\mathfrak{e}_{7(7)}$ valued differential operator ${\bf D}_{56}$ and ${\bf D}_{133}$ in the fundamental and the adjoint  representations, respectively, as \cite{Minimal}
\be \label{D4R44DM} 
 {\bf D}^{3}_{56} \cE_\grad{8}{2}{2} = - 9 {\bf D}_{56}  \cE_\grad{8}{2}{2} \ , \qquad {\bf D}^{3}_{133}  \cE_\grad{8}{2}{2}  = - 20 {\bf D}_{133} \cE_\grad{8}{2}{2} \ . 
\ee

\addtocontents{toc}{\protect\setcounter{tocdepth}{1}}
\subsubsection{$E_{7(7)}$ Eisenstein series}
\addtocontents{toc}{\protect\setcounter{tocdepth}{2}}

One can define solutions to these differential equations in terms of Eisenstein series defined as constrained Epstein series in the fundamental representation  \cite{Obers:1999um}. Let us consider a rank one charge vector $\Gamma$ in the ${\bf 56}$ of $E_{7(7)}$ such that the second derivative of the quartic invariant restricted to the adjoint representation vanishes. Acting with the scalar field one obtains that the central charges $Z(\Gamma)_{ij} = \cV_{ij}{}^I\Gamma_I$ satisfy to 
\be Z_{[ij} Z_{kl]} = \frac{1}{24} \varepsilon_{ijklpqrs} Z^{pq} Z^{rs} \ , \qquad Z_{ik} Z^{jk} = \frac{1}{8} \delta_i^j Z_{kl} Z^{kl} \ . \ee
The action of the covariant derivative on the central charges gives
\be \cD_{ijkl} Z^{pq} = 3 \delta^{pq}_{[ij} Z_{kl]} \ , \qquad \cD_{ijkl} Z_{pq} = \frac{1}{8} \varepsilon_{ijklpqrs} Z^{rs} \ . \ee
One computes then that 
\be \cD_{ijkl} |Z|^2 = 6 Z_{[ij} Z_{kl]} \ , \qquad \cD_{ijpq} \cD^{klpq} |Z|^2 = 6 Z_{ij} Z^{kl}+2 \delta_{ij}^{kl} |Z|^2   \ , \ee
with $|Z|^2 = Z_{ij} Z^{ij}$. Using moreover the intermediate step 
\be \cD_{ijpq} |Z|^2 \cD^{klpq} |Z|^2 = 2 Z_{ij} Z^{kl} |Z|^2 + \frac{1}{4} \delta_{ij}^{kl} |Z|^4 \ , \ee
one computes that 
\be \cD_{ijpq} \cD^{klpq} |Z|^{-2s} = 2s(s-2) Z_{ij} Z^{kl} |Z|^{-2s-2} + \frac{s(s-11)}{4} \delta_{ij}^{kl} |Z|^{-2s} \  . \ee
One gets therefore a solution to the second order equation \eqref{QuadraR4} associated to the $R^4$ type invariant  for $s=2$. One computes then that 
\bea&&  \cD_{jr[kl} \cD^{irmn} \cD_{pq]mn} |Z|^{-2s} \CR
&=& - \frac{1}{2} s(s-2)(s-4) \delta^i_j Z_{[kl} Z_{pq]} |Z|^{-2s-2} + \frac{1}{2} s( s^2-9s -40) \delta^i_{[k} Z_{pq} Z_{l]j} |Z|^{-2s-2} \ , \eea
and therefore the third equation  \eqref{CubicR} is automatically satisfied by $|Z|^{-2s}$. One computes moreover 
\be \cD_{ijpq} \cD^{pqrs} \cD_{rskl} |Z|^{-2s} = - 3 s(s-2)(s-4) Z_{ij} Z_{kl} |Z|^{-2s-2} - \frac{3}{2} s(s^2-15s + 8) Z_{[ij} Z_{kl]} |Z|^{-2s-2} \ . \ee
One concludes therefore that the function $|Z|^{-2s}$ solves to the cubic equation \eqref{CubicC} for $s=4$. In general one has moreover 
\be \Delta |Z|^{-2s} = 3s(s-9) |Z|^{-2s} \ . \ee
One formally obtains $E_{7(7)}(\mathds{Z})$ invariant functions by considering the sum over all integral charges satisfying to the rank one constraint 
\be  E_{\mbox{\DEVII000000s}} =  \sum_{\vspace{-2mm}\begin{array}{c}\scriptstyle \vspace{-4mm}  \Gamma\in \mathds{Z}^{56} \vspace{2mm}\\ \scriptscriptstyle I_4^{\prime\prime}(\Gamma)|_{\bf 133}=0\end{array}} |Z(\Gamma)_{ij} Z(\Gamma)^{ij}|^{-s} \ . \label{E56s} \ee
However this series does not converge for $s\le 9$, which includes the cases of interest. Using the theorem of \cite{Krutelevich}, the rank 1 integral charge vectors $\Gamma$ are in the $E_{7(7)}(\mathds{Z})$ orbit of an integer element of grad $3$ in the parabolic decomposition of $\mathfrak{e}_{7(7)}$ 
\bea \mathfrak{e}_{7(7)}&\cong& \overline{\bf 27}^\ord{-2} \oplus \scal{ \mathfrak{gl}_1 \oplus \mathfrak{e}_{6(6)} }^\ord{0} \oplus {\bf 27}^\ord{2} \CR
{\bf 56} &\cong& {\bf 1}^\ord{-3} \oplus {\bf 27}^\ord{-1} \oplus \overline{\bf 27}^\ord{1} \oplus {\bf 1}^\ord{3} \ , \eea
\ie that 
\be \{\Gamma \in  \mathds{Z}^{56}\, |\, I_4^{\prime\prime}(\Gamma)|_{\bf 133}=0\}  \cong \mathds{Z}^* \times E_{7(7)}(\mathds{Z})/ \scal{ E_{6(6)}(\mathds{Z}) \ltimes \mathds{Z}^{27}} \ . \ee
Using the property that the rank 1 charge vector (with unit grater common divider of all components) defines a character of $E_{7(7)}$ whose restriction to the Cartan subgroup is the exponential of the generator \DEVII0000001 in the appropriate basis  
\be \cV(\phi,V,a) \Gamma^\ord{3} = e^{3\phi} \Gamma^\ord{3} \ , \ee
one obtains that \eqref{E56s} coincides with the Langlands formula 
\be  E_{\mbox{\DEVII000000s}} =2 \zeta(2s)  \sum_{ g\in \frac{E_{7}(\mathds{Z})}{ E_{6}(\mathds{Z}) \ltimes \mathds{Z}^{27}} }   g\scal{ e^{-6s \phi} } \ , \ee
where $g$ acts on $e^{-6s \phi} $ through the non-linear realisation of $E_{7(7)}$ on the coset representative of $E_{7(7)}/\SU$ in the parabolic gauge. 
Using Langlands functional identities one shows that these Eisenstein series exist as functions and are related through  \cite{Green:2010kv} 
\be  E_{\mbox{\DEVII{\frac{3}{2}}000000}} \propto   E_{\mbox{\DEVII0000002}} \ , \qquad E_{\mbox{\DEVII{\frac{5}{2}}000000}} \propto   E_{\mbox{\DEVII0000004}} \ , \ee
such that these functions indeed satisfy to the differential equation associated to the $R^4$ and $\nabla^4 R^4$ type invariants 
\be \cE_\grad844 =  E_{\mbox{\DEVII{\frac{3}{2}}000000}}  \ , \qquad \cE_\grad822 =  \frac{1}{2} E_{\mbox{\DEVII{\frac{5}{2}}000000}}  \ , \ee
consistently with the conjecture that they define the exact low energy effective action in type II string theory \cite{Obers:1999um,Green:2010kv}. 

\subsection{Decompactification limit to seven dimensions}
\label{4to7}
Any supersymmetry invariant in seven dimensions, dimensionally reduces to a well defined supersymmetry invariant in four dimensions. It follows that the structure of the invariants in seven dimensions must be compatible with the differential equations we have derived in four dimensions. In this section we will solve these differential equations in the parabolic gauge associated to the dimensional reduction from seven to four dimensions, to exhibit the differential equations satisfied by the seven-dimensional scalar fields. But before to do this, let us review shortly some properties of the theory in seven dimensions.

\addtocontents{toc}{\protect\setcounter{tocdepth}{1}}
\subsubsection{Maximal supergravity in seven dimensions}
\addtocontents{toc}{\protect\setcounter{tocdepth}{2}}
In seven dimensions the scalar fields parametrise the symmetric space $SL(5)/SO(5)$, and the double cover $Sp(2)$ of $SO(5)$   is the R-symmetry group. The $SL(5)$ representative $V$ is defined such that it transforms with respect to rigid $SL(5)$ on the right and local $Sp(2)$ on the left 
\be
 V_{ij}{}^K(x) \rightarrow  L_i{}^k(x) L_j{}^l(x) V_{kl}{}^L(x) R_L{}^K \  , 
\ee
where  $i, j,\dots = 1, \dots 4$ are the indices in the fundamental representation of $Sp(2)$. The theory is defined in the linearised approximation in terms of the real scalar superfield $L^{i j, k l}$ 
\be L^*_{ij,kl} = \Omega_{ip} \Omega_{jq} \Omega_{km} \Omega_{ln} L^{pq,mn}\ , \ee
in the $[0,2]$ of $Sp(2)$, \ie
\be
 L^{i j , k l} = - L^{j i , k l} = - L^{i j , l k} = L^{k l, i j}  \ , \qquad \Omega_{i j} L^{i j, k l} = 0 \ , \quad L^{i [j, k l]} = 0\ , 
  \quad  \Omega_{j l} L^{i j, k l} = 0  \ , 
\ee
with $\Omega_{i j}$ the symplectic form of $Sp(2)$. This superfield satisfies to the linear constraint that its covariant derivative vanishes in the $[1,2]$ of $Sp(2)$, and its second derivative vanishes in the vector representation of $SO(1,7)$ times the $[2,0]$ of $Sp(2)$ and in the $SO(1,7)$ singlet in the $[0,1]$ of $Sp(2)$. In particular  
\be
D_{\alpha}^{i} L^{j k , l m} = \Omega^{i [j} \chi^{k], l m}_{\alpha} + \Omega^{i [l} \chi^{m], j k}_{\alpha} + \frac{1}{4} \Omega^{j k} \chi_{\alpha}^{i, l m} + \frac{1}{4} \Omega^{l m} \chi_{\alpha}^{i, j k} \ , 
\ee
where $\chi^{i, j k}_{\alpha}$ is the Dirac $Spin(1,6)$ spinor in the $[1,1]$ of $Sp(2)$. At mass dimension $1$ the field content includes the scalar field momentum $P_{a}^{i j , k l} = \partial_a L^{ij,kl}$ transforming in the $[0,2]$, the two-form field strength $F_{ab}^{i, j}$ in the $[2,0]$ and the three-form field strength $H_{a b c}^{i j}$ in the $[0,1]$ irreducible representation. At mass dimension $\frac{3}{2}$ there is the Rarita--Schwinger field strength  $\rho^{i}_{a b \alpha}$ in the $[0,1,1]$ irreducible representation of $Spin(1,6)$ and at mass dimension $2$ the Riemann tensor  $R_{a b, c d}$ in the $[0,2,0]$ of $SO(1,6)$. 
\vskip 10mm

\begin{figure}[htbp]
\def\xmin{5}
 \def\ymin{0}
 \begin{tikzpicture}[thick, scale = 1, transform canvas=={scale=1}]
\draw (\xmin,\ymin + 0.4) node{$L^{ij,kl}$};
\draw (\xmin + 2,\ymin + 0.4) node{$\chi_{\alpha}^{i,kl}$};
\draw (\xmin + 4,\ymin + 0.4) node{$F^{i,j}_{a b} / H_{a b c}^{i j}$};
\draw (\xmin + 6,\ymin + 0.4) node{$\rho^{i}_{a b \alpha}$};
\draw (\xmin + 8,\ymin + 0.4) node{$R_{ab,cd}$};
\draw (\xmin + 1,\ymin - 0.3) node{$D_{\alpha}^{i}$};

\draw[->,draw=black,thick] (\xmin,\ymin) -- (\xmin + 2,\ymin); 
\draw[->,draw=black,thick] (\xmin + 2,\ymin) -- (\xmin + 4,\ymin); 
\draw[->,draw=black,thick] (\xmin + 4,\ymin) -- (\xmin + 6,\ymin); 
\draw[->,draw=black,thick] (\xmin + 6,\ymin) -- (\xmin + 8,\ymin); 
\end{tikzpicture}
\caption{Supergravity multiplet in  seven dimensions}
\end{figure}

The $R^4$ type invariant can be defined in the linearised approximation in harmonic superspace, using harmonic variables $u^r{}_i,\, u_{ri}$ parametrising $Sp(2)/U(2)$, with $r=1,2$ of $U(2)$ \cite{Bossard:2009sy}, such that the superfield  
\be
W = u^{1}{}_{i} u^{2}{}_{j} u^{1}{}_{k} u^{2}{}_{l} L^{i j,k l}
\ee
satisfies the G-analyticity constraint
\be
u^{r}{}_{i} D_{\alpha}^{i} W = 0  \ . 
\ee
One can write generic invariants 
\be
 \int d^{16} \theta du F^{[0, 2n]}_{u} W^{4 + n} \sim W^{n\, [0,2n]} R^4 + \dots + W^{n-12\, [0,2n-24]} \chi^{16\, [0,24]} \ , 
\ee
with $F^{[0, 2n]}_{u}$  defined as the function of the inverse harmonic variables in the $[0,2n]$ irreducible representation of $Sp(2)$
\be
F^{[0, 2n]}_{u} = \prod_{m=1}^n u^{[i_m}{}_{1} u^{j_m]}{}_{2} u^{[k_m}{}_{1} u^{l_m]}{}_{2}  \ . 
\ee
This suggests the gradient expansion of the non-linear invariant 
\be \cL[\cE_\gra{4}{2}] = \sum_{n = 0}^{12} \cD_{[0,2n]}^n \cE_\gra{4}{2}\,   \cL^{[0,2n]}\ .   \ee

One $\nabla^4 R^4$ type invariant can be defined in the linearised approximation in harmonic superspace, using harmonic variables $u^1{}_i,\, u^r{}_i,\, u^4{}_i$ parametrising $Sp(2)/(U(1)\times Sp(1))$, with $r=1,2$ of $Sp(1)$ \cite{Bossard:2009sy}, such that the superfield  
\be
W^{rs} = u^{1}{}_{i} u^{1}{}_{k} u^{r}{}_{j} u^{s}{}_{l} L^{i j,k l}
\ee
satisfies to the $G-$analyticity constraint 
\be
u^{1}_{i} D_{\alpha}^{i} W^{r s} = 0 \ . 
\ee
One can write generic invariants 
\bea
&&  \int d^{24} \theta du F^{[4k, 2n]}_{u\,  r_1s_1,r_2s_2\dots r_n s_n} (W^{rs} W_{rs})^{2+k}  W^{r_1s_1} W^{r_2 s_2} \dots W^{r_n s_n} \CR
&\sim& W^{2k+n\, [4k,2n]} \nabla^4 R^4 + \dots + W^{2k+n-20\, [4k-24,2n-16]} \chi^{24\, [24,16]} \ ,
\eea
with
\be
F^{[4k, 2n]}_{u\,  r_1s_1,r_2s_2\dots r_n s_n} = \prod_{a=1}^{4k+2n}    \scal{ u^{i_a}{}_1 } \prod_{b=1}^{2n} u^{[j_b}{}_{r_b} u^{k_b]}{}_{s_b} \Big|_{[4k, 2n]} \ , \ee
projected to the $[4k, 2n]$ irreducible representation of $Sp(2)$. This suggests the gradient expansion of the invariant at the non-linear level 
\be \cL[\cE_\gra41] =  \sum_{n,k = 0}^{n + 2 k \leq 20} \cD_{[4k,2n]}^{n + 2k}\cE_\gra{4}{1}  \, \cL^{[4k,2n]} \  . \label{D4R47DGrad}  \ee

\addtocontents{toc}{\protect\setcounter{tocdepth}{1}}
\subsubsection{$E_{7(7)}/\SU$ in the parabolic gauge}
\addtocontents{toc}{\protect\setcounter{tocdepth}{2}}
We consider the graded decomposition of $\mathfrak{e}_{7(7)}$ associated to the Cartan element \DEVII0000100, \ie 
\be \label{133Graded}
\text{{\goth e}}_{7(7)} =  \bar{{\bf 5} }^\ord{-6} \oplus  ( {\bf 3} \otimes {\bf 5})^\ord{-4}  \oplus   ( \bar{ \bf 3} \otimes {\bf \overline{10}})^\ord{-2}  \oplus \scal{  \mathfrak{gl}_1 \oplus \mathfrak{sl}_3 \oplus \mathfrak{sl}_5 }^\ord{0} \oplus ( { \bf 3} \otimes { \bf 10})^\ord{2}
\oplus (\bar{ \bf 3} \otimes \bar{ \bf 5})^{(4)} \oplus { \bf 5}^\ord{6}\ , 
\ee
such that the grad zero component includes the product of the seven-dimensional duality group $SL(5)$ times the symmetry group $SL(3)$ associated to the compactification on $T^3$. The scalar fields $A^{A}_{IJ}$ in the $(\bf{3} \otimes \bf{10})$ of $SL(3) \times SL(5)$, with $A=1,2,3$ of $SL(3)$ and $I=1$ to $5$ of  $SL(5)$, are the scalar components of the seven-dimensional 1-forms potentials. The scalar fields $B^{I}_{A}$ in  the $(\bf{\bar{3}} \otimes \bf{\bar{5}})$ are the scalar components of the seven-dimensional $2$-form potentials, whereas the scalars  $C_{I}$ in the  $\bf{5}$ are dual to  the $2$-form component of the seven-dimensional $2$-form potentials. Due to the Chern-Simons terms in seven dimensions, the gauge invariant differentials are 
\bea
\nabla A^{A}_{I J} &=& d A^{A}_{I J} \ , \qquad \nabla B^{I}_{A} = d B^{I}_{A} + \frac{1}{4} \ve^{I K L P Q} \ve_{A C D} A^{C}_{K L} dA^{D}_{P Q} \ ,  \CR
\nabla C_{I} &=& d C_{I} + \frac{1}{\sqrt{2}} B^{K}_{A} d A^{A}_{IK} - \frac{1}{\sqrt{2}} A^{A}_{I K} dB^{K}_{A} + \frac{1}{12 \sqrt{2}} \ve^{J K L M N} \ve_{A B C} A^{A}_{I J} A^{B}_{K L} d A^{C}_{M N}  \ . \label{Nilbeins}
\eea
We define the nilpotent component of the $E_{7(7)}$ coset representative in the fundamental representation
\be\label{56Graded} 
{\bf 56} =\overline{\bf 3}^\ord{-5} \oplus {\bf 10}^\ord{-3} \oplus ({\bf 3}\otimes  \overline{\bf  5})^\ord{-1} \oplus (\overline{\bf 3}\otimes {\bf 5})^\ord{1} \oplus \overline{\bf 10}^\ord{3} \oplus {\bf 3}^\ord{5}\ , 
\ee
as
\be {\bf E} = \begin{pmatrix}
\ 0 \ & A^{A}_{K L} 	& \ \ve^{ACE} B_{E}^{K} \				& \ C_{K} \delta^{A}_{C} \				& \ 0 \								& \ 0 \ \\
\ 0 \ & 0 			& \ \frac{1}{2} \ve^{IJKPQ} A_{PQ}^{C} \	& \ \sqrt{2} B_{C}^{[I} \delta^{J]}_{K} \		& \ \frac{1}{2} \ve^{IJKLP} C_{P}	 \		& \ 0 \ \\
\ 0 \ & 0			& \ 0	\								& \ \sqrt{2} \ve_{A C E} A^{E}_{I K} \	& \ \sqrt{2} \delta_{I}^{[K} B_{A}^{L]}  \ 		& \ \delta_{A}^{C} C_{I} \ \\
\ 0 \ & 0			& \ 0	\								& \ 0	\							& \ \frac{1}{2} \ve^{IKLPQ} A_{PQ}^{A} \	& \ \ve^{A C E} B_{E}^{I} \ \\
\ 0 \ & 0			& \ 0	\								& \ 0	\							& \ 0 	\								& \ A_{IJ}^{C} \ \\
\ 0 \ & 0			& \ 0	\								& \ 0 \							& \ 0 \								& \ 0 \
\end{pmatrix} \ , \ee
and its semi-simple component 
\be \cU = \left(  \begin{array}{cccccc} \ e^{5\phi} \upsilon^a{}_C  \ & \ 0 \ & \ 0 \ & \ 0 \ & \ 0\ & \ 0 \\
 \ 0 \ & \ e^{3\phi} V^{\inv}{}_{[K}{}^{[i}  V^{\inv}{}_{L]}{}^{j]}  \ & \ 0 \ & \ 0 \ & \ 0\ & \ 0 \\
  \ 0 \ & \ 0 \ & \ e^\phi \upsilon^{\inv}{}^C{}_a V_i{}^K  \ & \ 0 \ & \ 0\ & \ 0 \\
   \ 0 \ & \ 0 \ & \ 0 \ & \ e^{-\phi} \upsilon{}^a{}_C  V^{\inv}{}_{K}{}^{i}  \ & \ 0\ & \ 0 \\
    \ 0 \ & \ 0 \ & \ 0 \ & \ 0 \ & \ e^{-3\phi} V_{[i}{}^{[K}  V_{j]}{}^{L]}\ & \ 0 \\
     \ 0 \ & \ 0 \ & \ 0 \ & \ 0 \ & \ 0\ & \ e^{-5\phi} \upsilon^{\inv}{}^C{}_a  \end{array}\right)\, ,\ee
such that the coset representative is 
\be \cV = \cU \exp( {\bf E}) \ . \ee
We use the notation that the $SL(5)$ indices $K,L$ and the $SL(3)$ index $C$ are contracted on the right-hand side through the left action of $E_{7(7)}$, whereas $I,J$ and $A$ are contracted on the left-hand-side through the right action of $E_{7(7)}$. The same convention is used for the $SO(3)$ indices  $a, b$ contracted on the left and $c, d$ on the right, and for the $SO(5)$ indices $i , j$ contracted on the left and $k, l$ on the right, through the respective right and left actions of $SU(8)$. We apologise to the reader for using now on $i,j$ as vector indices of $SO(5)$, whereas we were using them as $Sp(2)$ indices in the preceding discussion. Here $\upsilon^a{}_C$ and $V_i{}^K$ are respectively $SL(3)$ and $SL(5)$ coset representatives for the scalar fields parametrising respectively $SL(3)/SO(3)$ and $SL(5)/SO(5)$. 

The vielbeins and the spin-connexion on $E_{7(7)}/\SU$ are defined respectively from the projections of the Maurer--Cartan form to the ${\bf 70}$ and the $\mathfrak{su}(8)$ representations as
\be
 d \cV \cV^{-1} = d \cU \cU^{-1} + \cU d \exp({\bf E})  \exp({-\bf E}) \cU^{-1}  =  P + B
\ee
and the metric on $E_{7(7)}/\SU$  reads 
\bea ds^2 = \frac{1}{6} \tr\,  P P &=& 60 d\phi^2 + 2 P_{ab} P^{ab} + 2 P_{ij} P^{ij} + e^{4\phi} M^{IK} M^{JL} \mu_{AB} \nabla{A}_{IJ}^A \nabla{A}_{KL}^B \CR
&&+ e^{8\phi} M^{\inv}_{IJ} \mu^{\inv AB} \nabla{B}_A^I  \nabla{B}_B^J + e^{12\phi} M^{IJ}  \nabla{C}_I  \nabla{C}_J \ , 
\eea
where the matrices $M^{I K} = V_{i}{}^{I} V^{i K}$ and $\mu_{A B} = \upsilon^{a}{}_{A} \upsilon_{a B}$ are symmetric by construction, and $\nabla A,\,  \nabla B,\, \nabla C$ are defined in \eqref{Nilbeins}. The derivatives dual to these differentials satisfying to
\be\begin{split}
\cD_{A}^{I J}{}^{\upmu}  \nabla_{\upmu} A^{B}_{K L} &= \delta_{A}^{B} \delta_{K L}^{I J}\ , \\  \cD^{A}_{I}{}^{\upmu}\nabla_{\upmu} A^{B}_{K L}  &= 0 \ , \\ \cD^{I}{}^{\upmu}   \nabla_{\upmu} A^{B}_{K L}  &= 0 \ , 
\end{split}\qquad\begin{split}
\cD_{A}^{I J}{}^{\upmu}  \nabla_{\upmu} B^{J}_{B}  &= 0 \ ,  \\ \cD^{A}_{I}{}^{\upmu} \nabla_{\upmu} B^{J}_{B}  &= \delta_{B}^{A} \delta_{I}^{J} \ , \\ \cD^{I}{}^{\upmu}\nabla_{\upmu} B^{J}_{B}  &= 0 \ , 
\end{split}\qquad\begin{split}
\cD_{A}^{I J}{}^{\upmu}   \nabla_{\upmu} C_{J} &= 0 \ , \\ \cD^{A}_{I}{}^{\upmu} \nabla_{\upmu} \nabla_{\upmu} C_{J}  &= 0 \ , \\ \cD^{I}{}^{\upmu} \nabla_{\upmu} C_{J}  &= \delta^{I}_{J} \ , 
\end{split}\label{Orthobase}\ee
are defined as
\bea
\cD_{A}^{K L} &=& \frac{\partial}{\partial A^{A}_{K L}} - \frac{1}{6 \sqrt{2}} \ve^{K L M N P} \ve_{A B C} A^{B}_{M N} A^{C}_{P I} \frac{\partial}{\partial C_{I}} + \frac{1}{4} \ve_{A B C} \ve^{K L M N P}  A^{B}_{M N} \frac{\partial}{\partial B^{P}_{C}} \ , \CR
\cD^{A}_{I} &=& \frac{\partial}{\partial B^{I}_{A}} - \frac{1}{\sqrt{2}} A^{A}_{I J} \frac{\partial}{\partial C_{J}} \ , \qquad \cD^{I} = \frac{\partial}{\partial C_{I}} \ . 
\eea

We are interested in finding functions of the scalar fields defining invariants in four dimensions that lift to seven dimensions. Therefore we will consider functions that depend only on the seven-dimensional scalar fields $t$ parametrising $SL(5) / SO(5) $ and the Kaluza--Klein dilaton $\phi$ that must appear at a specific power determined from
\be \int d^4 x \sqrt{-g} e^{-6(3+n)\phi} \cE(t) \nabla^{2n} R^4 \sim \int d^7 x \sqrt{-g} \cE(t) \nabla^{2n} R^4  \ , \label{Roxi} \ee
for the invariant to be diffeomorphism invariant in seven dimensions. With this restricted ansatz for the function, the differential operator ${\bf D}_{56} = {\bf E}^{\mu} (\partial_{\mu}  - B_{\mu})$ is block diagonal in the decomposition \eqref{56Graded}, \ie 
\begin{multline} {\bf D}_{56} = \text{diag} \biggl( 
\frac{1}{12} \delta^{a}_{c} \partial_{\phi}, 
\hspace{2mm} \frac{1}{20} \delta^{i j}_{k l} \partial_{\phi} - 2 \delta^{[i}_{[k} \cD^{j]}{}_{l]} , 
\hspace{2mm} \frac{1}{60} \delta^{c}_{a} \delta^{k}_{i} \partial_{\phi}  + \delta_{a}^{c} \cD_{i}{}^{k}, \\
\hspace{2mm} - \frac{1}{60} \delta^{a}_{c} \delta^{i}_{k} \partial_{\phi} - \delta_{c}^{a} \cD_{k}{}^{i},
\hspace{2mm} - \frac{1}{20} \delta^{k l}_{i j} \partial_{\phi} + 2 \delta^{[k}_{[i} \cD^{l]}{}_{j]} ,
\hspace{2mm} - \frac{1}{12} \delta^{c}_{a} \partial_{\phi} \biggr) \ . 
\end{multline}
Now we want to compute the action of the second order derivative  ${\bf D}^2$ 
\be
 {\bf D}_{56}^2 \cE = {\bf D}_{56}^{\upmu} \partial_{\upmu}  \scal{ {\bf D}^{\upnu}_{56} \partial_{\upnu}  \cE} - {\bf D}^{\upmu} [{\bf B}_{\upmu}, {\bf D}^{\upnu}_{56}] \partial_{\upnu}\label{D2Formula} \cE 
\ee
on a function of $\phi,t$ defined on $GL(5)/SO(5)$. Note that the spin-connexion decomposes into 
\be
 {\bf B} ={\bf B}_{\mathfrak{so}(3)\oplus \mathfrak{so}(5)} +  {\bf M}^{ij}_{a} e^{2\phi} V_{i}{}^{I} V_{j}{}^{J}  \upsilon^a{}_A \nabla A^{A}_{I J}  + {\bf M}^{a}_{i} e^{4\phi} V^\inv_I{}^i \upsilon^{\inv A}{}_a \nabla B^I_A + {\bf M}^{i} e^{6\phi} V_i{}^I  \nabla C_I \ee
where the matrices ${\bf M}^{ij}_{a},\,{\bf M}^{a}_{i} ,\,  {\bf M}^{i}  $ are constant tensors and ${\bf B}_{\mathfrak{so}(3)\oplus \mathfrak{so}(5)} $ is the spin-connexion on $SL(3)/SO(3)\times SL(5)/SO(5)$, such that its contribution in \eqref{D2Formula} simply replaces all the partial derivatives on $SL(3)/SO(3)\times SL(5)/SO(5)$ by covariant derivatives. Moreover, using \eqref{Orthobase} one obtains that
\bea {\bf D}^{\upmu} \otimes {\bf B}_\upmu &=&  {\bf D}^{\upmu}_{\mathfrak{so}(3)\oplus \mathfrak{so}(5)} \otimes {\bf B}_{\upmu\, \mathfrak{so}(3)\oplus \mathfrak{so}(5)} \CR
&& + \scal{ {\bf \tilde{M}}_{kl}^{c} e^{-2\phi} V^\inv_K{}^k V^\inv_J{}^L \upsilon^{\inv C}{}_c \cD_C^{KL}  + {\bf \tilde{M}}_{c}^{k}  e^{-4\phi} V_k{}^K \upsilon^c{}_C \cD^C_K + {\bf \tilde{M}}_{k} e^{-6\phi} V^\inv_K{}^k \cD^K } \CR
&& \hspace{15mm} \otimes \scal{{\bf M}^{ij}_{a} e^{2\phi} V_{i}{}^{I} V_{j}{}^{J}  \upsilon^a{}_A \nabla A^{A}_{I J}  + {\bf M}^{a}_{i} e^{4\phi} V^\inv_I{}^i \upsilon^{\inv A}{}_a \nabla B^I_A + {\bf M}^{i} e^{6\phi} V_i{}^I  \nabla C_I }\CR
&=& {\bf D}^{\upmu}_{\mathfrak{so}(3)\oplus \mathfrak{so}(5)} \otimes {\bf B}_{\upmu\, \mathfrak{so}(3)\oplus \mathfrak{so}(5)} + {\bf \tilde{M}}_{ij}^{a} \otimes {\bf M}^{ij}_{a} + {\bf \tilde{M}}_{a}^{i} \otimes {\bf M}^{a}_{i}  + {\bf \tilde{M}}_i \otimes {\bf M}^i \ , 
\eea 
 where the matrices ${\bf \tilde{M}}$ are also constant tensors. Defining $\cD_\upmu^\ord{0}$, the covariant derivative with respect to the grad zero  $\mathfrak{so}(3)\oplus \mathfrak{so}(5)$ spin-connexion, one obtains therefore that \eqref{D2Formula} simplifies to 
\be
{\bf D}^2_{56} \cE = {\bf D}^{\upmu}_{56} \cD^\ord{0}_{\upmu} \scal{ {\bf D}^{\upnu}_{56} \cD^\ord{0}_{\upnu} \cE}  -\scal{  {\bf \tilde{M}}_{ij}^{a}  [{\bf M}^{ij}_{a}, {\bf D}^{\upmu}_{56}]  + {\bf \tilde{M}}_{a}^{i}  [{\bf M}^{i}_{a}, {\bf D}^{\upmu}_{56}]  + {\bf \tilde{M}}^{i}  [{\bf M}_{i}, {\bf D}^{\upmu}_{56}] }\cD^\ord{0}_{\upmu} \cE  \ . 
\ee
On a function of $\phi,\, t$ on $GL(5)/SO(5)$, one computes in this way that ${\bf D}^2_{56}$ reduces to 
\begin{multline} {\bf D}^2_{56} = \text{diag} \biggl( 
 \delta^{a}_{c}\Scal{ \frac{1}{12^2} \partial^2_{\phi} + \frac{3}{8} \partial_\phi} , \\
\hspace{2mm}\delta^{i j}_{k l}\Scal{ \frac{1}{20^2} \partial_\phi^{\; 2} + \frac{11}{40} \partial_\phi }- \scal{ \sfrac{7}{2}  + \sfrac{1}{5} \partial_\phi}  \delta^{[i}_{[k} \cD^{j]}{}_{l]} \partial_{\phi} + 2 \delta^{[i}_{[k} \cD^{j] m} \cD_{l] m}+ 2 \cD^{[i}{}_{[k} \cD^{j]}{}_{l]} , \\
\hspace{2mm} \delta^{c}_{a} \Scal{  \delta^{k}_{i} \Scal{ \frac{1}{60^2} \partial_\phi^{\; 2} + \frac{29}{120} \partial_\phi } +\cD^{k}{}_{j} \cD^{j}{}_{i} + \scal{ \sfrac{3}{4} + \sfrac{1}{30} \partial_\phi}  \cD^{k}{}_{i}}, \\
\hspace{2mm}  \delta^{a}_{c} \Scal{  \delta^{i}_{k}\Scal{ \frac{1}{60^2} \partial_\phi^{\; 2} + \frac{29}{120} \partial_\phi }+  \cD^{i}{}_{j} \cD^{j}{}_{k} + \scal{ \sfrac{3}{4} + \sfrac{1}{30} \partial_\phi}  \cD^{i}{}_{k}},\\
\hspace{2mm}  \delta^{k l}_{i j}\Scal{ \frac{1}{20^2} \partial_\phi^{\; 2} + \frac{11}{40} \partial_\phi }- \scal{ \sfrac{7}{2}  + \sfrac{1}{5} \partial_\phi}  \delta^{[k}{}_{[i} \cD^{l]}{}_{j]} + 2 \delta^{[k}_{[i} \cD^{l]}{}_{m} \cD^{m}{}_{j]} ,\\
\hspace{2mm}  \delta^{c}_{a}\Scal{ \frac{1}{12^2} \partial^2_{\phi} + \frac{3}{8} \partial_\phi} \biggr) \ . 
\end{multline}
We can use this expression to solve the differential equation \eqref{R44DEq} for a function $\cE_\grad844 = e^{a\phi} \cE(t)$ on $GL(5)/SO(5)$. These equations give
\bea \label{R47D}
\Scal{ \frac{a^2}{12^2}  + \frac{3a}{8}} \cE(t) &=&  - \frac{9}{2}  \cE(t)  \ , \CR
 \Scal{ 2 \delta^{[i}_{[k} \cD^{j] m} \cD_{l] m}+ 2 \cD^{[i}{}_{[k} \cD^{j]}{}_{l]} } \cE(t) &=&  \Scal{ \frac{a}{5} + \frac{7}{2} }  \delta^{[i}_{[k} \cD^{j]}{}_{l]}  \cE(t) -\Scal{ \frac{a^2}{20^2}  + \frac{11\, a}{40} + \frac{9}{2}}  \delta^{i j}_{k l}    \cE(t) \ ,\CR
\cD^{k}{}_{j} \cD^{j}{}_{i} \cE(t) &=&  - \Scal{ \frac{a}{30}   + \frac{3}{4}}  \cD^{k}{}_{i} \cE(t) - \Scal{  \frac{a^2}{60^2}  +\frac{29\, a}{120}+ \frac{9}{2}}   \delta^{k}_{i} \, \cE(t) \ . \quad 
\eea
The first equation implies $a=-18$ or $a=-36$, but we are going to see that the second solution does not have a solution on $GL(5)/SO(5)$. The second equation implies that the second derivative of $\cE(t)$ vanishes in the irreducible representation $[2,0]$ of $Sp(2)$, \ie 
\be \cD_{[i}{}^{[k} \cD_{j]}{}^{l]} \cE(t) = - \frac{2}{3} \delta_{[i}^{[k} \cD_{j]}{}^p \cD_p{}^{l]} \cE(t) + \frac{1}{12} \delta_{ij}^{kl} \cD_{pq} \cD^{pq}  \cE(t) \ . \label{MinimalSL5} \ee
Using the commutation relation 
\be [ \cD_{ij}, \cD^{kl} ] \cD^{pq} = \frac{1}{2} \Scal{ \delta_{(i}^{(k} \delta^{l)(p} \cD_{j)}{}^{q)} - \delta_{(i}^{k)} \delta_{j)}^{(p} \cD^{q)(l}} \ , \ee
one computes that 
\be \cD_{i}{}^p \cD_p{}^{[k} \cD_{j}{}^{l]} = -\cD_{j}{}^{[k} \cD_{i}{}^p \cD_p{}^{l]} + \frac{1}{4} \delta_{i}^{[k} \cD_{j}{}^{l]}  - \frac{1}{16} \delta_{j}^{[k} \cD_{i}{}^{l]}\label{CubCon} \ . \ee 
Substituting \eqref{MinimalSL5} in \eqref{CubCon} and taking the trace over $i$ and $k$ one obtains 
\be \cD_i{}^k \cD_k{}^l \cD_l{}^j \cE(t) = \cD_i{}^j \Scal{ \frac{13}{20} \cD_{pq} \cD^{pq} + \frac{9}{16} }\cE(t) + \frac{1}{5} \delta_i^j \cD_k{}^l \cD_l{}^p \cD_p{}^k \cE(t) \ . \ee
Using the third differential equation in \eqref{R47D} in this equation one obtains 
\be(  5832 + 342 a + a^2 ) \cD_i{}^j \cE(t)= 0 \ , \ee
and therefore $a=-18$ only is possible, as required for a $R^4$ type invariant \eqref{Roxi}. Therefore we obtain 
\bea  \cD^{k}{}_{j} \cD^{j}{}_{i} \cE(t) &=&  -\frac{3}{20}  \cD^{k}{}_{i} \cE(t) -\frac{6}{25}  \delta^{k}_{i} \, \cE(t)  \ ,\CR
\cD_{[i}{}^{[k} \cD_{j]}{}^{l]} \, \cE(t) &=& \frac{1}{10}   \delta^{[i}_{[k} \cD^{j]}{}_{l]}  \cE(t) + \frac{3}{50} \delta^{i j}_{k l}    \cE(t)\ . \quad 
\eea
This function is one example of the generic class of functions for which the second order derivative restricted to the $[2,0]$ vanishes, and we write them 
\bea \cD_i{}^k \cD_k{}^j \, \cE_{\scriptscriptstyle [ s,0,0,0]}  &=&- \frac{3(4s-5)}{20} \cD_i{}^j  \,  \cE_{\scriptscriptstyle [ s,0,0,0]}  + \frac{2s(2s-5)}{25}  \delta_i^j \,  \cE_{\scriptscriptstyle [ s,0,0,0]} \ ,  \\ 
\cD_{[i}{}^{[k} \cD_{j]}{}^{l]} \, \cE_{\scriptscriptstyle [ s,0,0,0]} &=&  \frac{4s-5}{10}  \delta_{[i}^{[k} \cD_{j]}{}^{l]} \, \cE_{\scriptscriptstyle [ s,0,0,0]} - \frac{s(2s-5)}{50} \delta_{ij}^{kl} \, \cE_{\scriptscriptstyle [ s,0,0,0]} \label{1000}\ ,  \eea
where the notation refers to the property that the Eisenstein series $E_{\scriptscriptstyle [ s,0,0,0]}$ satisfies to these equations whenever it converges. 

The result is consistent with the conjectured exact low energy effective action in type II string theory. We just note here that the general solution depending on $\mathds{R}_+^* \times SL(5)/SO(5) \times SL(3)/SO(3)$ is such that one should have the expansion of the Eisenstein series at large volume modulus $V(T^3)= e^{-6\phi}$, 
\be E_{\mbox{\DEVII{\frac{3}{2}}000000}}= e^{-18\phi} E_{\scriptscriptstyle [ \frac{3}{2},0,0,0]}(t)  - \frac{4}{\pi}  \, e^{-20\phi} E_{\scriptscriptstyle [2,0]}(g_{T^3})  + \mathcal{O}(e^{-e^{-2\phi}})  \ . \ee

We will now analysis the differential equations \eqref{D4R44DM} relevant for the $\nabla^4 R^4$ type invariant. For this, we need in particular to compute the third order differential operator  ${\bf D}^3_{56}$
\be
{\bf D}^3_{56} \cE = {\bf D}^{\upmu}_{56} \cD^\ord{0}_{\upmu} \scal{ {\bf D}^{2}_{56}  \cE}  - {\bf \tilde{M}}_{ij}^{a}  [{\bf M}^{ij}_{a}, {\bf D}^{2}_{56}  \cE]  - {\bf \tilde{M}}_{a}^{i}  [{\bf M}^{i}_{a},  {\bf D}^{2}_{56}  \cE]  - {\bf \tilde{M}}^{i}  [{\bf M}_{i},  {\bf D}^{2}_{56}  \cE]   \ . \label{SimplifyCom}
\ee
One computes that on a function of $\phi$ and $t$ on $GL(5)/SO(5)$, it reduces to
\begin{multline} {\bf D}^3_{56} = \text{diag}\Biggl( \Scal{  \frac{1}{12^{3}} \partial_{\phi}^{3} + \frac{97}{1440} \partial^{2}_{\phi} + \frac{3}{4} \partial_{\phi} -  \frac{3}{2} \cD_{ps} \cD^{ps} }\delta^{a}_{c} , \\
  \delta^{i j}_{k l} \Scal{  \frac{1}{20^{3}} \partial_{\phi}^{3} + \frac{61}{2400}  \partial^{2}_{\phi} + \frac{1}{5} \partial_{\phi} -   \cD^{sp} \cD_{sp}}  + \Scal{ 11 + \frac{3}{10} \partial_{\phi} } \delta^{[i}_{[k} \cD^{j]}{}_{s} \cD^{s}{}_{l]} +\Scal{ \frac{21}{2} + \frac{3}{10} \partial_{\phi} } \cD^{[i}{}_{[k} \cD^{j]}{}_{l]}  \\
 \quad  + \Scal{ - \frac{3}{200} \partial_{\phi}^{2} - \frac{63}{40} \partial_{\phi} - \frac{119}{8} } \delta^{[i}_{[k} \cD^{j]}{}_{l]} -  2 \delta^{[i}_{[k} \cD^{j]}{}_{s} \cD^{s}{}_{p} \cD^{p}{}_{l] }  - 2 \cD^{[i}{}_{[k} \cD^{j]}{}_{p} \cD^{p}{}_{l]}  - 4 \cD^{[i}{}_{p} \cD^{p}{}_{[k} \cD^{j]}{}_{l]} , \\
 \delta_{a}^{c}  \biggl( \delta_{i}^{k}\Scal{  \frac{1}{60^{3}} \partial_{\phi}^{3} + \frac{49}{7200} \partial^{2}_{\phi}  + \frac{1}{20} \partial_{\phi} - \frac{1}{4} \cD^{p s} \cD_{p s} } + \Scal{ \frac{1}{20} \partial_{\phi} + 2 }  \cD_{i}{}^{p} \cD_{p}{}^{k} \biggr . \\
 \quad  \biggl . +\Scal{  \frac{3}{60^2} \partial_{\phi}^{2} + \frac{37}{80} \partial_{\phi} + \frac{63}{16}} \cD_{i}{}^{k} + \cD_{i}{}^{s} \cD_{s}{}^{p}  \cD_{p}{}^{k}\biggr)  , \dots \Biggr) 
\end{multline}
where the dots stand for the conjugate representations that are identical to the ones written explicitly up the sign. Using the same ansatz $\cE_\grad822 = e^{a\phi} \cE(t)$, one obtains combining these equations that 
\bea &&  6\cD^{[i}{}_{[k} \cD^{j]}{}_{p} \cD^{p}{}_{l]}  \cE(t) - \frac{3(a+35)}{10} \cD^{[i}{}_{[k} \cD^{j]}{}_{l]}\cE(t) \CR
&=& \delta^{[i}_{[k} \biggl(  \Scal{ 15 + \frac{2a}{5}} \cD^{j]}{}_p \cD^p{}_{l]}  - \Scal{ \frac{33}{4} + \frac{13\, a}{20} + \frac{a^2}{75}} \cD^{j]}{}_{l]} -\Scal{  \frac{9a}{20} + \frac{17 a^2}{600} + \frac{a^3}{2250} }  \delta_{l]}^{j]} \biggr) \cE(t) \ , \qquad  \eea
which implies that the tensor structure of the first term must necessarily reduce. Considering the general solution of such a system depending on four variables associated to a maximal abelian subgroup of $SL(5)$, we find that there is a two-parameter family of equations with this structure, such that 
\bea  &&  6\cD^{[i}{}_{[k} \cD^{j]}{}_{p} \cD^{p}{}_{l]}  \cE_{\scriptscriptstyle [ s,s^\prime,0,0]}  + \frac{3(2s+4s^\prime-5)}{10} \cD^{[i}{}_{[k} \cD^{j]}{}_{l]}\cE_{\scriptscriptstyle [ s,s^\prime,0,0]}  \CR
&=& \delta^{[i}_{[k} \biggl(  \frac{2s+4s^\prime-5}{5} \cD^{j]}{}_p \cD^p{}_{l]}  + \Scal{ 6 \Scal{ \frac{3(2s+4s^\prime-5)}{20}}^2  + \frac{(s+2)(s-3)}{2}} \cD^{j]}{}_{l]} \biggr . \CR
&& \hspace{40mm} \biggl .  - \frac{2s+4s^\prime-5}{80} \Scal{\frac{9(2s+4s^\prime-5)^2}{25}  + 4s^2 -4s-9}  \delta_{l]}^{j]} \biggr)  \cE_{\scriptscriptstyle [ s,s^\prime,0,0]}  \ , \CR
&&\biggl( \cD^i{}_k \cD^k{}_l \cD^l{}_j  + \frac{2s+4s^\prime-5}{5} \cD^i{}_k \cD^k{}_j -\Scal{ \frac{3( 2s+4s^\prime-5)^2}{400} + \frac{2s^2-2s-3}{8}} \cD^i{}_j\biggr)  \,   \cE_{\scriptscriptstyle [ s,s^\prime,0,0]}  \CR
&=&  \frac{2s+4s^\prime-5}{160} \Scal{ \frac{9( 2s+4s^\prime-5)^2}{25} - 4 s^2 + 4 s-9} \delta^i_j\,   \cE_{\scriptscriptstyle [ s,s^\prime,0,0]} \ . \label{st00}
\eea
Note that $\cE_{\scriptscriptstyle [ s,s^\prime,0,0]}$ and $\cE_{\scriptscriptstyle [ 1-s,s^\prime+s-1/2,0,0]}$ satisfy to the same equations, so we will not consider them as independent solutions, unless $s^\prime$ or $s$ vanishes. Indeed, if  $s^\prime=0$ the function satisfies to the stronger equations \eqref{1000}, whereas for $s=0$ it is proportional to the function satisfying to
\be \cD_i{}^k \cD_k{}^j \,  \cE_{\scriptscriptstyle [ 0,0,s,0]}  =   \frac{4s-5}{20} \cD_i{}^j  \,  \cE_{\scriptscriptstyle [ 0,0,s,0]}  + \frac{3s(2s-5)}{25}  \delta_i^j \,  \cE_{\scriptscriptstyle [ 0,0,s,0]} \ , \label{0010} \ee
for $s=\frac{5}{2}-s^\prime$. Again the notation we use refers to the property that the corresponding Eisenstein series $E_{\scriptscriptstyle [ s,s^\prime,0,0]} $ and $E_{\scriptscriptstyle [ 0,0,s,0]}$ satisfy to the same equations when they converge. This way we find only three independent solutions, \ie 
\bea
{\bf D}^3_{56} e^{-30\phi} \cE_{\scriptscriptstyle [ \frac{5}{2},0,0,0]} &=& - 9 {\bf D}_{56}e^{-30\phi} \cE_{\scriptscriptstyle [ \frac{5}{2},0,0,0]} \ , \CR
{\bf D}^3_{56} e^{-30\phi} \cE_{\scriptscriptstyle [0,0, \frac{5}{2},0]} &=& - 9 {\bf D}_{56}e^{-30\phi} \cE_{\scriptscriptstyle [0,0, \frac{5}{2},0]} \ , \CR
{\bf D}^3_{56} e^{-36\phi} \cE_{\scriptscriptstyle [4,-\frac{1}{2},0,0]} &=& - 9 {\bf D}_{56} e^{-36\phi} \cE_{\scriptscriptstyle [4,-\frac{1}{2},0,0]} \ . 
\eea
We already see that the two first solutions correspond to the seven-dimensional $\nabla^4 R^4$ type invariant, whereas the second would correspond to the $\nabla^6 R^4$ invariant. The first equation in \eqref{D4R44DM} is indeed also satisfied for the $\nabla^6 R^4$ type invariant that descends from ten dimensions, and the type IIB $3$-loop invariant in ten dimensions indeed defines a function solving \eqref{st00} for $s=4,\, s^\prime=-\frac{1}{2}$. 

Now we want to check the second equation in \eqref{D4R44DM}. However the computation of the commutator terms of the ${\bf M}$ matrices \eqref{SimplifyCom} becomes rather tedious in the adjoint representation, and we will only fix the coefficients on the general covariant ansatz by demanding that the knows solutions indeed satisfy the equations, \ie 
\bea\label{133D4R47} 
{\bf D}^3_{133} e^{- 18 \phi} \cE_{[\frac{3}{2} 0 0 0]} &=& - 14 {\bf D}_{133}e^{- 18 \phi} \cE_{[\frac{3}{2} 0 0 0]}  \ , \CR
 {\bf D}^3_{133} e^{- 30 \phi} \cE_{[\frac{5}{2} 0 0 0]} &=& - 20 {\bf D}_{133} e^{- 30 \phi} \cE_{[\frac{5}{2} 0 0 0]} \ , \CR 
 {\bf D}^3_{133} e^{- 30 \phi} \cE_{[ 0 0 \frac{5}{2} 0]} &=& - 20 {\bf D}_{133} e^{- 30 \phi} \cE_{[0 0 \frac{5}{2} 0]} \ . 
\eea
It appears that the system of equations for the coefficients is over-constrained, and it is a non-trivial check that one can indeed find a solution. The positive grad component \eqref{133Graded} of the differential operator ${\bf D}_{133}$ restricted to a function on $GL(5)/SO(5)$ is 
 \be  {\bf D}_{133} = \text{diag} \biggl( \cD_{i}{}^{k} + \frac{1}{10} \delta_{i}^{k} \partial_{\phi} ,\delta_{a}^{c}\Scal{  -  \cD_{k}{}^{i} + \frac{1}{15}  \delta_{k}^{i} \partial_{\phi}} , 
\delta_{c}^{a}  \Scal{  2 \delta_{[k}^{[i} \cD^{j]}{}_{l]} + \frac{1}{30} \delta_{kl}^{ij} \partial_{\phi} }, \dots
 \biggr) \ , 
\ee
and we obtain after calibrating the coefficients such that \eqref{133D4R47} are all satisfied that 
 \begin{multline} {\bf D}^3_{133} = \text{diag} \Biggl( \cD_{i}{}^{p} \cD_{p}{}^{q} \cD_{q}{}^{k} + \biggl( \frac{19}{2} + \frac{3}{10} \partial_{\phi} \biggr) \cD_{i}{}^{p} \cD_{p}{}^{k}  
+\biggl( \frac{3}{100} \partial^2_{\phi} +\frac{81}{40} \partial_{\phi} + \frac{217}{16} \biggr) \cD_{i}{}^{k} \\
+ \delta_{i}^{k}  \biggl( \frac{1}{10^3} \partial^3_{\phi} +\frac{31}{300} \partial^2_{\phi} + \frac{1}{5} \partial_{\phi} - \frac{7}{4} \cD_{p q} \cD^{p q} \biggr) , \\
 \delta_{a}^{c}\biggl(   - \cD_{k}{}^{p} \cD_{p}{}^{q} \cD_{q}{}^{i} + \biggl( \frac{11}{2} + \frac{1}{5} \partial_{\phi} \biggr) \cD_{k}{}^{p} \cD_{p}{}^{i}  
+\biggl(-  \frac{1}{75} \partial^2_{\phi} - \frac{23}{20} \partial_{\phi} - \frac{37}{16} \biggr) \cD_{k}{}^{i}\biggr .  \\
\biggl . +  \delta_{k}^{i}  \biggl(\frac{1}{15^3} \partial^3_{\phi} + \frac{47}{900} \partial^2_{\phi} - \frac{1}{30} \partial_{\phi} - \frac{5}{4} \cD_{p q} \cD^{p q} \biggr) \biggr), \\
\delta_{c}^{a}\biggl( 2  \delta^{[i}_{[k} \cD^{j]}{}_{s} \cD^{s}{}_{p} \cD^{p}{}_{l] }  + 2  \cD^{[i}{}_{[k} \cD^{j]}{}_{p} \cD^{p}{}_{l]} +4  \cD^{[i}{}_{p} \cD^{p}{}_{[k} \cD^{j]}{}_{l]} +  \biggl( \frac{1}{5} \partial_{\phi} + 4\biggr)  \delta^{[i}_{[k} \cD^{j]}{}_{s} \cD^{s}{}_{l]}\biggr .    \\
 +
\biggl( \frac{1}{5} \partial_{\phi} +\frac{9}{2}\biggr) \cD^{[i}{}_{[k} \cD^{j]}{}_{l]} + \biggl(\frac{1}{150} \partial^{2}_{\phi} + \frac{29}{20} \partial_{\phi} -\frac{27}{8} \biggr)\delta^{[i}_{[k} \cD^{j]}{}_{l]} \\
\biggl  . + \biggl( \frac{1}{30^3} \partial^3_{\phi}+\frac{19}{30^2} \partial^2_{\phi}-\frac{1}{15} \partial_{\phi} - \frac{1}{2} \cD^{pq} \cD_{pq}\biggr)  \delta^{ij}_{kl}\biggr) , \dots \Biggr) \ . 
\end{multline}
The dots stand for the zero and negative grad components. In the same way, one finds that the component in the adjoint of $SL(5)$ of ${\bf D}^3_{133} -\lambda {\bf D}_{133}  $ admits the two components 
\bea&&  \cD^{{\rm \scriptscriptstyle Adj} \; 3}_{(i j)}{}^{[k l]} +   \delta_{(i}{}^{[k} \cD_{j)}{}^{p} \cD_{p}{}^{l]}  +2 \Scal{  \frac{3}{5} \partial_\phi  - \frac{41}{16} - \lambda} \delta_{(i}{}^{[k} \cD_{j)}{}^{l]}  \ , \CR
&& \cD^{{\rm \scriptscriptstyle Adj} \; 3}_{[i j]}{}^{(k l)}+   \delta_{[i}{}^{(k} \cD_{j]}{}^{p} \cD_{p}{}^{l)}  + 2\Scal{  \frac{3}{5} \partial_\phi  - \frac{13}{16} - \lambda} \delta_{[i}{}^{(k} \cD_{j]}{}^{l)} \ ,  \eea
using the equations 
\bea
 \cD^{{\rm \scriptscriptstyle Adj} \; 3}_{(i j)}{}^{[k l]} \cE_{[s,0,0,0]}  &=& \biggl(s(2s-5) +\frac{15}{8} \biggr)  \delta_{(i}{}^{[k} \cD_{j)}{}^{l]} \cE_{[s,0,0,0]}\ ,  \CR
 \cD^{{\rm \scriptscriptstyle Adj} \; 3}_{[i j]}{}^{(k l)} \cE_{[s,0,0,0]}  &=& \biggl(  s(2s-5) -\frac{13}{8} \biggr)  \delta_{[i}{}^{(k} \cD_{j]}{}^{l)} \cE_{[s,0,0,0]} \ , \CR
 \cD^{{\rm \scriptscriptstyle Adj} \; 3}_{(i j)}{}^{[k l]} \cE_{[0,0,s,0]} &=& \biggl( s(2s-5) + \frac{7}{8} \biggr)  \delta_{(i}{}^{[k} \cD_{j)}{}^{l]} \cE_{[0,0,s,0]} \ , \CR
 \cD^{{\rm \scriptscriptstyle Adj} \; 3}_{[i j]}{}^{(k l)} \cE_{[0,0,s,0]} &=& \biggl(s(2s-5) - \frac{21}{8} \biggr) \delta_{[i}{}^{(k} \cD_{j]}{}^{l)} \cE_{[0,0,s,0]} \ . 
\eea 
Using these equations, one finds then indeed that $ e^{-36\phi} \cE_{\scriptscriptstyle [4,\mbox{-}\frac{1}{2},0,0]}$ is not a solution, and  we have therefore the two unique solutions corresponding to the $\nabla^4 R^4$ type invariant
\be \cE_\grad822  = e^{- 30 \phi} \cE_{[\frac{5}{2} 0 0 0]}+ e^{- 30 \phi} \cE_{[ 0 0 \frac{5}{2} 0]}\  . \ee
\addtocontents{toc}{\protect\setcounter{tocdepth}{1}}
\subsubsection{$\nabla^4 R^4$ threshold function in seven dimensions}
\addtocontents{toc}{\protect\setcounter{tocdepth}{2}}
Consistently with the analysis in \cite{Minimal}, we find that there are only two classes of $\nabla^4 R^4$ type invariants in seven dimensions. The first class is associated to the linearised invariants discussed above, with the gradient expansion \eqref{D4R47DGrad}, with the function $\cE_\gra41 = \cE_{{[0,0,\frac{5}{2},0]}} $, that admits the correct gradient expansion as a consequence of \eqref{0010}. This equation indeed implies that the  order $n$ derivative is only non-vanishing in the representations $[4p,2q]$ for $2p+q\le n$, and is related to lower order derivatives when $2p+q< n$. The second solution $\cE_{[\frac{5}{2},0,0,0]}$ was shown in  \cite{Minimal} to correspond to a chiral invariant in six dimensions, which explains that the corresponding seven-dimensional invariant cannot be defined as a harmonic superspace integral in the linearised approximation. 

The solution to \eqref{1000} can be defined in terms of a vector $Z_i = V_i{}^I n_I$ such that 
\be \cD_{ij} Z^k = \frac{1}{2} \delta^k_{(i} Z_{j)} - \frac{1}{10} \delta_{ij} Z^k \ . \ee
One computes that the function $(Z_i Z^i)^{-s}$ solves \eqref{1000}. However the associated Epstein series 
\be E_{[s,0,0,0]} = \sum_{n\in \mathds{Z}^5} (Z_i(n) Z^i(n))^{-s} \ , \ee
diverges at $s=\frac{5}{2}$, and one must consider the regularised series \cite{Green:2010wi}
\be \hat{E}_{[\frac{5}{2},0,0,0]} = \lim_{\epsilon\rightarrow 0} \Scal{ E_{[\frac{5}{2}+\epsilon,0,0,0]}- \frac{4\pi^2}{3\epsilon}} \ , \ee
that satisfies the equation 
\be \cD_{ik} \cD^{jk}  \hat{E}_{[\frac{5}{2},0,0,0]}  = -\frac{3}{4} \cD_i{}^j  \hat{E}_{[\frac{5}{2},0,0,0]} +\frac{8\pi^2}{15} \delta_i^j \ .  \ee
Given any contravariant vector $m^I$ with $Z(m)^i = V^\inv_I{}^i m^I$, one computes that 
\be \cD_{ik} \cD^{jk} \mbox{ln}\scal{  Z(m)^l Z(m)_l } = -\frac{3}{4} \cD_i{}^j  \mbox{ln}\scal{ Z(m)^l Z(m)_l  } + \frac{2}{5} \delta_i^j \ ,  \ee
such that the relevant function to define the string theory Wilsonian action in \eqref{7Dresume} is 
\be \cE'_{\scriptscriptstyle \frac{1}{4}} = \hat{E}_{[\frac{5}{2},0,0,0]}  -\frac{4\pi^2}{3} \mbox{ln}\scal{  Z(m)^l Z(m)_l }  \ . \ee
As explained in \cite{Minimal}, this additional function defines a consistent anomaly for the continuous $SL(5)$ Ward identity, because the $\mathfrak{sl}_5$ variation of this function solves \eqref{1000} for $s=\frac{5}{2}$ by construction, whereas the function itself does not. We therefore conclude that this contribution comes from the 2-loop supergravity amplitude \cite{Bern:1998ug}. 

Similarly, the solution to \eqref{0010} can be defined in terms of an antisymmetric tensor  $Z^{ij} = V^\inv_I{}^i V^\inv_J{}^j n^{IJ}$ satisfying to the constraint 
\be n^{[IJ} n^{KL]} = 0 \ , \qquad Z^{[ij} Z^{kl]} = 0 \ , \ee
such that 
\be \cD_{ij} Z^{kl} =  -\delta^{[k}_{(i} Z_{j)}{}^{l]}  + \frac{1}{5} \delta_{ij} Z^{kl} \ . \ee
One computes that the function $(Z_{ij} Z^{ij})^{-s}$ solves \eqref{0010}. However the associated Eisenstein series 
\be E_{[0,0,s,0]} = \sum_{n\in \mathds{Z}^{10}|n\wedge n=0} (Z_{ij}(n) Z^{ij}(n))^{-s} \ , \ee
diverges at $s=\frac{5}{2}$, and one must consider the regularised series \cite{Green:2010wi}
\be \hat{E}_{[0,0,\frac{5}{2},0]} = \lim_{\epsilon\rightarrow 0} \Scal{ E_{[0,0,\frac{5}{2}+\epsilon,0]}- \frac{2\pi^5}{9\epsilon}} \ , \ee
that satisfies to the equation 
\be \cD_{ik} \cD^{jk}  \hat{E}_{[\frac{5}{2},0,0,0]}  = \frac{1}{4} \cD_i{}^j  \hat{E}_{[\frac{5}{2},0,0,0]} +\frac{2\pi^5}{15} \ .  \ee
For a given covariant rank one antisymmetric tensor $m_{IJ}$, one computes similarly that for $Z_{ij} = V_i{}^I V_j{}^J m_{IJ}$ 
\be \cD_{ik} \cD^{jk} \mbox{ln}\scal{ Z(m)_{ij} Z(m)^{ij} }= \frac{1}{4} \cD_i{}^j  \mbox{ln}\scal{ Z(m)_{ij} Z(m)^{ij} }+\frac{3}{10} \ ,  \ee
such that the relevant function to define the string theory Wilsonian action in \eqref{7Dresume} is 
\be \cE_\gra41 = \frac{6}{\pi^3} \hat{E}_{[\frac{5}{2},0,0,0]}- \frac{8\pi^2}{3}   \mbox{ln}\scal{ Z(m)_{ij} Z(m)^{ij} } \ . \ee
In the same way, the additional function defines a consistent anomaly for the continuous $SL(5)$ Ward identity. The specific $m_{IJ},\, m^I$ that define the logarithm function of the scalar appearing in the 2-loop supergravity amplitude must depend of the specific parametrisation of the symmetric space $SL(5)/SO(5)$, and this ambiguity amounts to a choice of renomalisation scheme. 

\subsection{Decompactification limit to eight dimensions}
To make link with the analysis of section \ref{8Dd4R4Explicit}, we will now solve equations \eqref{1000} and \eqref{0010} in the parabolic gauge associated to the large compactification radius limit, with the graded decomposition 
\be \mathfrak{sl}_5 \cong ( {\bf 2}\otimes \overline{\bf 3})^\ord{-5} \oplus \scal{ \mathfrak{gl}_1 \oplus \mathfrak{sl}_2 \oplus \mathfrak{sl}_3}^\ord{0} \oplus  ( {\bf 2}\otimes {\bf 3})^\ord{5} \ . \ee
We consider therefore the $SL(5)$ representative  $\cV$ in this gauge such that 
\be \label{A4:A1A2}
\cV = \left( \begin{array}{cc} \ e^{-3\phi} v^\inv_j{}^\alpha  \ &  \ 0 \ \\ 
\  e^{2\phi} V^a{}_K a_j^K & \ e^{2\phi} V^{a}{}_J \ \end{array} \right) \ , 
\ee
with the indices $\alpha, \beta$ running from 1 to 2 of the  local $SO(2)$, $i, j$ from 1 to 2 of the rigid $SL(2)$, $a, b$ from 1 to 3 of the local $SO(3)$ and $I, J$ from 1 to 3 of the rigid $SL(3)$. We decompose the Maurer--Cartan form into symmetric and antisymmetric components 
\be  d \cV \cV^{-1} = P + B  \ee
to obtain the symmetric traceless scalar momentum 
 \be {\bf P} =  \left( \begin{array}{cc} - 3 d\phi \delta_\alpha^\beta - P_\alpha{}^\beta \ &  \ \frac{1}{2} e^{5\phi} v_\alpha{}^i V^b{}_I  da^I_i  \ \\ 
 \ \frac{1}{2} e^{5\phi} v^{\beta i} V_{aI}  da^I_i  \ & \  2 d\phi \delta_a^b + P_a{}^b  \ \end{array} \right) \ ,  \ee
and the antisymmetric spin-connexion ${\bf B}$. The metric on the symmetric space is defined as
\be  ds^2 \equiv 2 \tr {\bf P}^2 = 60 d \phi^2  + e^{10 \phi} \mu^{i j} M_{I J} da_{i}^{J} da_{j}^{J} + 2 P_{\alpha \beta} P^{\alpha \beta} + 2 P_{ab} P^{a b} \quad \ . \ee
The differential operator in tangent frame  ${\bf D}$  is
\be {\bf D}  = \left( \begin{array}{cc}-  \frac{1}{20} \partial_\phi  \delta_\alpha^\beta - {\cD}_\alpha{}^\beta \ &  \ \frac{1}{2} e^{-5\phi} v^\inv_{i\alpha} V^{\inv I b} \partial_I^i   \ \\ 
 \ \frac{1}{2} e^{-5\phi} v^\inv_{i}{}^\beta V^{\inv I}{}_a \partial_I^i   \ & \ \frac{1}{30} \partial_\phi \delta_a^b +\cD_a{}^b  \ \end{array} \right) \ ,  \ee
with by construction 
\be
{\bf D}^\upnu \, 2 \tr  {\bf P}_\upmu  {\bf P}_\upnu = {\bf P}_\upmu  \ . 
\ee
As in the last section one defines the second order differential operator 
\be {\bf D}^2  \cE =  {\bf D}^\upmu \partial_\upmu \scal{ {\bf D}^\upnu \partial_\upnu \cE} -  {\bf D}^\upmu [ {\bf B}_\upmu ,  {\bf D}^\upnu ]  \partial_\upnu \cE \ , \ee
which we compute to be 
\bea {\bf D}^2 &=&  \left( \begin{array}{c} \ \scal{  \frac{1}{400} \partial_\phi^{\; 2} + \frac{1}{16} \partial_\phi + \frac{1}{2} \cD_{\gamma\delta} \cD^{\gamma\delta}  }   \delta_\alpha^\beta +{\cD}_\alpha{}^\beta \scal{ \frac{3}{4} + \frac{1}{10} \partial_\phi }  + \frac{1}{4} e^{-10\phi} v^\inv_{i\alpha} v^\inv_{j}{}^{\beta} M^{\inv IJ} \partial_I^i \partial_J^j  \  \\ 
-  \frac{1}{8} e^{-5\phi} \Scal{ v^\inv_i{}^\beta V^{\inv I}{}_a \scal{ 1 + \frac{1}{15} \partial_\phi}  -4 v^\inv_i{}^\beta V^{\inv I}{}_c \cD_a{}^c + 4 v^\inv_i{}^\gamma  V^{\inv I}{}_a  \cD_\gamma{}^\beta }   \partial_I^i \  \end{array} \right .   \CR
 &&   \qquad   \left .  \begin{array}{c} -  \frac{1}{8} e^{-5\phi} \Scal{ v^\inv_{i \alpha} V^{\inv I b} \scal{ 1 +\frac{1}{15} \partial_\phi}  - 4 v^\inv_{i\alpha} V^{\inv I c} \cD_c{}^b +4 v^\inv_{i\gamma} V^{\inv I b}  \cD_\alpha{}^\gamma }   \partial_I^i \   \\ 
 \ \scal{ \frac{1}{900}  \partial_\phi^{\; 2} + \frac{1}{24} \partial_\phi  }  \delta_a^b + \cD_a{}^c \cD_c{}^b + \cD_a{}^b \scal{ \frac{1}{2} + \frac{1}{15} \partial_\phi} + \frac{1}{4} e^{-10\phi} V^{\inv I}_a V^{\inv J b} M^\inv_{ij} \partial_I^i \partial_J^j  \  \end{array} \right ) \  \qquad   \eea
 \addtocontents{toc}{\protect\setcounter{tocdepth}{1}}
 \subsubsection{$\cE_{[s,0,0,0]}$ solution}
\addtocontents{toc}{\protect\setcounter{tocdepth}{2}}
 We shall consider first the solution for a function on $\mathds{R}_+^* \times SL(2)/SO(2)\times SL(3)/SO(3)$. In this case \eqref{1000} reduces to 
 \bea \frac{1}{10} {\cD}_\alpha{}^\beta \scal{\partial_\phi +15-6 s  }  \,  \cE  &=&   - \scal{ \textstyle  \frac{1}{400} \partial_\phi^{\; 2} + \frac{10-3s}{100} \partial_\phi + \frac{1}{2} \cD_{\gamma\delta} \cD^{\gamma\delta} - \frac{2s(2s-5)}{25} }  \delta_\alpha^\beta \,  \cE\ ,  \CR
 \cD_a{}^c \cD_c{}^b \, \cE +\cD_a{}^b \scal{ \textstyle \frac{12s-5}{20} +  \frac{1}{15} \partial_\phi } \cE&=&  -  \scal{\textstyle   \frac{1}{900} \partial_\phi^{\; 2} +\frac{6s+5}{300} \partial_\phi- \frac{2s(2s-5)}{25}  }  \delta_a^b \cE \ . \hspace{10mm}  \label{1000Decom}
 \eea
The two sides of the first equation must vanish separately, therefore one concludes that either the function does not depend on the complex scalar field $\tau$ parametrising $SL(2)/SO(2)$ and  $\cE  = e^{-4s\phi} \cE(t)$ or $\cE  = e^{8(2s-5)\phi} \cE(t)$ , or $\cE = e^{3(2s-5) \phi} \cE(\tau,t)$ and 
\be \Delta_{SL(2)} \cE(\tau,t) =\frac{(2 s-3)(2s-5)}{4} \cE(\tau,t) \ . \ee 
 We note moreover that a function satisfying to the second equation must satisfy to \eqref{EisenQuadra} for some $s^\prime$, \ie 
\be \cD_a{}^c \cD_c{}^b  \cE_{s^\prime} = - \frac{4 s^\prime-3}{12}  \cD_a{}^b \cE_{s^\prime}  + \frac{s^\prime ( 2 s^\prime-3)}{9}  \delta_a^b  \cE_{s^\prime} \ .  \ee
Only for $s^\prime = 0$ or $\frac{3}{2}$, one can have an additional constant term \cite{Minimal}. Using this one finds the unique compatible solution 
\be \cE =  e^{-4s\phi} \cE_s(t)  + e^{3(2s-5) \phi} \cE_{\frac{5}{2}-s} (\tau) \ . \ee
The dependence in $a^I_i$ can be determined for each Fourier momentum $e^{iq_I^i a^I_i}$ separately. The equation in the ${\bf 10}$ implies the 1/2 BPS constraint \cite{Ferrara:1997ci}
\be \varepsilon_{ij} q^i_I q^j_J = 0 \ , \ee
and one can define the invariant mass $|Z(q)| = \sqrt{ \mu^{\inv}_{ij} M^{\inv IJ} q^i_I q^j_J }$. The rank one $2\times 3$ matrix then factorises, such that one can define 
\be q^{i}_I = p^i m_I \ , \label{pqDef} \ee
from which we can define the invariant mass $|\xi|= \sqrt{ \mu^\inv_{ij} p^i p^j}$. In principle one could expect a dependence in the $SL(2)$ factor of $SL(3)$ that leaves invariant $q^i_I$, but this is forbidden by the differential equation. One finds the general solution of  \eqref{1000} (which vanishes at large $e^{-5\phi} |Z|$)
as a function of $|Z|,\, |\xi|$ and $ \phi$ \be \cE_q =  e^{-6\phi} \biggl(  e^{ \phi} \frac{|Z(q)|}{|\xi(p)|^2\hspace{-1.5mm}} \biggr)^{s-\frac{3}{2}}    K_{s-\stfrac{3}{2}}(e^{-5\phi} |Z(q)|) e^{i q_I^i a^I_i} \ . \ee
These results are in agreement with the constant term formula for the corresponding Eisenstein series \cite{Green:2010wi}, and one computes using Poisson summation formula
\bea   E_{[s,0,0,0]}  &=&  e^{-4s\phi} E_{[s,0]}(t)  +\pi^{\frac{3}{2}}\frac{ \Gamma(s-\tfrac{3}{2})}{\Gamma(s)}  e^{3(2s-5) \phi} E_{[s-\stfrac{3}{2}]}(\tau) \CR
&& \qquad + \frac{2\pi^s}{\Gamma(s)} \sum_{n\in \mathds{Z}^2_*,\, m\in \mathds{Z}^3_*}    e^{-6\phi} \biggl(  e^{2 \phi} \frac{M^{\inv IJ} m_I m_J}{\mu^\inv_{ij} n^i n^j} \biggr)^{\frac{2s-3}{4}}  K_{s-\stfrac{3}{2}}(2\pi e^{-5\phi} |Z(m\otimes n)|) e^{2\pi i m_I n^i a^I_i} \CR
&=&  e^{-4s\phi} E_{[s,0]}(t)  +\pi^{\frac{4s-5}{2}}\frac{ \Gamma(\tfrac{5}{2}-s)}{\Gamma(s)}  e^{3(2s-5) \phi} E_{[\stfrac{5}{2}-s]}(\tau) \CR
&& \qquad + \frac{2\pi^s}{\Gamma(s)} \sum_{q\in \mathds{Z}^6|q\wedge q=0}    e^{-6\phi} \sum_{p|q}  \biggl(  e^{ \phi} \frac{|Z(q)|}{|\xi(p)|^2\hspace{-1.5mm}} \biggr)^{s-\frac{3}{2}}     K_{s-\stfrac{3}{2}}(2\pi e^{-5\phi} |Z(q)|) e^{2\pi i q_I^i a^I_i} \ . 
\eea
 Note that the dependence in the specific integral vector $p^i = r p^{\prime i}$ does not depend on the scalar fields, and defining $p^{\prime i}$, the solution to \eqref{pqDef} such that $p^{\prime 1}$ and $p^{\prime 2}$ are relative primes, one has 
\be \sum_{p|q}  \biggl(  e^{ \phi} \frac{|Z(q)|}{|\xi(p)|^2\hspace{-1.5mm}} \biggr)^{s-\frac{3}{2}}   = \biggl( \sum_{r|q}   r^{2-3s}    \biggr)      \biggl(  e^{ \phi} \frac{|Z(q)|}{|\xi(p^\prime)|^2\hspace{-1.5mm}} \biggr)^{s-\frac{3}{2}}   \ . \ee
For $s=\frac{3}{2}$, we get back the property that the solution behaves like an eight-dimensional threshold 
\be \int d^7 x \sqrt{-g} e^{-2(3+n)\phi} \cE(t) \nabla^{2n} R^4 \sim \int d^8 x \sqrt{-g} \cE(t) \nabla^{2n} R^4  \ \label{Roxi78} , \ee
 with 
 \be \cE =  e^{-6\phi} \cE_\frac{3}{2}(t)  + e^{-6 \phi} \cE_{1} (\tau) \ . \ee
 Although the value $s=\frac{3}{2}$ does not define a smaller representation of $SL(2)\times SL(3)$ as in lower dimensions, we see nonetheless that the Fourier modes simplify at this value, and become a function of $\phi$ and $|Z(q)|$ only. The expansion of the Eisenstein series  
\begin{multline}  {E}_{[\frac{3}{2},0,0,0]} = e^{-6\phi} \Scal{ \hat{E}_{[\frac{3}{2},0]}(t) + 2   \hat{E}_{[1]}(\tau) -20 \pi \phi}\\ + 4\pi  \sum_{q\in \mathds{Z}^6|q\wedge q=0}    e^{-6\phi} \biggl( \sum_{r|q} 1 \biggr)   K_{0}(2\pi e^{-5\phi} |Z(q)|) e^{2\pi i q_I^i a^I_i}  \ .  \end{multline} 
includes the additional function linear in the dilatons 
\be \cE =e^{-6\phi} \biggl(  10 \phi + \mbox{ln}\Bigl( \frac{M_{IJ} m^I m^J}{\mu^{ij} n_i n_j}\Bigr) \biggr)  \ , \ee
which is also an exact solution to \eqref{1000Decom}.  
 
For $s=\frac{5}{2}$, we get the function associated to the invariant that cannot be written as a harmonic superspace integral in the linearised approximation 
 \be \cE =  e^{-10\phi} \cE_\frac{5}{2}(t)  + \cE_{1} (\tau) \ . \ee
The regularised Eisenstein function decomposes as \cite{Green:2010wi}
\begin{multline}  \hat{E}_{[\frac{5}{2},0,0,0]} = e^{-10\phi} E_{[1,0]}(t) + \frac{4\pi}{3} \hat{E}_{[1]}(\tau) +8\pi^2 \phi\\ + \frac{8\pi^2}{3} \sum_{q\in \mathds{Z}^6|q\wedge q=0}    e^{-6\phi} \sum_{p|q}  \biggl(  e^{ \phi} \frac{|Z(q)|}{|\xi(p)|^2\hspace{-1.5mm}} \biggr)  K_{1}(2\pi e^{-5\phi} |Z(q)|) e^{2\pi i q_I^i a^I_i}  \ .  \end{multline}
 \addtocontents{toc}{\protect\setcounter{tocdepth}{1}}
 \subsubsection{$\cE_{[0,0,s,0]}$ solution}
\addtocontents{toc}{\protect\setcounter{tocdepth}{2}}

Let us now consider equation \eqref{0010}, which reduces on $\mathds{R}_+^* \times SL(2)/SO(2)\times SL(3)/SO(3)$ to 
 \bea \frac{1}{10} {\cD}_\alpha{}^\beta \scal{\partial_\phi +2s+5  }  \,  \cE  &=&   - \scal{ \textstyle  \frac{1}{400} \partial_\phi^{\; 2} + \frac{s+5}{100} \partial_\phi + \frac{1}{2} \cD_{\gamma\delta} \cD^{\gamma\delta} - \frac{3s(2s-5)}{25} }  \delta_\alpha^\beta \,  \cE\ ,  \CR
 \cD_a{}^c \cD_c{}^b \, \cE +\cD_a{}^b \scal{ \textstyle \frac{15-4s}{20} +  \frac{1}{15} \partial_\phi } \cE&=&  -  \scal{\textstyle   \frac{1}{900} \partial_\phi^{\; 2} +\frac{15-2s}{300} \partial_\phi- \frac{3s(2s-5)}{25}  }  \delta_a^b \cE \ . \hspace{10mm} 
 \eea
In the same way we get that the two sides of the first equation must vanish separately, such that either the function does not depend on the complex scalar field $\tau$ parametrising $SL(2)/SO(2)$ and  $\cE  = e^{-12s\phi} \cE(t)$ or $\cE  = e^{4(2s-5)\phi} \cE(t)$ , or $\cE = e^{-(2s+5) \phi} \cE(\tau,t)$ and 
\be \Delta_{SL(2)} \cE(\tau,t) =\frac{(2 s-1)(2s-3)}{4} \cE(\tau,t) \ . \ee 
 Using moreover that a function satisfying to the second equation must satisfy to \eqref{EisenQuadra} for some $s^\prime$, we get the general solution 
 \be \cE_{[0,0,s,0]}  = e^{-12s \phi} + e^{4(2s-5)\phi} \cE_{s-1}(t) + e^{-(2s+5)\phi} \cE_{s-\stfrac{1}{2}}(\tau) \cE_{2-s}(t)  \ . \ee
 This solution is consistent with the constant term formula \cite{Green:2010wi}, \ie 
 \begin{multline}  E_{[0,0,s,0]} =2 \zeta( 2s)\zeta(2s-1)  e^{-12s \phi} + \frac{\pi^2 \zeta(2s-3)}{(s-1)(s-\tfrac{3}{2})}  e^{4(2s-5)\phi} E_{[s-1,0]}(t) \\  +\frac{\pi^{2s-2}\Gamma(2-s)}{2\Gamma(s)}  e^{-(2s+5) \phi} E_{[s-\stfrac{1}{2}]}(\tau) E_{[2-s,0]}(t)  + \mathcal{O}(e^{- e^{-5 \phi}}) \ . \end{multline}

 For $s=\frac{5}{2}$ we  get the function associated to the invariant that can be written as a harmonic superspace integral in the linearised approximation 
 \be \cE_{[0,0,\stfrac{5}{2},0]}  =  e^{-30 \phi} + \cE_{\frac{3}{2}}(t) + e^{-10 \phi} \cE_{2}(\tau) \cE_{-\frac{1}{2}}(t)  \ . \ee

\subsection{Decompactification limit to ten dimensions}
The decompactification limit to type IIB supergravity in seven dimensions can be obtained in the same way as in \eqref{A4:A1A2} for the inverse matrix
 \be
\cV = \left( \begin{array}{cc} \ e^{3\phi} v_\alpha{}^j  \ &  \ e^{3\phi} v_\alpha{}^k B_k^J \ \\ 
\  0\ & \ e^{2\phi} V^{\inv J}{}_a \ \end{array} \right) \ , 
\ee
such that $v_\alpha{}^j(\tau)$ is now parametrised by the string coupling constant complex modulus $\tau$. One obtains therefore as in the last section that 
\bea   E_{[s,0,0,0]} &=&  e^{-6 s \phi} E_{[s]}(\tau)   +   \pi^{2s-\frac{5}{2}} \frac{ \Gamma(\tfrac{5}{2}-s)}{\Gamma(s)}  e^{(4s-10)\phi} E_{[\frac{5}{2}-s,0]}(g_{T^3})  \\
&& \quad + \frac{2\pi^{s}}{\Gamma(s)} \sum_{q\in \mathds{Z}^6|q\wedge q=0}    e^{-6\phi} \sum_{p|q}  \biggl(  e^{ \phi} \frac{|Z(q)|}{|\xi(p)|^2\hspace{-1.5mm}} \biggr)^{1-s}     K_{s-1}(2\pi e^{-5\phi} |Z(q)|) e^{2\pi i q_I^i B^I_i} \ , \nn
\eea
where the first term defines the exact $R^4$ and $\nabla^4 R^4$ type IIB threshold functions for $s=\frac{3}{2}$ and $s=\frac{5}{2}$, respectively. According to the Kaluza--Klein reduction 
\be \int d^7x \, e^{-6(3+k)\phi} \cE(\tau)  \nabla^{2k} R^4 \sim \int d^{10} x \,  \cE(\tau)  \nabla^{2k} R^4 \ , \ee
it follows that a $\cE_{[\stfrac{3}{2},0,0,0]} R^4$ type invariant in seven dimensions can lift to a $\cE_{[\stfrac{3}{2}]}(\tau) R^4$ type invariant in ten dimensions, and  a $\cE_{[\stfrac{5}{2},0,0,0]} \nabla^4 R^4$ type invariant can lift to a $\cE_{[\stfrac{5}{2}]}(\tau) \nabla^4 R^4$ type invariant. 

However 
 \be \cE_{[0,0,\stfrac{5}{2},0]}  = 1+ e^{-20\phi} \cE_{-1}(g_{T^3}) + e^{-5\phi} \cE_{-\stfrac{1}{2}}(\tau) \cE_{2}(g_{T^3})  \ , \ee
and there is no solution that lifts to ten dimensions such that no $\cE_{[0,0,\stfrac{5}{2},0]} \nabla^4 R^4$ type invariant in seven dimensions does lift to type IIB supergravity.  

To understand the decompactification limit to IIA supergravity, it is more convenient to take an explicit basis for the diagonal elements of the matrix $\cV\in SL(5)$, \ie 
\be \cV_1{}^1 =  y_7^{\, \frac{2}{5}} \ , \quad \cV_2{}^2 = \frac{y_7^{- \frac{1}{10}}}{\sqrt{r_8 r_B}} \ , \quad \cV_3{}^3 = y_7^{- \frac{1}{10}} \sqrt{ \frac{r_B}{r_8} } \ , \quad  \cV_4{}^4 = y_7^{- \frac{1}{10}} \sqrt{ \frac{r_8}{r_A} } \ , \quad  \cV_5{}^5 = y_7^{- \frac{1}{10}} \sqrt{ r_8 r_A  } \ ,\ee
where $y_7$ is the effective string coupling constant in seven dimensions, whereas $r_8$ and $r_B$ are the radii moduli in type IIB and $r_8$ and $r_A$ the radii moduli in type IIA. In this basis, the only solutions that lift to ten dimensions  are 
\bea \cE_{[\stfrac{3}{2},0,0,0]} &=& e^{-6\phi} \cE_\frac{3}{2}(t) = y_7^{-\frac{1}{5} } \Scal{\,  \frac{1}{y_7}  + r_8 r_B }\ ,  \CR
\cE_{[\stfrac{3}{2},0,0,0]} &=&  e^{-6 \phi} \cE_{1} (\tau) = y_7^{-\frac{1}{5} } \,  r_8  r_A  \ , \eea
with arbitrary coefficients, which shows that the eight-dimensional threshold $\cE_\frac{3}{2}(t) R^4$ includes both type IIA and IIB tree-level $R^4$ thresholds, and the 1-loop type IIB $R^4$ threshold, whereas $\cE_{1} (\tau)   R^4$ includes the type IIA $R^4$ threshold that lifts to eleven dimensions. Similarly, the only solutions that lift to ten dimensions are 
\bea \cE_{[\stfrac{5}{2},0,0,0]} &=& e^{-10\phi} \cE_\frac{5}{2}(t) = y_7^{-1} \Scal{ \, \frac{1}{y_7}  +y_7  ( r_8 r_B)^2  } \ ,  \CR
\cE_{[0,0,\stfrac{5}{2},0]} &=&  e^{-10 \phi} \cE_{2}(\tau) \cE_{-\frac{1}{2}}(t)= y_7^{-1 } \, y_7  ( r_8 r_A)^2   \ , \eea
with arbitrary coefficients, which shows that the eight-dimensional threshold $\cE_\frac{5}{2}(t)  \nabla^4 R^4$ includes both type IIA and IIB tree-level $\nabla^4 R^4$ thresholds, and the 2-loop type IIB $\nabla^4 R^4$ threshold, whereas $\cE_{2}(\tau) \cE_{-\stfrac{1}{2}}(t) \nabla^4 R^4 $ includes the type IIA $\nabla^4 R^4$ threshold.

In type IIB, supersymmetry implies  a second order Poisson equation on $SL(2)/SO(2)$, such that the two invariants must be in the same $SL(2)$ representation, whereas in type IIA supergravity there is only one scalar, and they are independent. 

\section*{Acknowledgment}
We thank K.S. Stelle for interesting discussions. This work was supported by the French ANR contract 05-BLAN-NT09-573739 and the ERC Advanced Grant no. 226371.

\end{document}